\documentclass[11pt]{book}
\usepackage{epsfig}
\usepackage{cite}
\usepackage{amssymb}
\usepackage{float}
\newcommand{\mbf}[1]{\mbox{\boldmath$#1$\unboldmath}}
\newcommand{\HopPaper}{I{}}
\newcommand{\VigoPaper}{II{}}
\newcommand{\QuenchPaper}{III{}}
\newcommand{\HopPRL}{IV{}}
\newcommand{\HopRMP}{V{}}

\newcommand{\HkIndexAPP}{A{}}

\title{Hopping in Disordered Media:\\
A Model Glass Former\\
and\\
A Hopping Model
}
\author{
Thomas B. Schr\o der (tbs@mmf.ruc.dk) \\
Department of Mathematics and Physics,\\ 
Roskilde University, DK-4000 Roskilde, Denmark}

\date{
Ph.D. thesis\\
Supervisor: Jeppe C. Dyre\\
\today}

\begin{document}
\maketitle

\setcounter{chapter}{1}
\thispagestyle{plain}
\section{Abstract}
Two models involving particles moving by ``hopping'' in 
disordered media are investigated:

A model glass-forming liquid is investigated by 
molecular dynamics under (pseudo-) equilibrium conditions. 
``Standard'' results such as 
mean square displacements, intermediate scattering functions, etc.
are reported. At low temperatures hopping is present in 
the system as indicated by a secondary peak in  the distribution
of particle displacements during a time interval $t$. 
The dynamics of the model is analyzed in terms of its 
potential energy landscape (potential energy as function of the $3N$
particle coordinates), and we present direct numerical evidence for a 
30 years old picture of the dynamics at sufficiently low temperatures.
Transitions between local  potential energy minima in configuration
space are found to involve particles moving in a cooperative string-like 
manner. 
 
In the symmetric hopping model particles are moving on a lattice by 
doing thermally activated hopping over energy barriers connecting
nearest neighbor sites. This model is analyzed in the extreme disorder
limit (i.e. low temperatures) using the Velocity Auto Correlation (VAC) 
method. The VAC method is developed in this thesis and has the advantage over
previous methods, that it can calculate a diffusive regime in 
finite samples using periodic boundary conditions. 
Numerical results using the VAC method are compared
to three analytical approximations, including the Diffusion Cluster
Approximation (DCA), which is found to give excellent agrement with the 
numerical results.

\tableofcontents

\chapter{Introduction}

\begin{figure}
\centerline {\epsfig{file=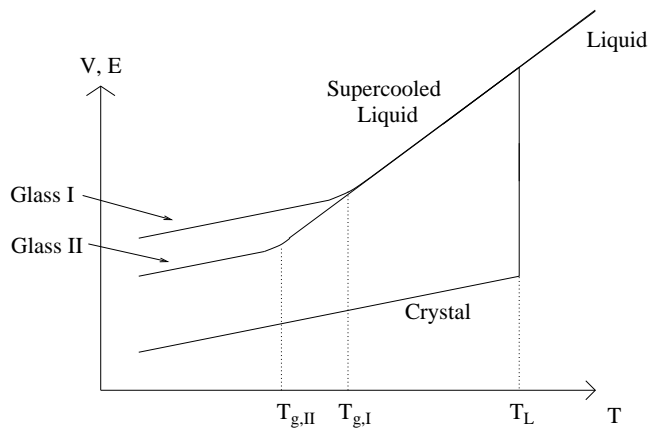, angle=0, width=8.5cm}}
\caption{
Schematic overview of the behavior of glass-forming liquids. 
Upon cooling crystallization is avoided and the liquid becomes ``supercooled'', 
which is a (pseudo-) equilibrium state (the crystal being the 
real equilibrium state).
Upon further cooling the liquid falls
out of equilibrium and undergoes a glass-transition at the temperature
$T_g$. 
$T_g$ depends on cooling
rate; 
glass II is obtained by a lower cooling rate than glass I.
}
\label{fig:Glass}
\end{figure}

This Ph.D. thesis deals with two models characterized by 
particles moving by ``hopping'' in disordered media. 
The first model is a simple model glass-former, investigated 
here by computer simulations using molecular dynamics \cite{Allen86}.
Glass forming liquids are liquids 
that upon cooling falls out of equilibrium 
and form a glass, see figure \ref{fig:Glass}. For general reviews on 
glass-forming liquids see \cite{Jones56,Harrison76,Brawer85,Debenedetti96,Ediger96}. 
For reviews on computer simulations of glass-forming liquids see
\cite{Angell81,Barrat91,Hiwatari91,Hansen93,Kob96}.

The normal liquid behavior is to crystallize upon cooling below the melting 
temperature, $T_L$. In a large class of liquids the 
crystallization can be avoided (e.g. by fast cooling) and the liquid becomes 
``supercooled''. A supercooled liquid is in a pseudo-equilibrium state;
The crystal is the real equilibrium state, but the supercooled state 
is typically stable on very long time scales. The  supercooled state 
is characterized by relaxation times increasing strongly with decreasing 
temperature. If cooled with a constant cooling rate this means that the 
liquid at some temperature falls out of equilibrium and undergoes a 
glass-transition and forms a glass, which is a disordered solid.
The glass-transition temperature, $T_g$, depends on the cooling
rate; a lower cooling rate results in a lower glass transition temperature.
By convention the ``laboratory'' glass transition is taken to be where
the relaxation time $\tau$ is on the order of 100 sec (i.e. the ``typical'' time
scale in the laboratory)\footnote{The relaxation time is related to the viscosity 
$\eta$ by $\eta=G_\infty \tau$, where $G_\infty$ is the instantaneous shear modulus.
$\tau \approx 100s$ typically corresponds to $\eta \approx 10^{13} \mbox{poise}$, which is
sometimes used as the definition of the laboratory glass transition.}.

Binary systems with particles 
differing about 20\% in diameter have been found to be good candidates 
for simple models of glass-forming liquids  in computer simulations 
\cite{Miygawa88,Roux89,Barrat90,Hiwatari91,Barrat91,Wahn91,Thirumalai93,Kob95a,Kob95b}. 
At sufficiently low  temperatures particles in these systems are found to move by 
``hopping'', i.e. they are localized for
a period of time, and then move more or less directly to another
position, where they again become localized 
\cite{Miygawa88,Roux89,Barrat90,Hiwatari91,Barrat91,Wahn91,Thirumalai93}.
Since hopping of some form is expected to play a increasingly dominant
role as temperature is lowered \cite{Goldstein69, Stillinger95}, systems 
exhibiting hopping at temperatures which can be reached under equilibrium
conditions  in computer simulations are of special interest.
In this thesis we investigate the dynamics of a binary Lennard-Jones
liquid, which has earlier been shown to exhibit hopping \cite{Wahn91}. 
This is done under (pseudo-) equilibrium conditions, i.e. above the 
glass transition temperature.

The model glass-former investigated in this thesis is \emph{not} a 
model of any particular liquid existing in the laboratory (or the 
real world for that matter). Like most models glass formers studied  
in computer simulations, 
it should be thought of as a (very) simple model exhibiting some of 
the complex behavior found in real glass forming liquids. The reason 
why it is interesting to investigate such a simple model is two-fold:
I) The  glass-transition is found to occur in liquids that are chemically 
very different, but the related phenomenology is found to be strikingly similar.
Since liquids with different chemical details behave in 
a similar way, it makes sense to study simple models ignoring the 
chemical details, in an attempt to understand the fundamental mechanisms.
II) Most theories for glass-forming liquids also treat these in a highly 
simplified manner, ignoring  the chemical details 
(for the same reasons given above).
Consequently computer simulations makes it possible to test these
theories on their own terms, i.e. without any additional approximations. 
One particular example of this is the Mode Coupling Theory (MCT) 
\cite{Gotze92,Gotze99}. When doing computer 
simulations of glass forming liquids it is rather natural to compare the results with the 
predictions of the MCT since these are plenty and detailed, \emph{and}
MCT deals with the temperature range accessible (at the present) to computer 
simulations under equilibrium conditions. It is not a goal in it self to test 
MCT in the present thesis, but some of the results
are compared to the predictions of the MCT. 

The second model investigated in the present thesis is a so-called
``hopping-model'', where the hopping behavior of particles is ``built
in''; In the symmetric hopping model 
\cite{Dyre88,Avramov93,Dyre94,Argyrakis95,Stein95,Dyre96} particles are 
moving on a lattice
by doing thermally activated hopping over random energy barriers. This
model is found to have interesting universal features in the 
extreme disorder limit, i.e. when  the temperature goes to zero. 
The symmetric hopping model has earlier been  treated 
as a model for frequency dependent conduction in glasses \cite{Dyre88,Dyre94}, 
with particles representing non-interacting charge carriers 
(ions or electrons). In that context the universality manifests itself
as follows: 
At low temperatures the shape of the conduction vs. frequency curve  
becomes independent of temperature, which is also what is found in experiments
(see eg. \cite{Howell74,Nowick94,Roling97}). In the present thesis the symmetric hopping
model is treated in the slightly more general context of diffusion
in disordered media. 

\newpage
\section{Outline}

This thesis consists of  3 main chapters: 

In chapter 3 we report results from the 
simulations of the model glass-former mentioned above, 
with emphasis on the dynamics. In this chapter ``standard''
results are reported, i.e. the measures commonly used to describe  
(glass-forming) liquids. All of the features found for this 
particular model has been reported earlier for other models, and
this chapter thus constitutes a ``review'' of the behavior of 
model glass-formers, illustrating the similarities  in the behavior of these. 

In chapter 4 the dynamics of the model glass-former described in chapter
3 is investigated in term of its ``potential energy landscape'', which is simply 
the potential energy as a function of the $3N$ particle coordinates.
The new concept of ``inherent dynamics'' is introduced, and the 
results from applying this concept to the model described in chapter 3 are
discussed.

Chapter 5 deals with the symmetric hopping model. A new numerical method
for investigating the model in the extreme disorder limit is developed.
Numerical results are presented and  compared with 3 analytical 
approximations. 

The three main chapter (3-5) contains individual conclusions.

\section{Acknowledgments}

The present Ph.D. thesis is the result of work done at Department of Mathematics and Physics 
(IMFUFA) at Roskilde University (RUC). I would like to thank the people at 
IMFUFA, especialy    
Jeppe C. Dyre (my supervisor), Niels B. Olsen, Tage E. Christensen, Johannes K. Nielsen
and Ib H\o st Pedersen.

The spring of 1998 I spent as a guest researcher at 
Center for Theoretical and Computational Materials Science (CTCMS) at 
National Institute of Standards and Technology (NIST), Maryland, USA.
I would like to thank NIST and CTCMS for the hospitality, and I would like to thank
the people I met and worked with there: 
Sharon C. Glotzer, Srikanth Sastry, Jack F. Douglas, Claudio Donati, 
and Paulo Allegrini.

\newpage

\section{Papers}

The following papers contain results obtained for the two models discussed
in the present thesis: 

\begin{itemize}
  \item{Paper I:} \emph{Effective one-dimensionality of universal ac hopping conduction 
              in the extreme disorder limit.} J.C. Dyre and T.B. Schr\o der. 
              Phys. Rev. B. {\bf54} 14884 (1996).
  \item{Paper II:} \emph{Hopping in a supercooled binary Lennard-Jones liquid.} 
        T.B. Schr\o der and  J.C. Dyre. J. Non-Cryst. Solids. {\bf235-237} 331 (1998).
  \item{Paper III:} \emph{Crossover to potential energy landscape dominated dynamics 
                          in a model glass-forming liquid }. 
          T.B. Schr\o der, S. Sastry,  J.C. Dyre and
          S.C. Glotzer. J. Chem. Phys. {\bf112} (2000) in press.\\
         (http:$\backslash \backslash$xxx.lanl.gov: cond-mat/9901271)
 \item{Paper IV:} \emph{Scaling and Universality of ac Conduction in Disordered Solids.}
             T.B. Schr\o der and  J.C. Dyre. Phys. Rev. Lett.  {\bf84} 310 (2000).
  \item{Paper V:} \emph{Universality of ac conduction in disordered solids.}
             J.C. Dyre and T.B. Schr\o der. Rev. Mod. Phys. {\bf72} (2000) in press. 
\end{itemize}

Paper  I reports numerical results for the symmetric hoping model, 
which was done mainly during my master thesis 
\cite{SchroederSpecialePhys,SchroederSpecialeDat}. Paper II-V report
results obtained as part of the present Ph.D. thesis. Paper
IV and V was written after the conclusion of my Ph.D. (summer 1999).

\chapter{A Model Glass-former}

A glass-forming binary Lennard-Jones liquid is  investigated by 
molecular dynamics under (pseudo-) equilibrium conditions. In this chapter we
report 'standard' results for this model liquid, i.e. pair-correlation
functions, mean square displacements, intermediate scattering functions, etc.
The main result is, that at low temperatures hopping is present in 
the system as indicated by a secondary peak in  $4\pi r^2G_s(r,t)$, 
where $G_s(r,t)$ is the van Hove self correlation function. It has not
been possible to identify a temperature range, where the asymptotic 
predictions of the ideal mode coupling theory are fulfilled.
Some of the results reported in this chapter are contained
in paper \VigoPaper ~and paper \QuenchPaper.

\section{Model and Method}
\label{sec:method}

\begin{figure}
\epsfig{file=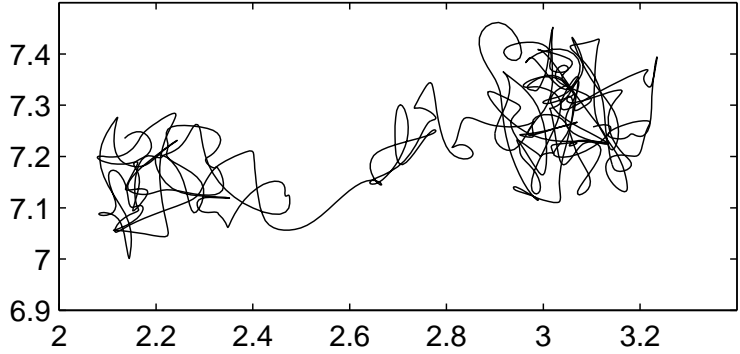, width=10cm}
\caption{
  Trajectory (in the x-z plane) of a ``hopping'' particle. 
  First, the particle stays relatively localized for a period of 
  time, seemingly oscillating around a position in the left side of the 
  figure. After this, the particle move more or less directly to a new
  position in the right of the figure, where it again become localized.
  The temperature is $T=0.59$, and the elapsed time is $\Delta t = 160$
  (see later for units of these numbers). 
} 
\label{fig:hop}
\end{figure}

The system investigated in the present work, has earlier been shown 
to exhibit hopping by Wahnstr\" om \cite{Wahn91}. By ``hopping'' is 
here meant, that particles behave like illustrated in figure \ref{fig:hop}; 
The particle shown stays relatively localized for a period of 
time, and then move some distance, where it again becomes localized.
The presence of hopping was the main motivation for investigating 
this system, and this kind of dynamics will be discussed in more detail 
in this and the following chapter.
 
In the work of Wahnstr\" om, the system was investigated both above
and below the glass-temperature, $T_g$, i.e. the system was allowed 
to fall out off equilibrium as it was cooled. 
As a consequence, it was unclear 
whether the hopping seen was a feature of the 
equilibrium liquid, or if it was a consequence of non-equilibrium 
dynamics. Here we attempt to keep the system equilibrated at all temperatures, 
i.e. the equilibration time is made longer and longer as the temperature
is lowered. 

The ``Wahnstr\" om system'' is a binary mixture of $N=500$ particles, 
with 50\% particles 
of type A, and 50\% particles of type B\footnote{The exact number of
particles used in the simulations are: 
$N_A=251$, and $N_B=249$}. The particles interact
via the pair-wise Lennard-Jones potential, where the parameters
depend on the types of the two particles involved ($\alpha$ and $\beta$): 
\begin{equation}
   V_{\alpha \beta}(r) = 4\epsilon_{\alpha \beta}
                      \left(
                        \left( \frac {\sigma_{\alpha \beta}}{r} \right) ^{12}
                        -
                        \left( \frac {\sigma_{\alpha \beta}}{r} \right) ^6
                      \right)
    \label{BLJpot}
\end{equation}
The forces are given by the negative gradient of the potential:
\begin{eqnarray}
   {\mathbf F}_{\alpha \beta}({\mathbf r}) 
   &=&  -\nabla V_{\alpha \beta}  
    =   -\frac{\partial V_{\alpha \beta}}{\partial r} 
         \frac{{\mathbf r}}{r} \\
   &=& \frac{48\epsilon_{\alpha \beta}}{(\sigma_{\alpha \beta})^2}
         \left(
            \left( \frac {\sigma_{\alpha \beta}}{r} \right) ^{14} -
            \frac{1}{2}\left( \frac {\sigma_{\alpha \beta}}{r} \right) ^8
         \right){\mathbf r}
\end{eqnarray}

$V_{\alpha \beta}(r)$ is characterized by a minimum at 
$V_{\alpha \beta}(2^{1/6}\sigma_{\alpha \beta}) = -\epsilon_{\alpha \beta}$, 
steep repulsion at shorter distances, and a weaker attraction at 
longer distances. 

The length-parameters used in the   Wahnstr\" om system 
are\footnote{We follow \cite{Kob94} 
and term the large particles  ``A'', and small particles ``B''}: 
$\sigma_{AA}=1$, $\sigma_{BB}=1/1.2\approx 0.833$
and $\sigma_{AB} = (\sigma_{AA}  + \sigma_{BB})/2$. The energy-parameters
are all identical: $\epsilon_{AA}=\epsilon_{AB}=\epsilon_{BB}=1$.
The masses of the particles are given by $m_A = 2$, and $m_B = 1$. 
The length of the sample was $L=7.28$, which gives a (reduced) density 
of $\rho = N/L^3 = 1.296$. Times are reported in units of 
$\tau \equiv (m_{B} \sigma_{AA}^2/48\epsilon_{AA})^{1/2}$ (This 
expression contains an error in paper \VigoPaper). 
When comparing simulations of a model-liquid like the one described
here with experimental results for real liquids, it is customary
to use ``Argon units'', i.e. parameters used when modeling 
Argon atoms by the Lennard-Jones potential; $\sigma = 3.4 \AA$, 
$\epsilon = 120K k_B$, and $\tau =  3\times10^{-13} sec$ \cite{Kob95a}.

The potential is ``cut and shifted'' \cite{Allen86} at 
$r=2.5\sigma_{\alpha \beta}$, 
which means that it is set to zero for $r\ge 2.5\sigma_{\alpha \beta}$,
and $|V_{\alpha \beta}(2.5\sigma_{\alpha \beta})|\approx 0.016$ is added
to the potential for $r\le 2.5\sigma_{\alpha \beta}$. This makes the 
potential continuous at $r=2.5\sigma_{\alpha \beta}$, while the 
force is not. The equations
of motion are integrated with periodic boundary conditions using the
Leap-Frog algorithm \cite{Allen86} (which is simply a discretizaition
of Newton's second law) with a time step of $0.01\tau$.
 
Three independent samples were used, each initiated by generating a 
random configuration followed by equilibration at a high temperature 
($T=5.0$). The cooling was done by controlling the total energy of the 
system, which was done by scaling  the velocities.
An equilibration run of length $t_{eq}$ was then performed, 
followed by a production run of the same length. If the 
samples was determined not to be equilibrated properly, 
$t_{eq}$ was doubled, by ``degrading'' the production 
run to be part of the equilibration run, and making a new
production run twice as long. This procedure was continued
until the samples was determined to be equilibrated. 
The criteria used for determining 
if the samples were  equilibrated will be discussed later.
Note, that by this procedure, it is the total energy 
($E_{tot}=E_{pot}+E_{kin}$) that is controlled. 
The reported temperatures and pressures are computed 
as time-averages over the instantaneous temperature and pressure
respectively \cite{Allen86}:
\begin{eqnarray}
   T(t) &\equiv& \frac{2E_{kin}(t)}{3N - 3}\\
   P(t) &\equiv& \rho T(t) + W(t)/V, 
                 ~~~W(t) \equiv \frac{1}{3}\sum_i\sum_{j>n} 
                 {\mathbf r}_{ij}\cdot{\mathbf F}_{ij}
\end{eqnarray}
where $W(t)$ is the virial and the summing is over all pair 
of particles.

Some of the results for the system described above, will be compared with 
results from a different binary Lennard-Jones mixture 
(the ``Kob \& Andersen system''), which has been 
investigated in a number of papers \cite{Kob94,Kob95a,Kob95b,Kob97,Donati98}.
This system is a 80:20 mixture, with $\sigma_{AA}=1$, $\sigma_{AB} = 0.8$,
$\sigma_{BB}=0.88$,
$\sigma_{AB} = 0.8$, $\epsilon_{AA}=1.0$, $\epsilon_{AB}=1.5$, 
$\epsilon_{BB}=0.5$, and $m_A = m_B = 1$.

\section{Static Results}

\begin{figure}
\epsfig{file=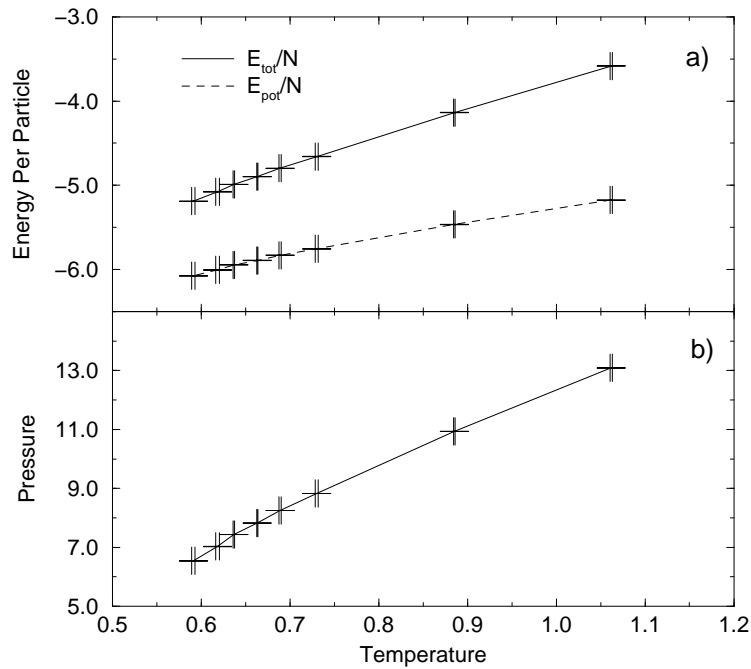, width=9.9cm}
\caption{a) Total and potential energy per particle as a function
of temperature.
b) Pressure vs. temperature. In both a) and b) (as in the following
if nothing else is mentioned)  error-bars are  estimated 
from deviation between three independent samples.
} 
\label{fig:EandPvsT}
\end{figure}

Figure \ref{fig:EandPvsT} shows as a function of temperature a)
the potential and total energy per particle, 
and b) the pressure vs. Temperature. The error-bars are estimated from deviations 
between the three independent samples, which are found to be in reasonable 
agreement. This is a necessary (but not sufficient) condition for the system(s)
to be equilibrated. Note that its the total energy
which was set to be equal for the three samples, and consequently there are
small deviations in the measured temperatures. 

The specific heat capacity at 
constant volume, $C_v$, is given by \cite{Allen86}:

\begin{equation} 
   C_v = \frac{1}{N}\frac{d E_{tot}}{dT}
   \label{Cv}
\end{equation}
From figure \ref{fig:EandPvsT}a we see that $C_v$ increases as the 
temperature is lowered. Note that there is no indication of a 
glass-transition, since this would be indicated by a sharp 
decrease in $C_v$ when cooling below the glass transition temperature, 
$T_g$, as seen in the results by Wahnstr\" ohm. The  heat capacity at 
constant volume will be discussed further in section \ref{sec:CvQ}.

\begin{figure}
  \epsfig{file=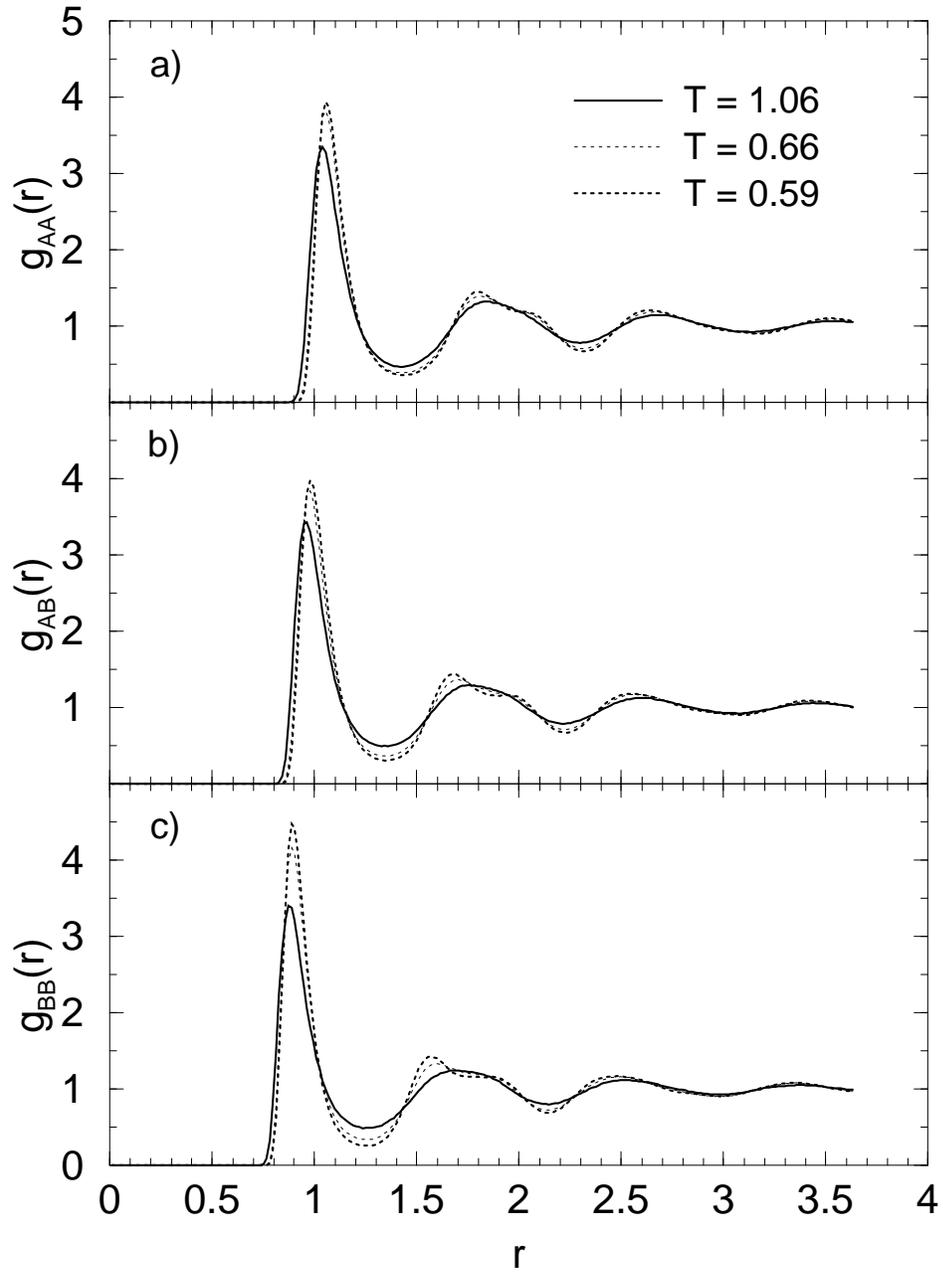, width=12.5cm}
  \caption{Pair correlation function at a high ($T=1.06$), 
   medium ($T=0.66$), and low temperature ($T=0.59$); 
   a) A-A correlation, $g_{AA}(r)$, 
   b) A-B correlation, $g_{AB}(r)$, 
   c) B-B correlation, $g_{BB}(r)$ 
   } 
   \label{fig:gr_all}
\end{figure}

In figure \ref{fig:gr_all} is shown the pair-correlation
functions, $g_{AA}(r)$, $g_{AB}(r)$, and $g_{BB}(r)$, at three of 
the temperatures simulated. The pair-correlation
function, $g_{\alpha \beta}(r)$, is the average relative density 
of particles of type $\beta$ around particles of type $\alpha$ \cite{Hansen86}.
It is normalized to be 1 at large distances, $r$, were the 
correlations disappears.
At the highest temperature ($T=1.06$) 
all three pair-correlation functions is seen to look like 'typical'
high temperature  pair-correlation functions, with a sharp 
first neighbor peak, a more rounded second neighbor peak, etc.
As the temperature is lowered the first neighbor peak becomes 
sharper and the second peak splits into two peaks. The splitting
of the second peak upon cooling is often seen in super-cooled liquids 
(see eg.\cite{Wahn91, Kob95a}).

The parameters used in the potential does not energeticly favor the 
mixture of A and B particles ($\epsilon_{AB}$ is not larger than 
$\epsilon_{AA}$ and $\epsilon_{BB}$), which might cause the system 
to phase separate at sufficiently low temperatures.
In the event of phase separation, we  would expect the area of the first 
peak in $g_{AB}(r)$ to decrease at the temperature where 
phase separation starts occurring.
There is no indication of this in figure \ref{fig:gr_all}.

\begin{figure}
\epsfig{file=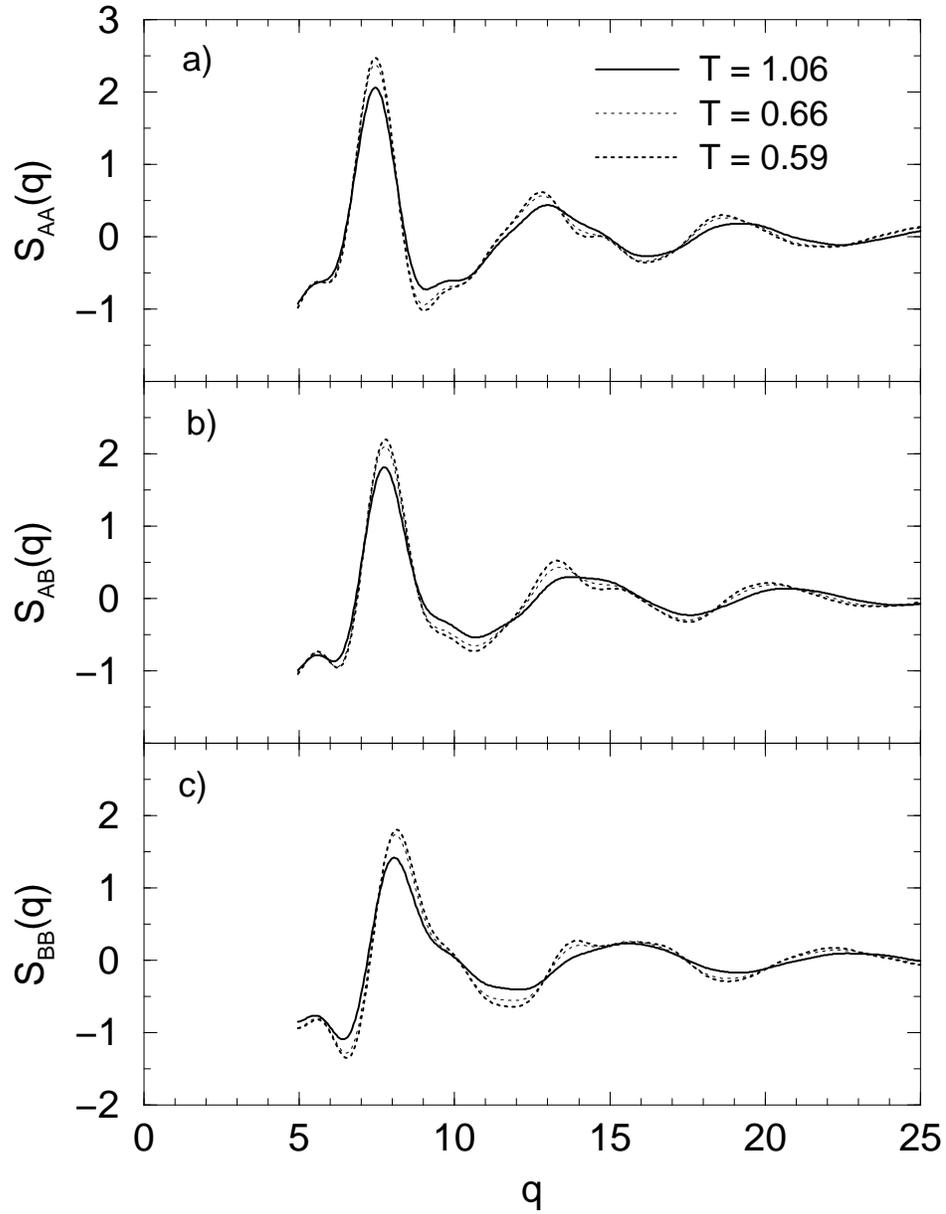, width=12.5cm}
\caption{Static structure factor at a high ($T=1.06$), 
medium ($T=0.66$), and low temperature ($T=0.59$); 
a) A-A correlation, $S_{AA}(q)$, 
b) A-B correlation, $S_{AB}(q)$, 
c) B-B correlation, $S_{BB}(q)$ 
} 
\label{fig:Sq_all}
\end{figure}

In figure \ref{fig:Sq_all} is shown the static structure factors, 
$S_{AA}(q)$, $S_{AB}(q)$, and $S_{BB}(q)$, corresponding
to the data shown in figure \ref{fig:gr_all}. 
The static structure factor is the (3 dimensional) Fourier transform 
of pair correlation function, and thus contains the same information.
However the finite sample size introduces some features in 
$S_{\alpha\beta}(q)$ which are clearly unphysical (see eg. \cite{Kob95b}).
By recalculating $S_{\alpha\beta}(q)$ from $g_{\alpha\beta}(r)$ with 
a cut-off in $r$ that is less than $L/2$ one can get an idea about
which features of $S_{\alpha\beta}(q)$ are due to finite-size effects.
Doing this shows, that at $q$-values to the left of the first peak
($q \approx 6$, depending on type of correlation)  
$S_{\alpha\beta}(q)$ is dominated by finite-size effects, which is
why most of it is not shown. Also the small ``wiggles'' seen at
$q \approx 10$ is a finite-size effect. The splitting of the 
peaks at $q \approx 15$ is however not a finite-size effect. 

\section{Mean Square Displacement}
\label{sec:R2}

\begin{figure}
\epsfig{file=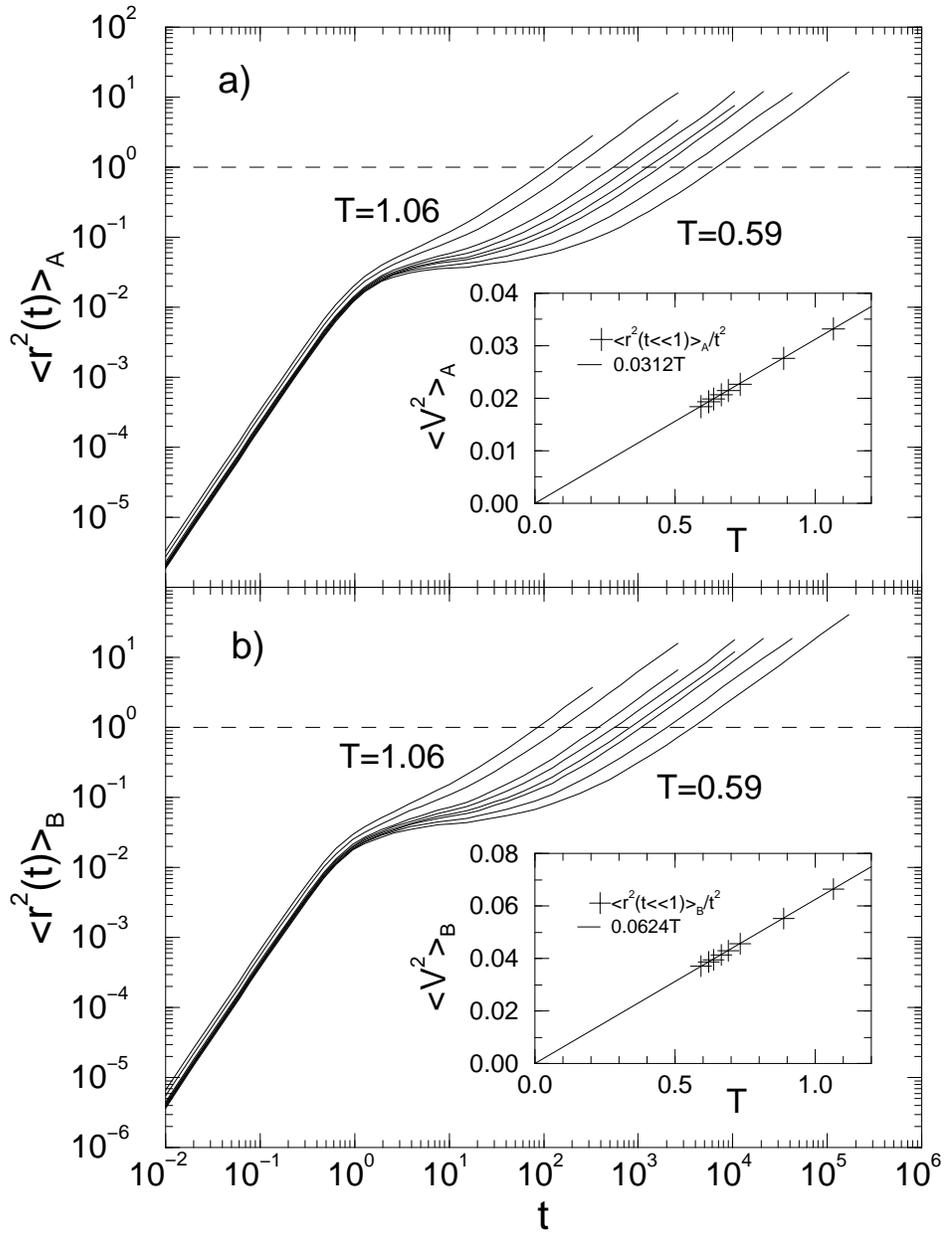, width=12.5cm}
\caption{
The mean square displacement of a) the large particles, 
$\langle r^2(t)\rangle_A$, and b)  the small  particles, 
$\langle r^2(t)\rangle_B$.
Insets show $\langle V^2\rangle_\alpha$ calculated from eq. 
\ref{V2} and \ref{V2fromT}, and demonstrates the agreement between 
the two equations.
The vertical dashed lines identifies $\langle r^2(t_1)\rangle_\alpha=1$. 
}
\label{fig:R2(t)}
\end{figure}

In figure \ref{fig:R2(t)} we show the mean square displacement of the a) 
A particles,
$\langle r^2(t)\rangle_A$, and b) B particles, $\langle r^2(t)\rangle_B$.
At short times ($t<1$) a ballistic regime 
($\langle r^2(t)\rangle_\alpha \propto t^2$) is seen for 
both A and B particles, 
at all temperatures. The ballistic regime is simply a consequence of
the velocities being constant at these short time scales:
\begin{equation}
  \langle r^2(t)\rangle_\alpha ~=~ \langle (Vt)^2\rangle_\alpha
                               ~=~ \langle V^2\rangle_\alpha ~t^2
  \label{V2}
\end{equation}
In the insets in figure \ref{fig:R2(t)} is shown $\langle V^2\rangle_\alpha$
calculated by eq. \ref{V2}. $\langle V^2\rangle_\alpha$ can also 
be calculated directly from the temperature:
\begin{equation}
  T = \frac{2E_{kin}}{3(N-1)} 
    = \frac{Nm_\alpha \langle (V/\tau)^2\rangle_\alpha }{3(N-1)}
    = \frac{48m_\alpha \langle V^2\rangle_\alpha}{3(1-1/N)}
  \label{V2fromT}
\end{equation}
In the insets in figure \ref{fig:R2(t)} is also shown 
$\langle V^2\rangle_\alpha$ as calculated from eq. \ref{V2fromT}.
The excellent agreement between eq. \ref{V2} and \ref{V2fromT} 
(with no fitting involved) is
not a big surprise, but it gives confidence in the argument leading
to eq.  \ref{V2}, and acts as a consistency check.

In figure \ref{fig:R2(t)} a diffusive regime 
($\langle r^2(t)\rangle_\alpha \propto t$) is 
seen at long times, for all temperatures. 
As the temperature is lowered the time scale at which the 
diffusive regime sets in increases, and a  plateau is seen to 
evolve between the ballistic and diffusive regimes. This 
behavior is typical for what is seen in simulations of
super-cooled liquids, 
and the plateau is 
argued to be associated with particles being trapped in 
local ``cages'' consisting of their neighbors 
\cite{Kob95a,Sciortino96}.

\begin{figure}
\epsfig{file=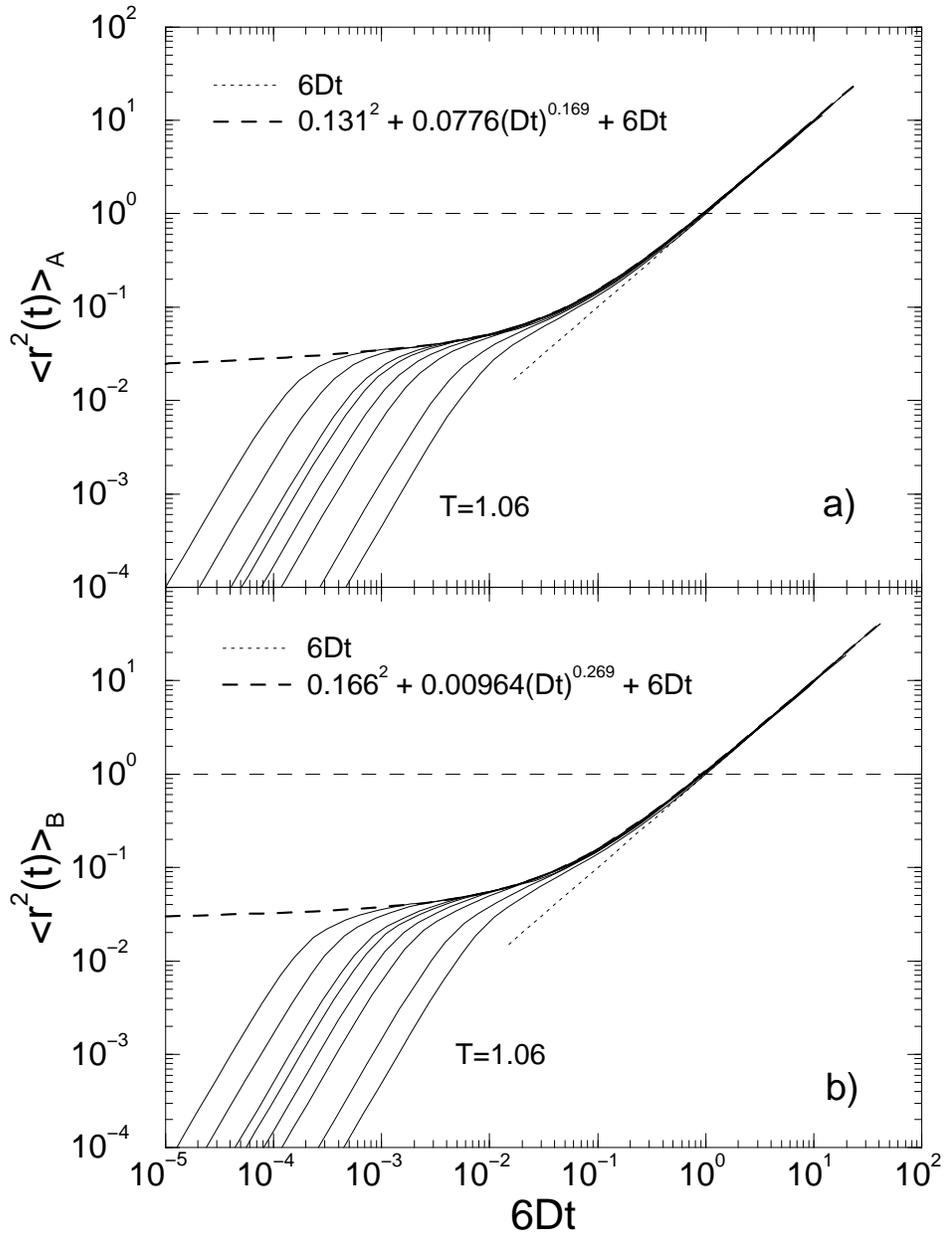, width=12.5cm}
\caption{
The mean square displacement plotted as function of $6Dt$
(same data as in figure \ref{fig:R2(t)}). 
For $\langle r^2(t)\rangle_\alpha \gtrsim 1$ the data for all
temperatures fall on the straight dotted line marking diffusive 
behavior (eq. \ref{R2Einstein}). The thick dashed lines are
fits to eq. \ref{R2Sweidler} 
%
}
\label{fig:R2Scaled}
\end{figure}

The vertical dashed lines  in  figure \ref{fig:R2(t)} 
identifies  the time $t_1$, defined
by $\langle r^2(t_1)\rangle_\alpha = 1$. The significance
of this time will be discussed in connection with figure \ref{fig:R2Scaled}.

Fitting $\langle r^2(t)\rangle_\alpha$ in the diffusive regime
to the Einstein relation:
\begin{equation}
  \langle r^2(t)\rangle_\alpha ~=~ 6Dt 
  \label{R2Einstein}
\end{equation}
we determine the diffusion coefficient, $D$, as a function 
of temperature (and particle type). We postpone the discussion
of the temperature dependence of $D$ to section \ref{TimesVsT}, 
where it will be  
discussed together with other measures of the time scales involved
(eg. $t_1$). Following Kob and Andersen \cite{Kob95a} we in figure 
\ref{fig:R2Scaled}  present the data for 
$\langle r^2(t)\rangle_\alpha$ as a function of $6Dt$.
In the diffusive regime the data for all temperatures should, 
according to eq. \ref{R2Einstein},
fall  on the line $6Dt$, which is seen to 
be the case for $\langle r^2(t)\rangle_\alpha \gtrsim 1$. In other
words the time $t_1$, defined by  
$\langle r^2(t_1)\rangle_\alpha=1$, can be viewed as marking 
the onset of the diffusive regime. The fact that the 
diffusive regime is reached at all temperatures 
(compare fig. \ref{fig:R2(t)}), is a necessary (but not sufficient)
condition, for the system(s) to be in equilibrium.

As is the case for the 
data presented by Kob and Andersen, the data in figure \ref{fig:R2Scaled} seems
to indicate a universal behavior; As the temperature is decreased
$\langle r^2(t)\rangle_\alpha$ follows the thick dashed curve in figure \ref{fig:R2Scaled} 
to lower and lower scaled times. The thick dashed curves are fits
to the fitting function used by Kob and Andersen \cite{Kob95a} 
(see table \ref{Tabel:Sweidler}):
\begin{equation}
  \langle r^2(t)\rangle_\alpha ~=~ r_c^2 + A(Dt)^b + 6Dt 
  \label{R2Sweidler}
\end{equation}
The parameter $b$ is the so-called ``von-Sweidler'' exponent, 
which is related to the dynamics at intermediate times, i.e. 
between the plateau and the diffusive regime. According
to the asymptotic predictions of the ideal mode coupling theory (MCT)
\cite{Gotze92,Gotze99}, 
the von-Sweidler exponent, $b$,  should be independent of the 
particle type. Kob and Andersen
argue that the two values they find, 0.48 for the A particles and
0.43 for the B particles, are within reasonable agreement. This 
is \emph{not} the case for the $b$-values found here;
$0.17\pm 0.03$ for the A particles and $0.27\pm 0.01$ for the B particles.

\begin{table}
\center{
\begin{tabular}{|c|c|c|c|} \hline
 Type & $r_c$ & $A$ & $b$ \\ \hline
 A    & $0.131\pm 0.013$ & $0.071\pm 0.002$ & $0.17\pm 0.03 $\\ \hline
 B    & $0.166\pm 0.003$ & $0.096\pm 0.003$ & $0.27\pm 0.01 $\\ \hline
 \end{tabular}
}
\caption{Parameters found by fitting equation \ref{R2Sweidler} to 
the ``universal'' curves in figure \ref{fig:R2Scaled} (thick dashed lines).
Error-bars are 68\% confidence intervals, as reported by the fitting
routine in Gnuplot.
}
\label{Tabel:Sweidler}
\end{table}


\newpage
\section{The van Hove Correlation Function} 

To characterize the dynamics in more detail, we calculate the van 
Hove self correlation function \cite{Hansen86} 
(where the sum is over particles 
of type $\alpha$):
\begin{equation} 
    G_{s\alpha}({\bf r},t) = 
    {1 \over N_\alpha} \sum_{i = 1}^{N_\alpha} 
   \left<\delta({\bf r}_i(t)-{\bf r}_i(0) - {\bf r})\right>,
   \label{Gs}
\end{equation}
which is  the probability density of a particle of 
type $\alpha \in \{A,B\}$ being displaced by the vector ${\bf r}$, 
during the time interval $t$. In an isotropic system $G_{s\alpha}({\bf r},t)$
does not depend on the direction of ${\mbf r}$, and the probability 
distribution of a particles 
being displaced the 
\emph{distance} ${r}$ is $4\pi r^2G_{s\alpha}({ r},t)$.
The mean square displacement, $\langle r^2(t)\rangle_\alpha$, is 
given by the second moment of $4\pi r^2G_{s\alpha}({ r},t)$, 
and the later thus gives a more detailed view of the dynamics.

\begin{figure}
\epsfig{file=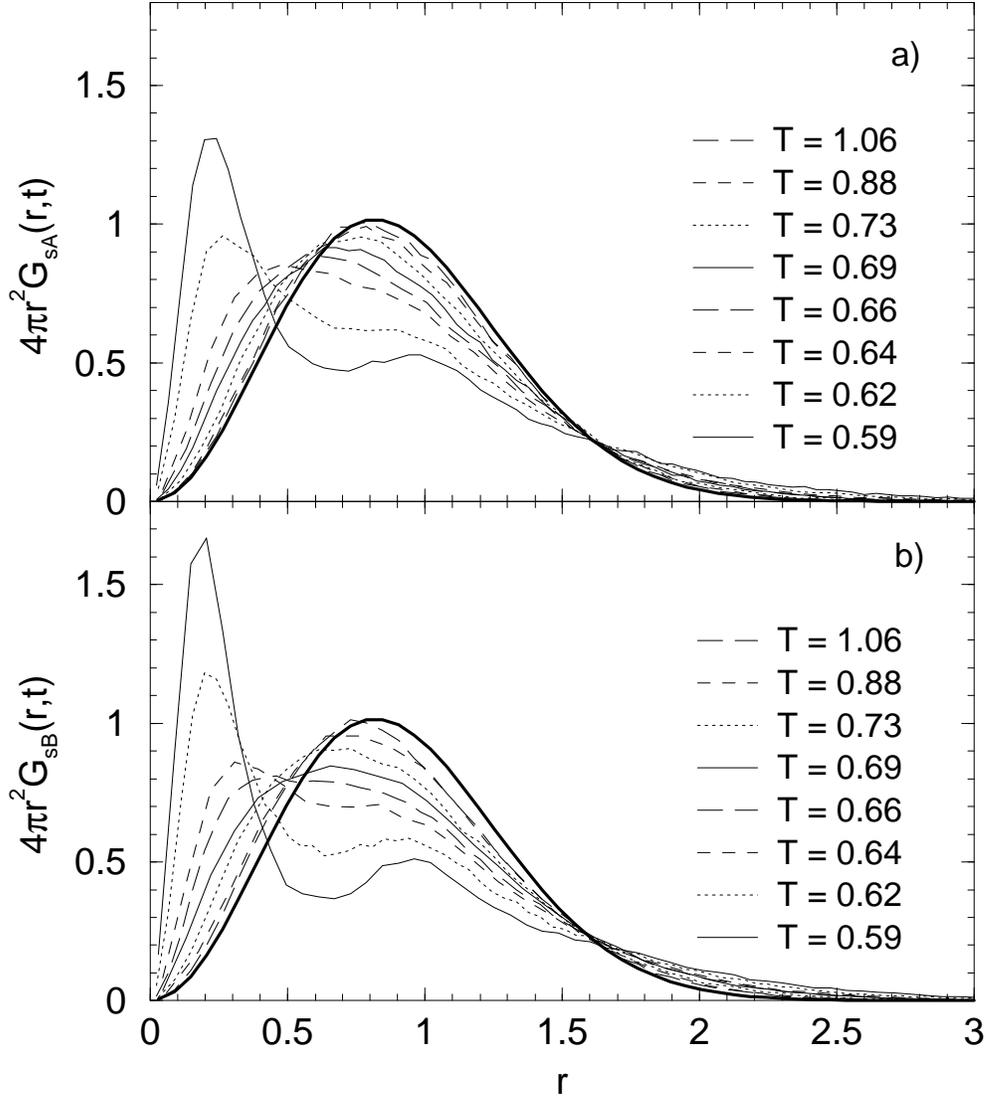, width=13cm}
\caption{
The distribution of particle displacements, 
$4\pi r^2G_{s\alpha}(r,t_1)$, where $t_1$ is defined by 
$\langle r^2(t_1)\rangle_\alpha=1$, see figure \ref{fig:R2(t)}. 
At high temperatures the Gaussian approximation (thick
curve) is reasonable fulfilled, whereas at the lowest
temperature a secondary peak is present, indicating 
that hopping is present in the system (see text). The small particles
(type B) are seen to have a larger tendency to exhibit hopping, compared
to the large particles. 
} 
\label{fig:GsAt1}
\end{figure}

In figure \ref{fig:GsAt1} is plotted $4\pi r^2G_{sA}(r,t_1)$ and
$4\pi r^2G_{sB}(r,t_1)$, i.e. the distribution of distances moved 
by A and B particles respectively in the time interval $t_1$, 
which in the previous section was demonstrated to be at 
the onset of the diffusive regime.
The thick curve is the Gaussian approximation:
$$G_{s}(r,t) = \left(\frac{3}{2\pi\langle r^2(t)\rangle }\right)^{3/2} 
                \exp\left(-\frac{3r^2}{2\langle r^2(t)\rangle }\right)$$
which is seen to be reasonably fulfilled at high
temperatures. As the temperature is lowered the results starts 
graduately deviating from the Gaussian approximation, and 
a shoulder builds at the average inter particle distance, $r\approx 1.0$, 
which at T=0.59 becomes a well-defined second peak.  The second
peak, observed also in other model liquids, is interpreted
\cite{Roux89,Barrat90,Hiwatari91,Barrat91,Wahn91,Thirumalai93,Kob95a} 
as single particle hopping (see figure \ref{fig:hop}), i.e. the particles 
stay localized in their local cages a certain time (first peak), 
then moves more or less directly out to distances approximately
equal to the inter particle distance (second peak) (This behavior 
was also reported in \cite{Miygawa88}). 
To what extent the hopping behavior is 
related to activated hopping over energy barriers will be discussed in
chapter \ref{EnergyLandscape}.

Note that when analyzing the dynamics in more detail by 
means of the van Hove self correlation function (fig. \ref{fig:GsAt1}), the 
universality seen in the mean square displacement (fig. \ref{fig:R2Scaled})
is completely gone; The way the system ``achieves'' the 
diffusive behavior (i.e. the dynamics at $t_1$) changes qualitatively 
from high temperature (Gaussian) to low temperature (Hopping).

\newpage

\begin{figure}
\epsfig{file=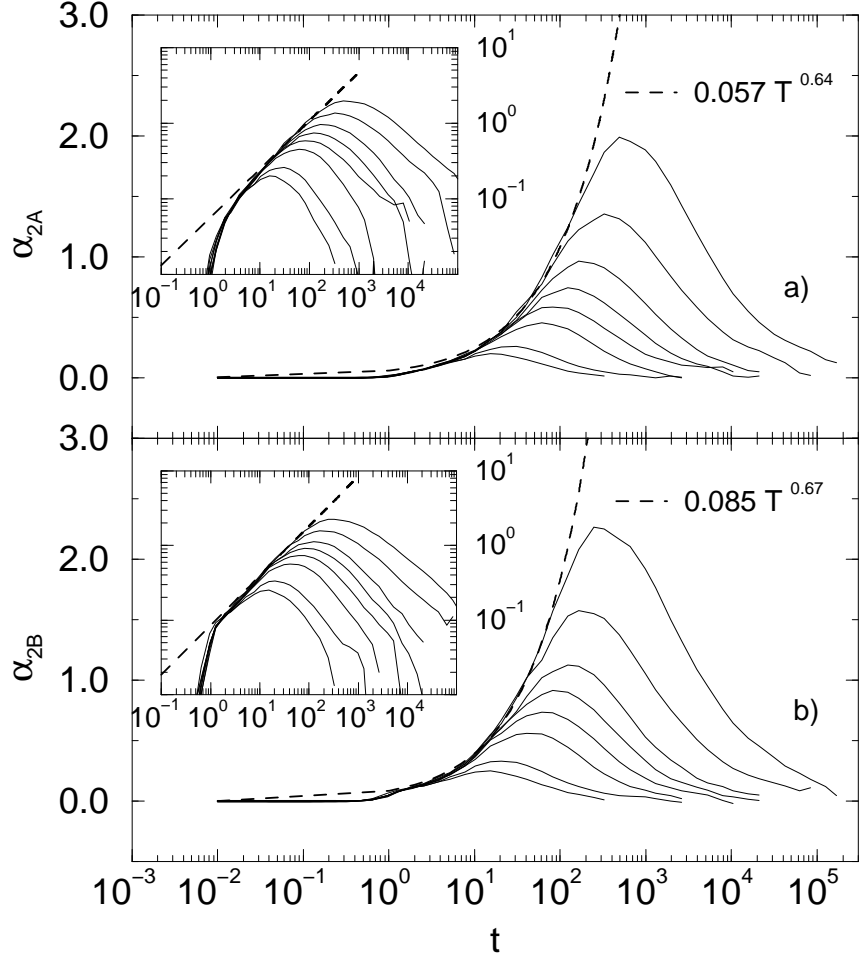, width=11.5cm}
\caption{The non-Gaussian parameters, $\alpha_{2A}(t)$ and $\alpha_{2B}(t)$.
The dashed line is a fit to a power-law in the ``universal'' regime 
before the maximum.
} 
\label{fig:Alpha2AB}
\end{figure}

The deviation from Gaussian behavior is often analyzed in terms
of the Non-Gaussian parameter 
\cite{Rahman64,Odagaki91,Miygawa91,Kudchadkar95,Kob95a,Sciortino96,Horbach97,Kob97}:
\begin{equation} 
    \alpha_2(t) = 
       \frac{\langle r^4(t)\rangle - \langle r^4(t)\rangle_{Gauss}}
            {\langle r^4(t)\rangle_{Gauss}}
    =  \frac{3\langle r^4(t)\rangle}{5\langle r^2(t)\rangle^2} -1
\end{equation}
which is the relative deviation between the measured $\langle r^4(t)\rangle$
and what it would  be for a given value of $\langle r^2(t)\rangle$, 
if $G_{s}(r,t)$ was Gaussian: 
$\langle r^4(t)\rangle_{Gauss} = 5\langle r^2(t)\rangle^2/3$. In figure 
\ref{fig:Alpha2AB} is shown $\alpha_2$ for the A and  B particles respectively.

Like in \cite{Kob95a,Sciortino96} we see a universal behavior in the 
way $\alpha_{2A}(t)$
increases, before it reaches its maximum. Following \cite{Sciortino96} a 
power-law fit is done in this regime, from which is found exponents
of $0.64$ and $0.67$ for the A and B particles respectively.
The exponent found in \cite{Sciortino96} is $0.4$. 
As noted in \cite{Kob95a} the universality seen in  $\alpha_{2}(t)$
is different from the one found in the 
behavior of $\langle r^2(t)\rangle_\alpha$ (fig. \ref{fig:R2Scaled});
The universality seen in $\alpha_{2}(t)$  does not involve 
any scaling. 

\begin{figure}
\epsfig{file=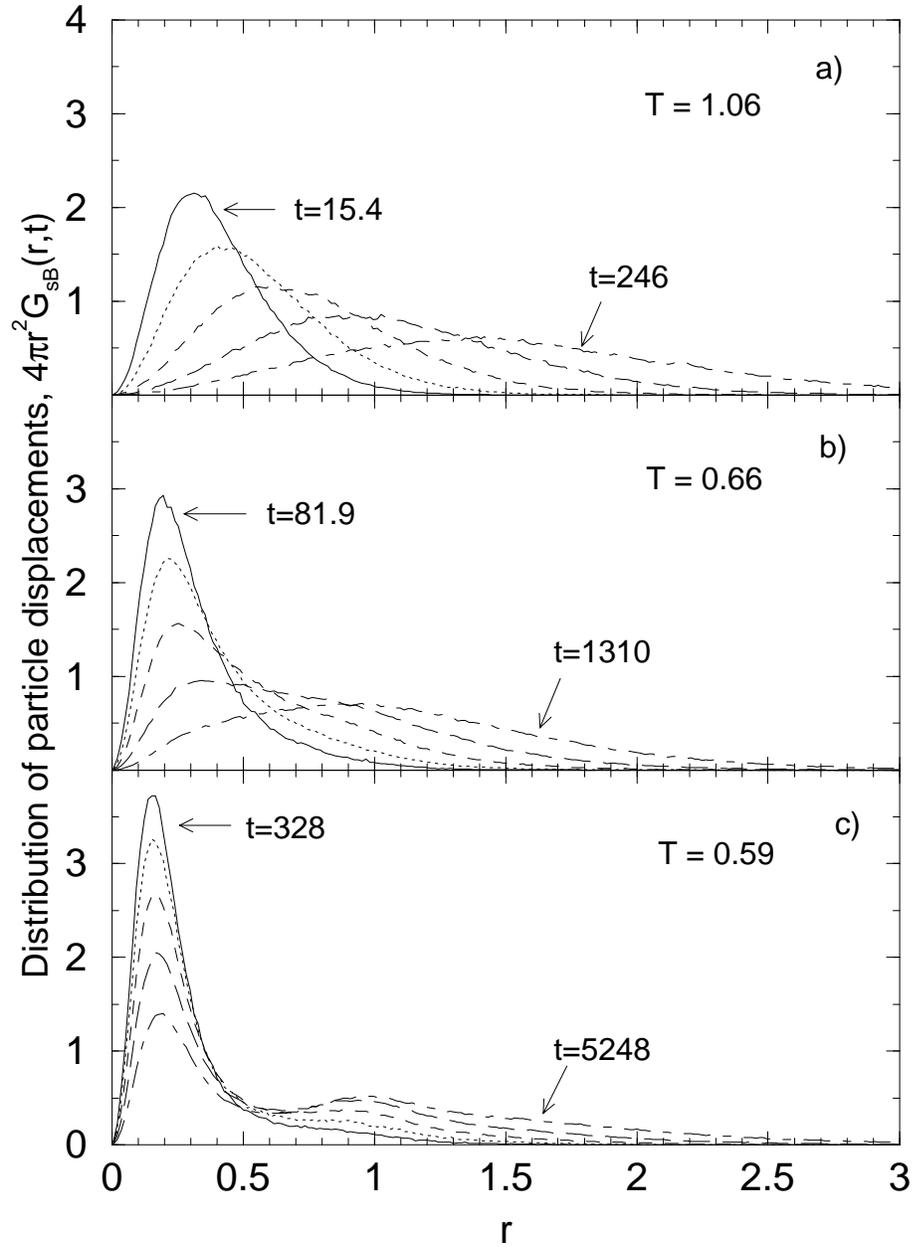, width=12cm}
\caption{Time development of the distribution of particle 
displacements for the B particles, $4\pi r^2G_{sB}(r,t)$, 
at the same temperatures as in figure \ref{fig:gr_all}. 
At each temperature $4\pi r^2G_{sB}(r,t)$
is shown at a time close to $t^*$ (smallest time indicated by arrows), 
and 2,4,8, and 16 times that time.
} 
\label{fig:VHBt_star}
\end{figure}

The time $t^*$ which is defined
as the position of the maximum in $\alpha_2(t)$, can be interpreted
as the time where the dynamics deviates most from Gaussian 
behavior \cite{Odagaki91,Kob97}. In figure \ref{fig:VHBt_star} is 
shown for three temperatures, the time development of $4\pi r^2G_{sB}(r,t)$.
At each temperature $4\pi r^2G_{sB}(r,t)$
is shown at a time close to\footnote{The reason that $t^*$ itself is
not used here, has to do with at what time-steps configurations are stored, 
and thus which time differences are easy accessible} 
$t^*$ (indicated by the arrows), and 2,4,8, and 16 times that time.
At $T=1.06$ the distribution of particle displacements is seen to 
be characterized by a single  peak, which ``spreads out'' without
any qualitative change in the shape, as time increases. This
is the typical behavior for liquids at high temperatures \cite{Hansen86}.
At $T=0.66$ the behavior is seen to be qualitatively similar, 
except for a weak indication of a shoulder 
 at $r\approx 1.0$. 
The time development of $4\pi r^2G_{sB}(r,t)$
at $T=0.59$ is qualitatively different from what is found at higher 
temperatures; The first peak 
decreases while the position is almost constant, and at the same time 
as the second peak starts building up. This 
 supports the ``hopping'' interpretation of the second peak 
given above; The  particles  escape their local 
``cages'' (the first peak), by ``suddenly appearing'' at 
approximately the inter particle distance (second peak). After the
last time shown here, the second peak starts decreasing.
This is interpreted as a consequence of particles hopping again.

With the regards to the interpretation of $t^*$ mentioned above, one 
should note that for $T=0.59$ the second peak in $4\pi r^2G_{sB}(r,t)$ builds
 up \emph{after} the time $t^*$, i.e. after $\alpha_2(t)$ has its maximum. 
A more detailed analysis of the non-Gaussian behavior should 
include the higher order analogues of $\alpha_2(t)$ 
(involving  higher moments of $4\pi r^2G_{s\alpha}(r,t)$) 
\cite{Rahman64}, but this approach will not be pursued further here.

\newpage

The van Hove distinctive correlation function, $G_{D\alpha\beta}(r,t)$,
is the time dependent generalization of the pair correlation
function \cite{Roux89}:
\begin{equation} 
     G_{D\alpha\beta}({\bf r},t) = 
    {1 \over \sqrt{N_\alpha N_\beta}} 
    \sum_{i = 1}^{N_\alpha} \sum_{j = 1}^{N_\beta}
   \left<\delta({\bf r}_j(t)-{\bf r}_i(0) - {\bf r})\right>,
   \label{GD}
\end{equation}
where the first sum is over particles of type $\alpha$, the 
second sum is over particles of type $\beta$, and $i\neq j$
if $\alpha = \beta$ (this last term is contained in $G_{s\alpha}(r,t)$, 
see equation \ref{Gs}). $ G_{D\alpha\beta}(r,t)$ is the relative 
density at time $t$ of $\beta$-particles at ${\bf r}$, given that 
there at time $t=0$ was a $\alpha$-particle at ${\bf r}=0$ 
(normalised to be 1 if there is no correlation). 

\begin{figure}
\epsfig{file=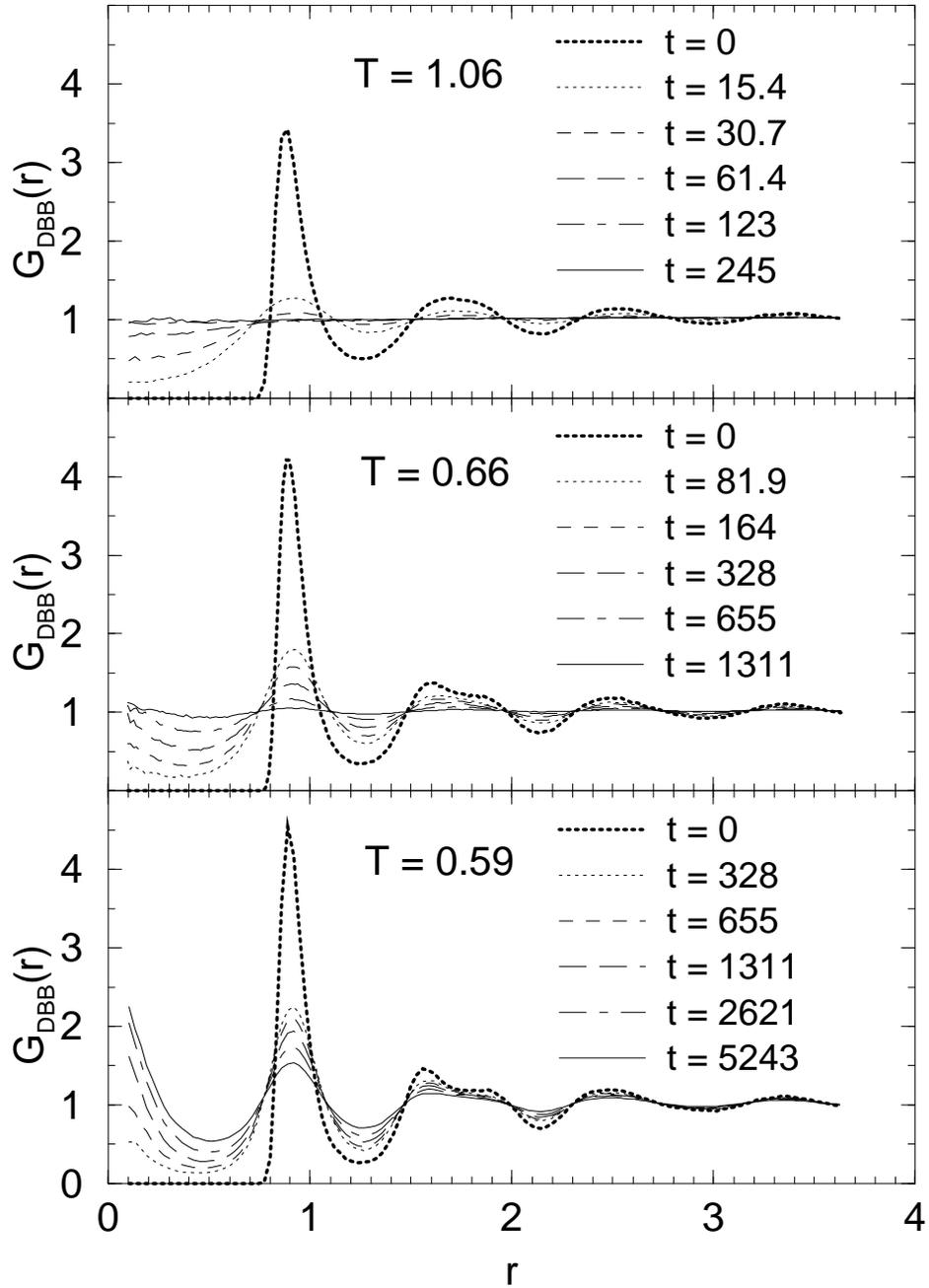, width=12.5cm}
\caption{
Time development of $G_{DBB}(r,t)$ for the same temperatures as in
figure \ref{fig:VHBt_star}. For each temperature, the 
times used here are the same as in figure \ref{fig:VHBt_star}, \emph{and}
$t=0$ (thick dotted curve), where $G_{DBB}(r,t=0)$ equals the 
pair correlation function, $g_{BB}(r)$.
} 
\label{fig:VHDBB}
\end{figure}

In figure \ref{fig:VHDBB} is shown $G_{DBB}(r,t)$
for the same temperatures and times as in figure \ref{fig:VHBt_star}.
Included in figure \ref{fig:VHDBB} is also 
$G_{DBB}(r,t=0) = g_{BB}(r)$.
Except for the lowest values of  $r$, the features of the pair correlation
function is seen to approach the long-time value 
$G_{D\alpha\beta}(r,t) = 1$ in a smooth
manner, which is the typical high-temperature behavior \cite{Hansen86}. 
As the temperature is decreased, a significant ``overshoot'' 
develops at the small $r$-values\footnote{the data for $r<0.1$ is removed, 
since the statistical noise dominates the data in this range}, 
indicating that there is an 
excess probability
that when a particle jumps, it  does so to a position previously occupied 
by another particle.
Thus, while $G_{sB}(r,t)$ (fig. \ref{fig:VHBt_star}a)
shows that particles has a tendency to jump a distance 
approximately equal to the inter-particle distance, $G_{D\alpha\beta}(r,t)$
give the further information,  that they have a tendency to jump to a
position previously occupied by another particle. 
Similar (but less pronounced behavior) was found in \cite{Roux89}.

In a scenario where the dynamics is dominated by particles jumping
to positions previously occupied by other particles, one should expect
that moving particles make up correlated string-like objects in the liquid.
This kind of behavior is found by Muranaki and Hiwatari in a soft
sphere system \cite{Muranaka97}, and by Donati et. al. 
in the Kob \& Andersen system \cite{Donati98}. This type of behavior
 will be discussed further in chapter 
\ref{EnergyLandscape}.

\newpage

\section{The Intermediate Scattering Function}
\label{sec:Fs}

The self part of the intermediate scattering function, $F_{s}(q, t)$,
is the (3 dimensional) Fourier transform of the van Hove
self correlation function \cite{Hansen86}:

\begin{eqnarray}
    F_{s}({\mathbf q}, t) &\equiv& 
      \int 
	 G_{s}({\mathbf r}, t) e^{-i \mathbf q \cdot \mathbf r} 
      d \mathbf r ~=~ 
      \langle 
         \cos{{\mathbf q} ({\mathbf r}_j(t) - {\mathbf r}_j(0))} 
      \rangle   \label{FsDef}
\end{eqnarray}

\begin{figure}
\epsfig{file=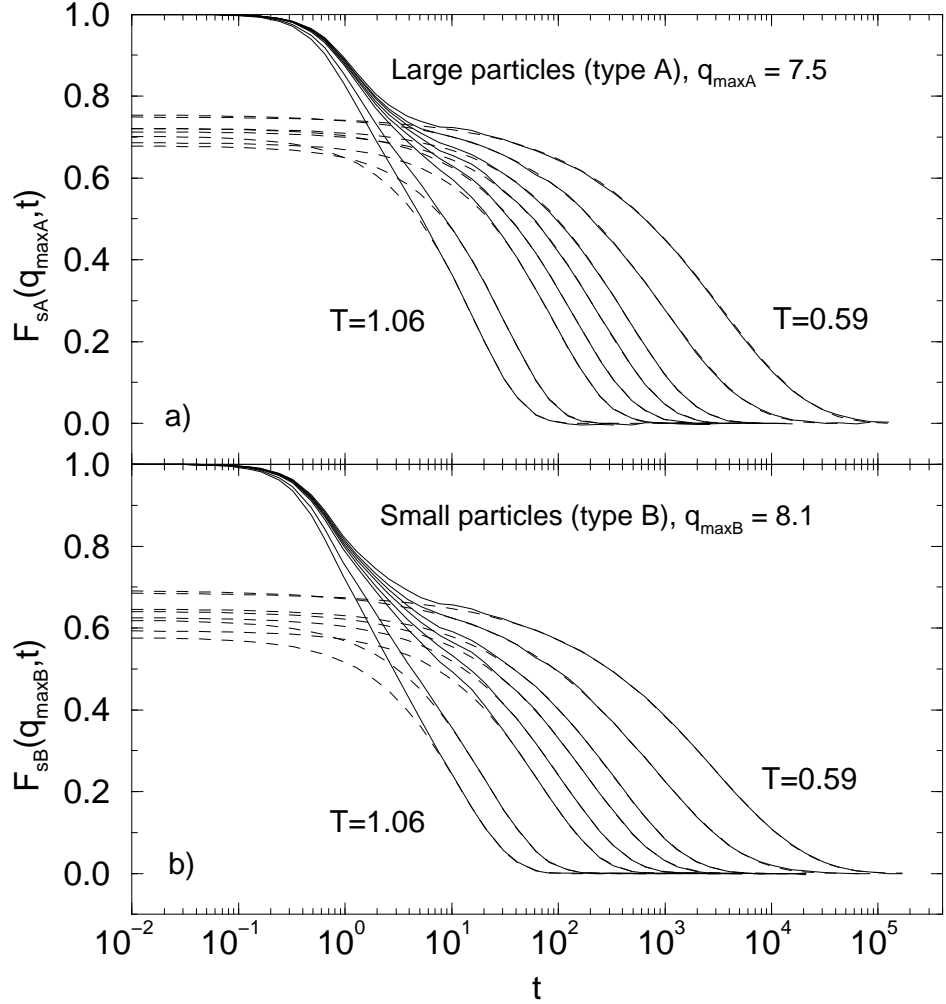, width=12.5cm}
\caption{ a) The self part of the intermediate scattering function
for the A particles, $F_{sA}(q_{maxA}=7.5, t)$.
The dashed  lines are fits to
stretched exponentials, $f(t) = f_c\exp(-(t/\tau_{\alpha})^\beta)$. 
The fitting was performed for $t>10$ for the 2 highest temperatures, 
and for $t>30$ for the rest of the temperatures. 
b) $F_{sB}(q_{maxB}=8.1, t)$, otherwise as above.
}
\label{fig:Intermediate}
\end{figure}

$F_{s}(q, t)$ is a relaxation function, 
normalized to be 1 at $t=0$, and it goes to zero for $t\rightarrow \infty$.
Figure \ref{fig:Intermediate} shows the self part of 
the intermediate scattering function for the A particles
and B particles; $F_{sA}(q_{maxA}=7.5, t)$ and $F_{sB}(q_{maxB}=8.1, t)$.
$q_{maxA} = 7.5$ and $q_{maxA} = 8.1$ are here the positions of the first 
peak in the static structure factor for the A-A correlation ($S_{AA}(q)$), 
and the the B-B correlation ($S_{BB}(q)$), respectively 
(see figure \ref{fig:Sq_all}).
As the temperature is lowered a two-step
relaxation develops, which is what is typically seen\footnote{
In most experiments, and some simulations, this is seen as two 
peaks (one for each relaxation step) in the imaginary part of 
the generalized susceptibility: 
$\chi''({\mathbf q},\omega) \equiv \omega \pi F({\mathbf q},\omega)$,
where $F({\mathbf q},\omega)$ is the Fourier transform of $F({\mathbf q},t)$,
see eg. \cite{Yonezawa94a,Kob95b}.
}
in glass-forming liquids, both in experiments \cite{Cummins97}
and in simulations \cite{Barrat90,Wahn91,Yonezawa94a,Matsui94,Kob94,Kob95b,Kudchadkar95,Teichler95,Sciortino96,Wahn97,Horbach97}. 
The initial relaxation is a 
consequence of the particles oscillating in their cages, while the 
the long-time (alpha) relaxation is a consequence of particles 
escaping their cages \cite{Yonezawa94a,Kudchadkar95,Sciortino96}.
The alpha relaxation is often found to be 
well approximated by stretched exponentials, 
$f(t) = f_c\exp(-(t/\tau_\alpha)^\beta)$ 
\cite{Barrat90,Wahn91,Kob94,Kob95b,Kudchadkar95,Sciortino96}, 
which is also the case in figure \ref{fig:Intermediate}, where
fits to stretched exponentials are shown as dashed lines.

\begin{figure}
\epsfig{file=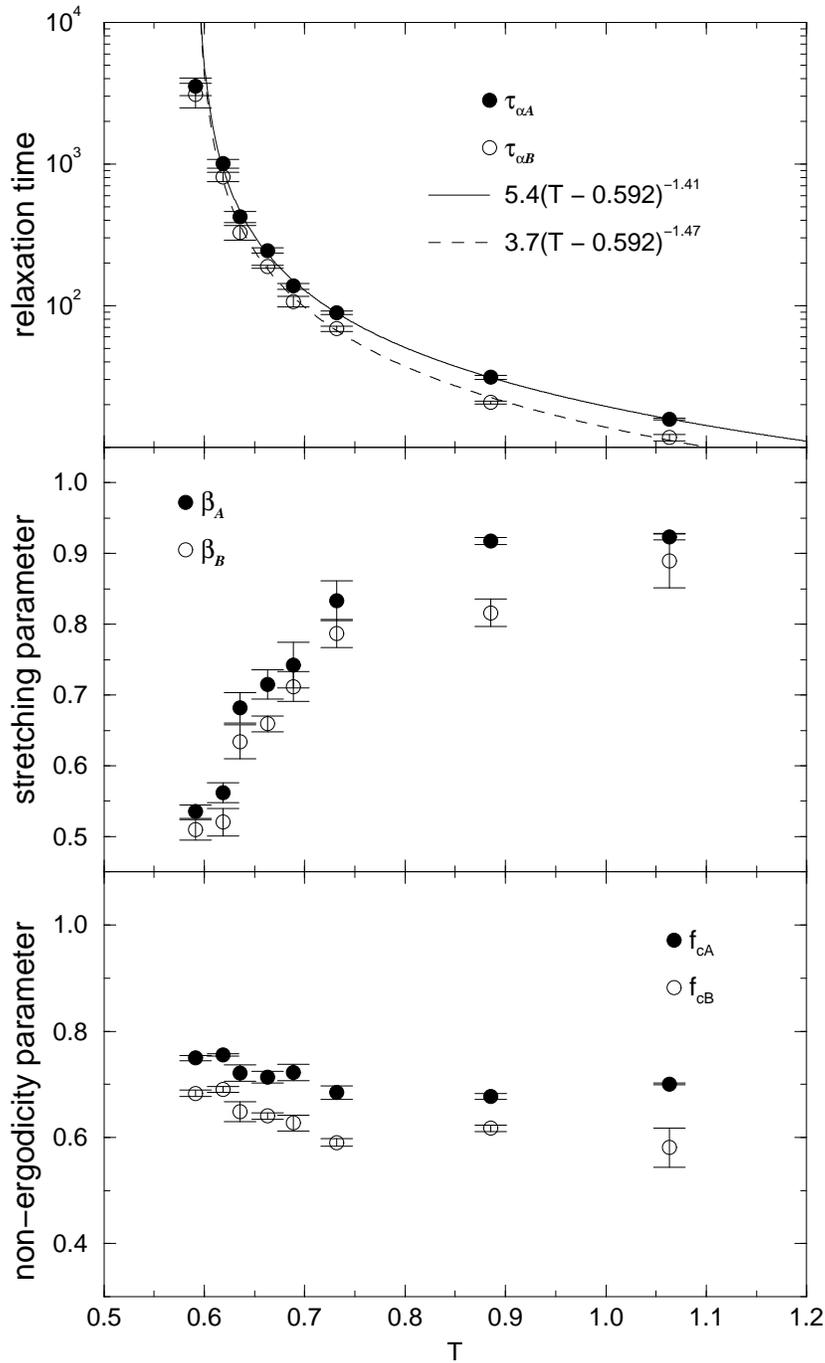, width=11cm}
\caption{ 
Fitting parameters used in figure \ref{fig:Intermediate}.
a) Relaxation times $\tau_{\alpha A}$ and $\tau_{\alpha B}$. 
The lines are fits to $\tau_\alpha \propto (T - T_c)^{-\gamma}$ (see text).  
b) Stretching parameters $\beta_A$ and $\beta_B$. 
c) Non-ergodicity parameters, $f_{cA}$ and $f_{cB}$.  
}
\label{fig:FsP}
\end{figure}

In Fig.~\ref{fig:FsP} we show as a function of temperature 
the three fitting parameters used 
in figure \ref{fig:Intermediate}; a) relaxation times, $\tau_\alpha$,
b) stretching parameters, $\beta$,  and c) non-ergodicity 
parameters $f_c$. As expected the relaxation times, $\tau_{\alpha A}$
and $\tau_{\alpha B}$ is seen to increase dramatically as the 
temperature is lowered. The asymptotic prediction of the ideal
mode-coupling theory (MCT) is $\tau_{\alpha} = \tau_0(T - T_c)^{-\gamma}$ 
\cite{Gotze92,Gotze99}. The divergence of $\tau_{\alpha}$ at $T_c$ 
predicted by the  ideal MCT  is never seen in 
practice. This is argued to be consequence of relaxation
by hopping taking over close to $T_c$ 
\cite{Gotze92,Wahn91,Kob95a,Cummins98,Gotze99}. 
This  means that close to $T_c$ the power-law is expected to break down 
and it should not be fitted in this regime. Unfortunately we
have no independent estimate of $T_c$, and furthermore it is not known 
how close to $T_c$ the power-law is expected to hold. Since
we have already demonstrated, that hopping is present in the 
system at the lowest temperatures, we can expect that these are 
close to $T_c$, which means that we might run into the problem described 
above. We note also, that the power-law is an \emph{asymptotic} prediction, 
i.e. it is expected to break down at high temperatures, but again it is
not known how far from $T_c$ this is supposed to happen.

As discussed above, the temperature range (if any) where the asymptotic
predictions of ideal MCT is supposed to hold, is 
not known. Consequently the following approach for fitting 
$\tau_{\alpha} = \tau_0(T - T_c)^{-\gamma}$ has been applied; 
First the fit was done for all 8 temperatures ($T_{cut}=0$), 
then by excluding  the lowest temperature ($T_{cut}=0.60$), 
and then by excluding the two  lowest temperatures ($T_{cut}=0.63$). 

The results of the fitting-procedure described above, is 
seen in table \ref{table:FsPowerFits}. For both the A and B 
particles the parameters found for
$T_{cut}=0.60$ and $T_{cut}=0.63$ agree within the error-bars, while they 
disagree with the parameters for $T_{cut}=0$. Thus one might argue
that the parameters found for $T_{cut}=0.60$ and $T_{cut}=0.63$ 
describes the ``true'' power-law, while the fit found using $T_{cut}=0$
deviates as a consequence of fitting too close to $T_c$. Thus our best 
attempt at a power-law fit is the one achieved for $T_{cut}=0.60$ 
(i.e. by excluding the lowest temperature), which is the fit shown 
in figure \ref{fig:FsP} for the A and B particles respectively.
The values estimated in this manner for $T_c$ and $\gamma$ for 
the A and B particles are identical within the error-bars, as
predicted by the ideal MCT. The temperature
dependence of $\tau_\alpha$ will be discussed further in 
section \ref{TimesVsT}. 

\begin{table}
\center{
\begin{tabular}{|c|c|c|c|c|} \hline
 Type &$T_{cut}$ & $\tau_0$          & $T_{c}$ & $\gamma$  \\ \hline \hline
 A    &$0.00$    & $4.2\pm 0.6$ & $0.571 \pm 0.004$ & $1.71 \pm 0.10$ \\ \hline
 A    &$0.60$    & $5.4\pm 0.5$ & $0.592 \pm 0.004$ & $1.41 \pm 0.07$ \\ \hline
 A    &$0.63$    & $5.3\pm 0.4$ & $0.586 \pm 0.008$ & $1.46 \pm 0.09$ \\ \hline \hline
 B    &$0.50$    & $2.9\pm 0.5$ & $0.571 \pm 0.004$ & $1.77 \pm 0.11$ \\ \hline 
 B    &$0.60$    & $3.7\pm 0.4$ & $0.592 \pm 0.005$ & $1.47 \pm 0.09$ \\ \hline
 B    &$0.63$    & $3.6\pm 0.4$ & $0.584 \pm 0.010$ & $1.52 \pm 0.12$ \\ \hline \hline
 \end{tabular}
}
\caption{Parameters found by fitting $\tau_{\alpha} = \tau_0(T - T_c)^{-\gamma}$
to all 8 temperatures ($T_{cut}=0$), excluding  the lowest 
temperature ($T_{cut}=0.60$), and  by excluding the two  
lowest temperatures ($T_{cut}=0.63$). The error-bars indicate the 
68.3\% confidence interval, as reported by Gnuplot.
}
\label{table:FsPowerFits}
\end{table}

The  numerical data for
$\tau_{\alpha}(T)$ does not by itself give strong evidence for a
dynamical transition at the estimated $T_c=0.592 \pm 0.004$;
nothing special seems to happen at that temperature. 
However,  the agreement with the lowest temperature ($T = 0.591 \pm 0.002$), 
where we clearly see hopping (figure \ref{fig:GsAt1}), 
gives us confidence that there \emph{is} a dynamical transition 
close to the estimated $T_c$.  

The stretching parameters, $\beta_A$ and $\beta_B$, are seen in figure
\ref{fig:FsP}b to decrease from values close to 1 (i.e. exponential relaxation)
at high temperatures, 
to values close 0.5 at the lowest temperatures. The values found at 
high-temperatures are somewhat dependent on the time-interval 
used for the fitting, and the real error-bars 
are thus bigger than the ones shown (which are estimated from 
deviations between the three independent samples, with the 
fitting done in the same time-intervals). However focusing 
on the medium to low temperatures ($T\lesssim 0.73$), there can be no 
doubt that $\beta$ is increasing as function of temperature. 
This is in contradiction with the asymptotic predictions of the ideal MCT, 
which predicts $\beta$ to be constant. The ideal MCT does not directly 
predict the long-time relaxation to be stretched exponentials, 
but it predicts it to exhibit time-temperature super-position (TTSP)
\cite{Gotze92,Gotze99}. 
This mean that the \emph{shape} of the long-time relaxation should 
be independent of temperature, which for stretched exponentials means
that $\beta$ should be constant.

\begin{figure}
\epsfig{file=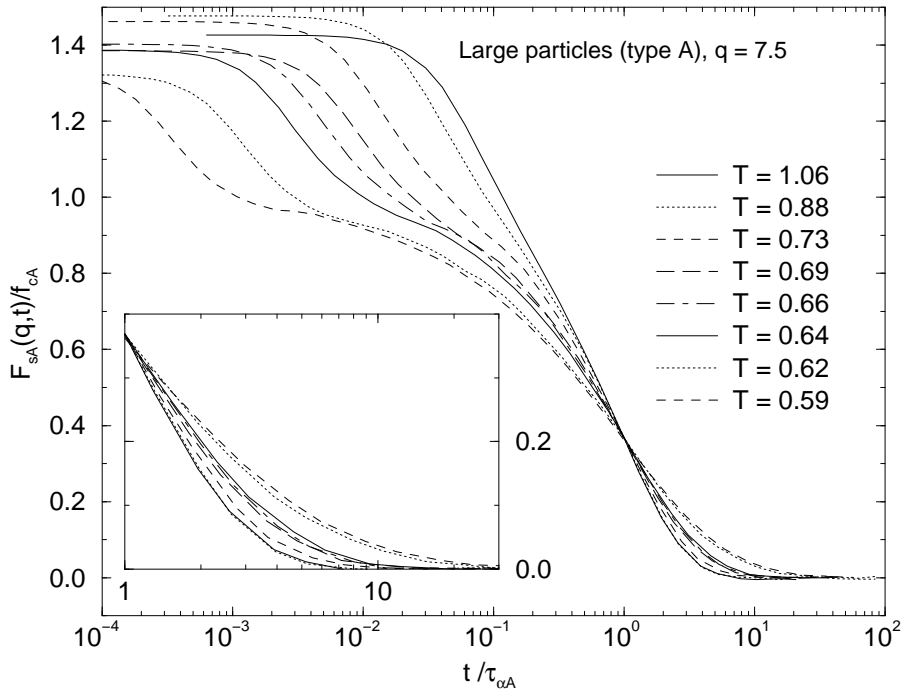, width=12cm}
\caption{ (a) The self part of the intermediate scattering function
for the A particles, $F_{sA}(q=7.5, t)$, scaled to test if 
the long-time relaxation exhibits time-temperature super-position.
This is not the case.
}
\label{fig:IntermediateScaled}
\end{figure}

The question of whether $\beta$ is constant or increases towards 1  
has led to controversy 
about what is the right fitting procedure \cite{Cummins98, Patkowski98}.
Another way of checking for TTSP, is by scaling the 
data appropriately and look for approach to an universal curve.
This is done in figure \ref{fig:IntermediateScaled} where
$F_{sA}(q_{max},t)/f_{cA}$ is plotted versus $t/\tau_{\alpha A}$ which
should be identical for any temperature range where TTSP holds. 
No such
temperature range can be identified. One might argue, that the 
scaling approach used in figure \ref{fig:IntermediateScaled} relies
on the values of $f_c$ estimated from the fits to stretched exponentials, 
which at the high temperatures is  dependent on the 
time-range chosen for the fitting. Scaling $F_{sA}(q_{max},t)$ 
 to agree at $F_{sA}(q_{max},t)=e^{-1}$ (i.e. without dividing by $f_{c}$), 
like done in \cite{Kob95b} does \emph{not} change the conclusion
above; there is no indication of TTSP. 

The fact that $F_{s\alpha}(q, t)$ decays to zero 
(figure \ref{fig:Intermediate}), and that the parameters describing the 
alpha relaxation is in reasonably agreement for the three independent
samples (as illustrated by the error-bars in  figure \ref{fig:FsP}), 
are necessary conditions for the liquid to be in equilibrium. 
Of all tests for equilibrium applied, this was the one that was found
to be most sensitive, i.e. requiring the longest equilibration times. 
  
\begin{figure}
\epsfig{file=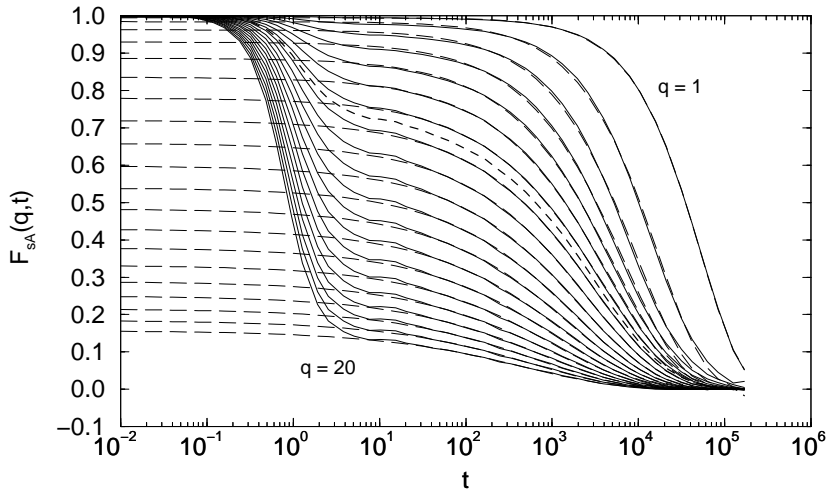, width=11cm}
\caption{ The self part of the intermediate scattering function
for the A particles, $F_{sA}(q, t)$, at $T=0.59$ for  $q$-values $1, 2, ..., 20$, 
and $q=q_{max}=7.5$ (thick dashed line). Thin dashed lines are fits
to stretched exponential, where the fitting was done for $t\ge 30$.
}
\label{fig:IntermediateQ}
\end{figure}

\begin{figure}
  \epsfig{file=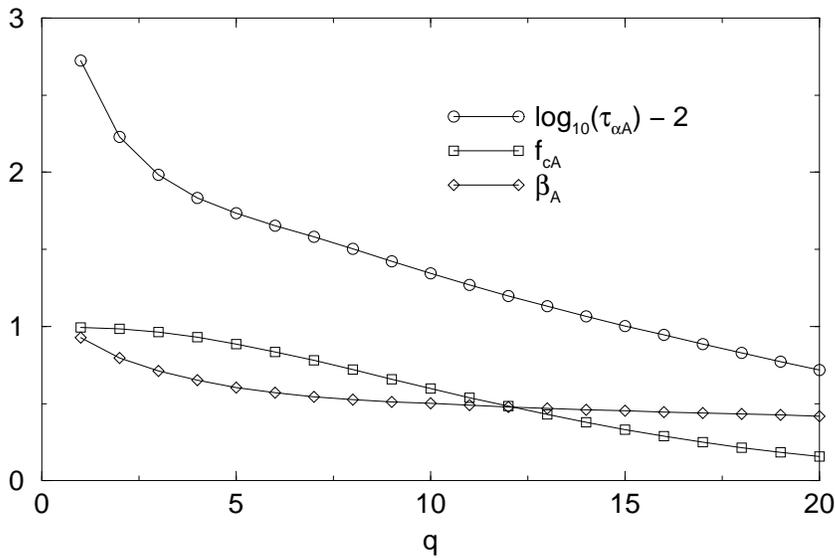, width=11cm}
  \caption{q-dependence of parameters found by fitting stretched 
  exponentials to $F_{sA}(q, t)$, at $T=0.59$, see figure \ref{fig:IntermediateQ}
  }
\label{fig:IntermediateQParams}
\end{figure}

In figure \ref{fig:IntermediateQ} is plotted $F_{sA}(q, t)$ at $T=0.59$ for
$q$-values $1, 2, ..., 20$, and $q=q_{max}=7.5$ (thick dashed line).
Also shown in figure \ref{fig:IntermediateQ} are fits to stretched 
exponentials (dashed lines). The fitting parameters used in \ref{fig:IntermediateQ}
is plotted in figure \ref{fig:IntermediateQParams}. At the lowest $q$-values
the fits are not perfect, but except for that the fits are reasonable.  
Note that in the time-range were the second peak builds up in 
$4\pi r^2G_{sA}(r,t)$, i.e. from $t^*\approx 6\cdot 10^2$ to 
$t_1\approx 7\cdot 10^3$, figure \ref{fig:IntermediateQ} shows no clear   
indication of this. Since $F_{s\alpha}(q, t)$ is the 
Fourier transform of $G_{s\alpha}(r, t)$, and thus contains the same 
information, it \emph{is} possible to extract the information 
about the hopping from $F_{s\alpha}(q, t)$. However since we already 
have an excellent indication of the hopping in $4\pi r^2G_{s\alpha}(r,t)$, 
and $G_{D\alpha\beta}(r,t)$, this will not be pursued further here.

\newpage

\section{Time Scales}
\label{TimesVsT}

\begin{figure}
\epsfig{file=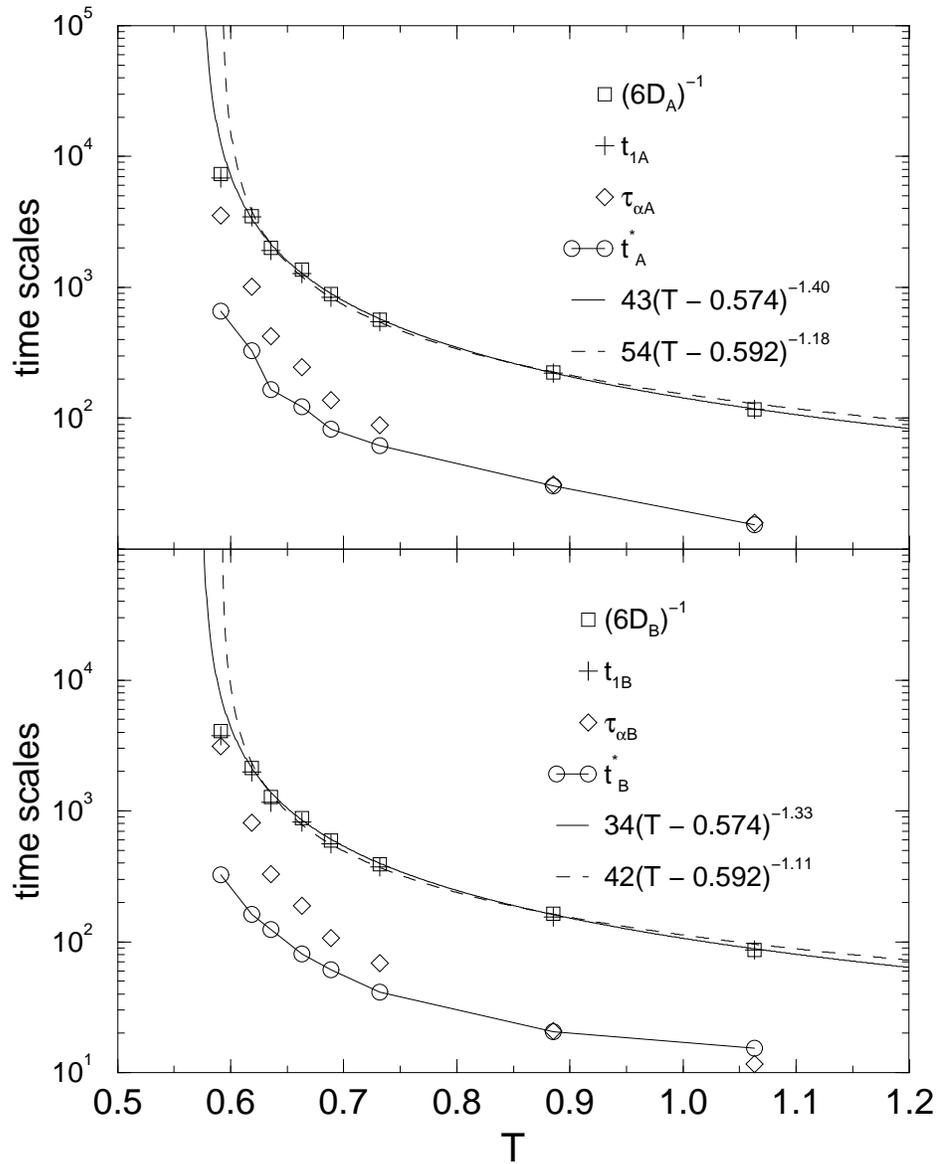, width=12.5cm}
\caption{ 
Time-scales vs. temperature; 
$(6D)^{-1}$ 
(squares, from figure \ref{fig:R2(t)}), 
$t_1$ (plus, from figure \ref{fig:R2(t)}), 
$t_\alpha$ (diamonds, from figure \ref{fig:FsP}),
and $t^*$ (circles, from figure \ref{fig:Alpha2AB}).
The lines are power-law fits to $(6D)^{-1}$ (see text).
}
\label{fig:TimeScales}
\end{figure}

In this section, we compare the measures of time-scales found in the 
previous sections.
In figure  \ref{fig:TimeScales} is plotted $(6D)^{-1}$ 
(squares, from figure \ref{fig:R2(t)}), 
$t_1$ (plus, from figure \ref{fig:R2(t)}), 
$t_\alpha$ (diamonds, from figure \ref{fig:FsP}),
and $t^*$ (circles, from figure \ref{fig:Alpha2AB}). If $t_1$ is in the diffusive
regime, the relation $(6D)^{-1}=t_1$ should hold, see equation \ref{R2Einstein}.
This is seen in figure \ref{fig:TimeScales} to hold to a good
approximation; the largest deviation is 8\%, with $t_1$ being 
smaller than $(6D)^{-1}$, thus confirming that $t_1$ is a good estimate 
of the onset of the diffusive regime, as argued in section \ref{sec:R2}.
$t^*$ is found to be roughly a factor $10$ smaller than  $t_1$, and
$(6D)^{-1}$ (the ratio changes from roughly 6 at high temperatures, 
to  roughly 11 at low temperatures). Also shown in figure \ref{fig:TimeScales}
are attempts to fit $(6D)^{-1}$ to a power-law. 
According to the asymptotic predictions 
of ideal MCT,   $(6D)^{-1}$  should have the same temperature dependence 
as  $\tau_\alpha$ (apart from a constant factor), which means that we should 
find  the  same $T_c$ and $\gamma$. Consequently the fitting was done in the 
temperature range which gave the ``best'' power-law fit to 
$\tau_\alpha$, i.e. by excluding the lowest temperature. This 
fit is shown, for the A and B particles respectively, as the 
full curves in figure \ref{fig:TimeScales}, and the parameters 
and error-bars are given in table \ref{table:DPowerFits}. The 
fit to the data are reasonable, but the estimated $T_c$ deviates
from the one found from $\tau_\alpha$, which is in
contradiction with MCT. The dashed curves in figure \ref{fig:TimeScales}
are the results of power-law fits where $T_c$ was set to have 
the value found for $\tau_\alpha$, $T_c = 0.592$. This also 
fits the data reasonable, but now the exponents $\gamma$ are different
from the ones found from $\tau_\alpha$, thus illustrating that 
$\tau_\alpha$ and $(6D)^{-1}$  has different temperature dependence
(as can be seen directly in figure \ref{fig:TimeScales}).
Also Kob and Andersen find different temperature dependence for the
diffusion and the relaxation \cite{Kob94,Kob95a}.

\begin{table}
\center{
\begin{tabular}{|c|c|c|c|c|c|} \hline
 Type &$T_{cut}$ & $\tau_0$          & $T_{c}$ & $\gamma$ \\ \hline \hline
 A    &$0.60$    & $43\pm 3$ & $0.574 \pm 0.005$ & $1.40 \pm 0.09$ \\ \hline 
 A    &$0.60$    & $53\pm 4$ & $0.592 $         & $1.18 \pm 0.03$ \\ \hline \hline
 B    &$0.60$    & $34\pm 2$ & $0.573 \pm 0.003$ & $1.33 \pm 0.05$ \\ \hline 
 B    &$0.60$    & $42\pm 3$ & $0.592 $         & $1.11 \pm 0.03$ \\ \hline 
 \end{tabular}
}
\caption{
Parameters found by fitting $(6D)^{-1} = \tau_0(T - T_c)^{-\gamma}$,
excluding  the lowest temperature ($T_{cut}=0.60$). The second set of
parameters for each type of particles is for $T_c$ being forced to 
have the value found for $\tau_\alpha$, $T_c = 0.592$.
The error-bars indicate the 
68.3\% confidence interval, as reported by Gnuplot.
}
\label{table:DPowerFits}
\end{table}

\begin{figure}
\epsfig{file=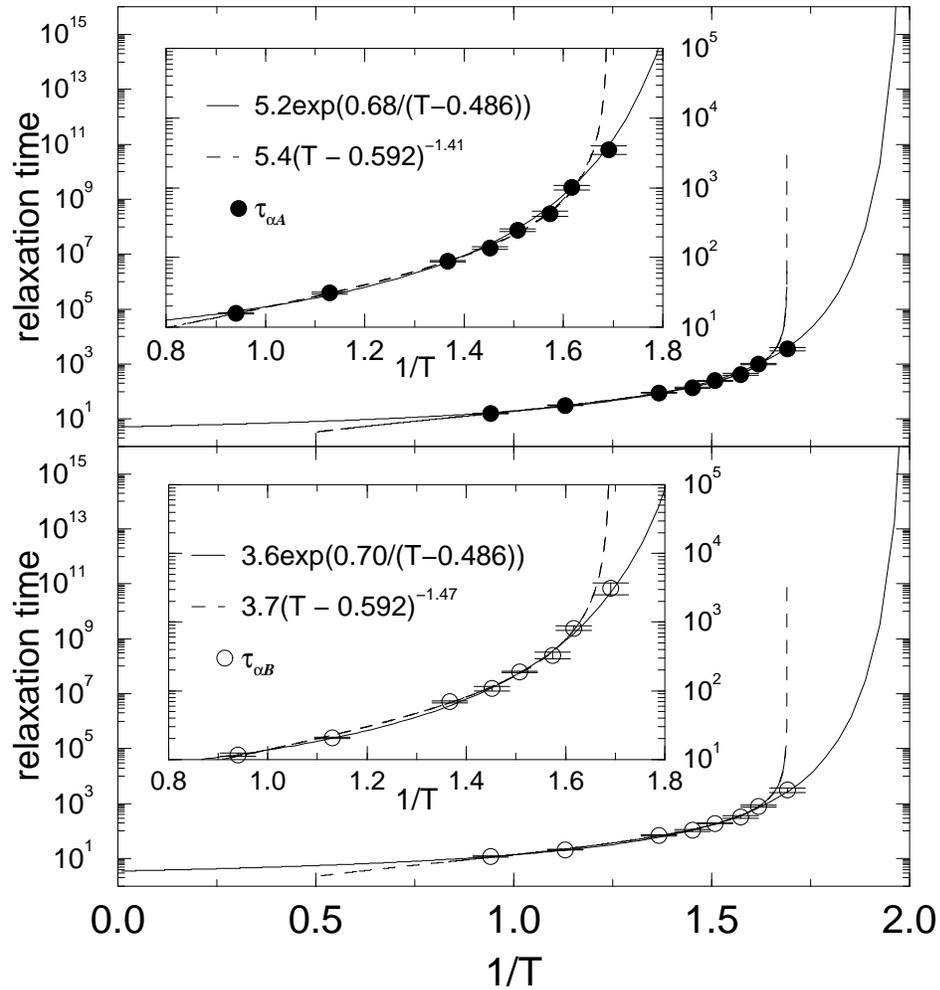, width=12.5cm}
\caption{
Arrhenius plot of the relaxation times, $\tau_\alpha$. Full curves: fits
to the Vogel-Fulcher law (see table \ref{table:VogelF}), extrapolated 
to the laboratorie glass-transition (see text). Dashed curves: power-law
fits from figure \ref{fig:FsP}.
}
\label{fig:TauVF}
\end{figure}

Figure \ref{fig:TauVF} shows the alpha relaxation times $\tau_{\alpha A}$
and $\tau_{\alpha B}$, in the so-called ``Arrhenius-plot'', i.e. as a function of
$1/T$ and with a logarithmic y-axis. In this plot, a Arrhenius
dependence of the relaxation time, $\tau_{\alpha} = \tau_0\exp(E/T)$ would
give a straight line, which is clearly not the case. Or in the 
terminology of Angel \cite{Angell85}, the liquid is ``fragile''
(non-Arrhenius), as opposed to ``strong'' (Arrhenius).
Included as full lines in figure \ref{fig:TauVF} are fits to 
the Vogel-Fulcher law,  $\tau_{\alpha} = \tau_0\exp(A/(T-T_0))$, 
which is often seen  to fit the relaxation times over
a range of temperatures in experiments.
The fits are extrapolated, to include the
time $\tau_{\alpha}\approx 3\cdot 10^{15}$ which in Argon units
(see section \ref{sec:method}) corresponds to $\tau_{\alpha}\approx 10^3$ 
seconds, i.e. including 
the time scales involved in  the laboratory glass-transition 
(see introduction). This extrapolation over 12 decades in time 
(from information covering 2.5 decades), should of course not 
be taken too seriously, but is included here to illustrate the large
differences in time scales. Included as dashed lines in figure \ref{fig:TauVF} 
are the power-law fits from figure \ref{fig:FsP}.

\begin{table}
\center{
\begin{tabular}{|c|c|c|c|c|c|} \hline
 Type &$\tau_0$     & $A$ & $T_0$ \\ \hline 
 A    &$5.2\pm 0.9$ & $0.68 \pm 0.07$ & $0.486 \pm 0.009$ \\ \hline
 B    &$3.5\pm 0.7$ & $0.70 \pm 0.08$ & $0.486 \pm 0.010$ \\ \hline
 \end{tabular}
}
\caption{Parameters found by fitting the relaxation times, 
$\tau_{\alpha A}$ and $\tau_{\alpha B}$ to the Vogel-Fulcher law,  
$\tau_{\alpha} = \tau_0\exp(A/(T-T_0))$.
}
\label{table:VogelF}
\end{table}

\newpage
\section{Finite Size Effects}

To estimate  finite size-effects we in this section present 
results from a sample with $N=1000$ particles which is otherwise 
equivalent with the samples described in the previous sections 
(i.e. same density and total energy per particle).

\begin{figure}
\epsfig{file=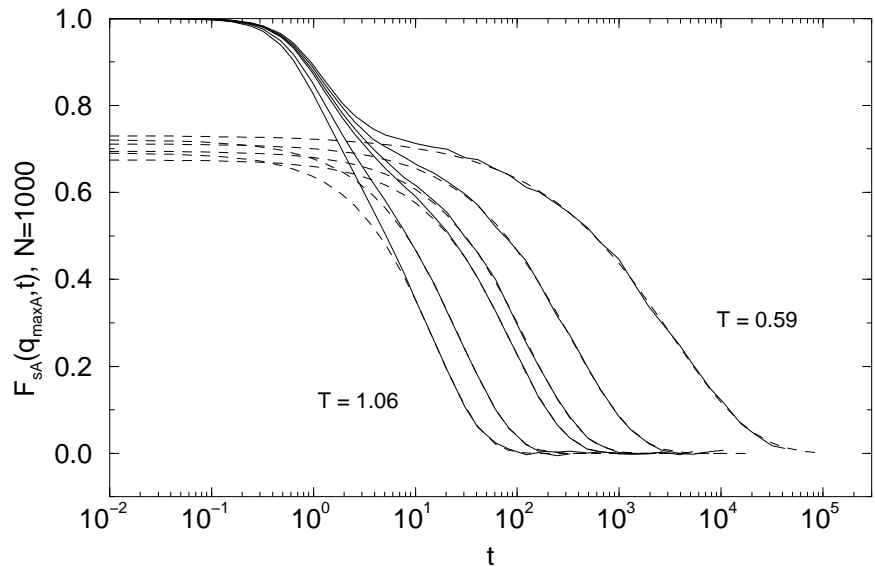, width=11.5cm}
\caption{ a) The self part of the intermediate scattering function
for the A particles, $F_{sA}(q_{maxA}=7.5, t)$ for a sample with N=1000
particles (compare figure \ref{fig:Intermediate}, $T=0.62$ (second lowest) 
and $T=0.66$ (fourth lowest) are missing).
}
\label{fig:IntermediateN1k}
\end{figure}

\begin{figure}
\epsfig{file=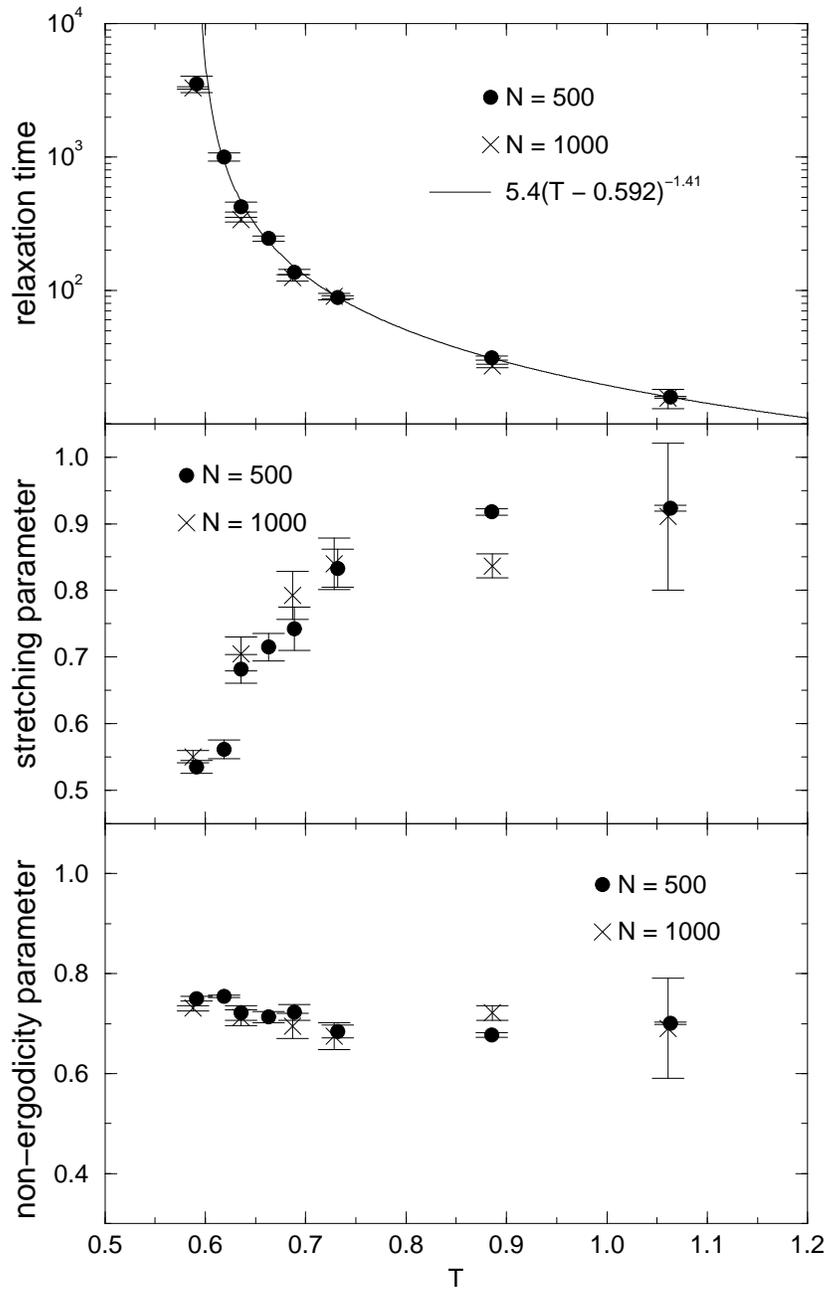, width=11cm}
\caption{ 
Fitting parameters used in figure \ref{fig:IntermediateN1k} for the N=1000 sample.
a) Relaxation times $\tau_{\alpha A}$. 
Power-law reproduced from fig. \ref{fig:Intermediate}a.
b) Stretching parameters $\beta_A$. 
c) Non-ergodicity parameters, $f_{cA}$.  
}
\label{fig:FsPN1k}
\end{figure}

Figure \ref{fig:IntermediateN1k} shows the self part of 
the intermediate scattering function for the A particles for 
the N=1000 system, together with fits to stretched exponential
(dashed lines), using the same fitting procedure as for the N=500 samples
(figure 
\ref{fig:Intermediate}). The resulting fitting parameters are shown 
in figure \ref{fig:FsPN1k} together with those found for N=500 
(figure \ref{fig:FsP}). The results for the two system sizes are 
seen to be in reasonable agreement, indicating that the 
results in the previous sections do \emph{not} depend strongly 
on system size. In \cite{Kim99} a decrease in $\tau_\alpha$ by a factor of $\approx 30$ 
was found at low temperatures for a soft sphere system 
by going from $N=108$ to $N=10000$. 
At the moment it is unclear if this much larger 
change in relaxation time is due to the larger difference in system size,
starting from a smaller system, or other differences between the 
two systems.

\section{Conclusions}

The original motivation for investigating the binary Lennard-Jones
system described here, was that it was already demonstrated
to exhibit hopping \cite{Wahn91}. The first goal of the present work was 
to test if the hopping was still there if the cooling was done under 
(pseudo-) equilibrium conditions.
Unfortunately there is not a single  
condition that is known to be sufficient for the system to be equilibrated.
The different possibilities include: No drift in static 
properties (temperature, potential energy, pressure, etc.), 
long time dynamics being diffusive, relaxation functions such as
$F_{s}(q, t)$ decaying to zero. These are all necessary conditions, 
but none of them are (known to be) sufficient.
In the present work, it was found that the most 
sensitive condition (i.e. requiring the longest equilibration time), 
was that $F_{s}(q, t)$ decays to zero, \emph{and} that it does so
in the same manner for the three independent samples, i.e. with 
the relaxation time and stretching exponent being within reasonable
agreement. 
Assuming that this condition is sufficient, it is concluded that the 
liquid is  in equilibrium  at all the temperatures presented here, 
and thus that the 
hopping found \emph{is} a feature of the equilibrium liquid. 

Although it was not a goal in itself to test the ideal mode
coupling theory (MCT), the results of the simulations was compared to the 
asymptotic predictions of ideal MCT.
This was done for two reasons; I) It provides a convenient way 
of comparing results with other simulations 
(e.g. Kob and Andersen \cite{Kob94,Kob95a,Kob95b}). 
II) It is ``common'' belief that ideal MCT breaks down when hopping dynamics
takes over, and the critical mode coupling temperature $T_c$ thus 
constitutes an estimate of when hopping should start to dominate the 
dynamics. At first hand the last point seems to work fine; the best 
attempt at a power-law fit to $\tau_\alpha$ gives $T_c = 0.592\pm 0.004$, 
which is close to the lowest of the temperatures simulated, where we 
clearly see hopping (figure \ref{fig:GsAt1}). The fact that hopping also 
is present at the second lowest temperature, might then be taken as an 
indication that dynamical transition from the ``MCT regime'' to the 
``hopping regime'' is not a sharp transition, but takes place over
some temperature interval. However, the big problem with this scenario
is the failure to identify any temperature range, where the 
asymptotic predictions of ideal MCT holds; There is no indication of 
TTSP, and the temperature dependence
of the time-scales for diffusion and relaxation are clearly different.

In the full version of ideal MCT (as opposed to the asymptotic 
predictions used in the present work), the exponents of the 
power-laws discussed here4 are not 
free parameters, but are determined from the static structure factor 
\cite{Gotze92,Kob97b,Gotze99}. Calculating the exponents from the 
static structure 
factor is however a difficult numerical task, which has only been 
done for a few systems \cite{Kob97b}. This has not been attempted
for the system used here.
In an  attempt to include the hopping dynamics in MCT,
the so-called  extended mode coupling theory has been 
developed introducing  an ``hopping-parameter'', as a extra fitting parameter \cite{Gotze99}.
It has not been attempted to apply extended MCT to the 
system investigated here.

\chapter{Inherent Dynamics}
\label{EnergyLandscape}

The dynamics of the model glass-forming liquid described in 
the previous chapter, is here analyzed in terms of its 
``inherent structures'', i.e. local minima in the potential energy. 
In particular, we compare the self part of the intermediate scattering
function, $F_s(q,t)$, with its inherent counterpart $F_s^I(q,t)$ 
calculated on a time series of  inherent structures. $F_s^I(q,t)$
 is defined as $F_s(q,t)$ except that the particle 
coordinates at time $t$, 
are substituted with the particle coordinates
in the corresponding inherent 
structure, found by
quenching the equilibrium configuration at time $t$. 
We find that the long time relaxation of $F_s^I(q,t)$ can be fitted 
to stretched exponentials, as is the case for  $F_s(q,t)$.
Comparing the fitting parameters from  $F_s(q,t)$ and  $F_s^I(q,t)$ 
we conclude, that below a transition temperature, $T_x$, the dynamics
of the system can be  separated into thermal vibrations around
inherent structures 
and transitions between these.

The main conclusions of this chapter can be found in paper \QuenchPaper . 
In the present chapter we introduce the concepts of ``energy landscape'' and
``inherent dynamics'', followed by a  summary of paper  \QuenchPaper .
Following this
the data shown in paper  \VigoPaper ~ and some additional data
is discussed, and a conclusion is given.

\newpage
\section{The Potential Energy Landscape}

\begin{figure}
\epsfig{file=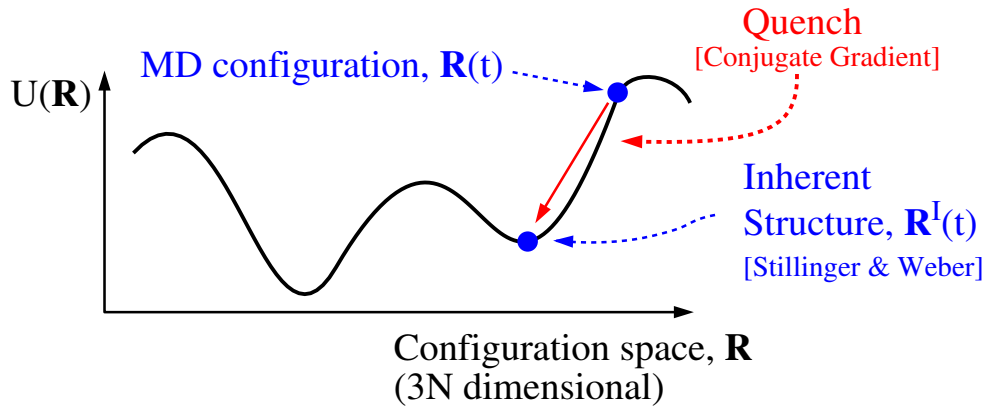, angle=-90, width=13.0cm}
\caption{
Basic concepts in the potential  energy landscape. The energy landscape,  $U(\mathbf R)$,
contains a large number ($\propto \exp{(kN)}$ \cite{Stillinger82}) of inherent structures, 
which are local minima  in $U(\mathbf R)$.
Any configuration $\mathbf R(t)$ can be mapped by a quench (local minimization 
of the potential energy), into a corresponding inherent structure, $\mathbf R^I(t)$. 
The fact that the $3N$-dimensional configuration space is drawn here as 1-dimensional 
can be misleading; the  configuration space \emph{is} $3N$-dimensional, and should be
thought of as such (difficult as it might be). One might think 
of the x-axis in this plot as a particular direction in 
configuration space, connecting two inherent structures. It 
should \emph{not} be thought of as an order-parameter 
(Readers who find this last sentence strange can safely ignore it).
}
\label{fig:NRG}
\end{figure}

The potential energy landscape \cite{Goldstein69,Stillinger82,Stillinger83,Stillinger88,Arkhipov94,Stillinger95,Heuer97,Wallace97,Sastry98,Angelini98,Angelini99,Schultz98} of an 
atomic system (i.e. particles without internal 
degrees of freedom) is simply the potential energy, 
as a function of the $3N$ particle coordinates. Letting $\mathbf R$
denote the $3N$ dimensional vector describing the state point of the system 
(i.e. its  position in  the $3N$ dimensional configuration space), 
we write the  energy landscape as $U(\mathbf R)$, see figure \ref{fig:NRG}. 
The behavior of the system can be viewed in terms of the state point, 
$\mathbf R(t)$, moving on  the energy landscape surface, $U(\mathbf R)$. 
This surface contains  a 
large number of minima, termed ``inherent structures'' by Stillinger and 
Weber \cite{Stillinger83}. The inherent structures are characterized by 
zero gradients in the potential, and they are thus
mechanically stable configurations. 
The inherent structures are  separated
by saddle points acting as energy barriers. 
Each inherent structure has a basin of attraction, in which a local minimization
of the potential energy (a ``quench'') will map the state point $\mathbf R(t)$
to the corresponding inherent structure  $\mathbf R^I(t)$.

So far all we have done is defining the energy landscape. 
For a system like the binary Lennard-Jones mixture investigated in 
the previous chapter we 
know $U(\mathbf R)$ exactly; its simply the sum of the pair-potentials 
(equation \ref{BLJpot})\footnote{The fact that we know $U(\mathbf R)$ does
not necessarily mean that we can find all the inherent structures. This 
can only be done for small systems ($N$ less than approxemately $30$ depending 
 on other factors such as density and the potential), see eg. \cite{Heuer97}. 
This is not the approach we take here.}. Of course having a well-defined quantity
only helps, if it can tell us something about what we are interested in, 
i.e. in this case  the dynamics of a glass-forming liquid. This is the main 
question we deal with in this chapter. First we note that for 
the simulations presented in the previous chapter, the dynamics is
governed by Newton's second law, which using the concepts introduced above 
can be written as:
\begin{eqnarray}
   \frac{d^2}{dt^2}\mathbf R(t) = -\mathbf M^{-1} \nabla U(\mathbf R(t))
\end{eqnarray}
where $\mathbf M$ is a ($3N$x$3N$) diagonal matrix, with the appropriate 
masses on the diagonal. In this sense the dynamics is \emph{defined} by 
the energy landscape. This is however obviously not telling us anything
new. What we want to know is, can we think of the dynamics in
terms of the following scenario; the dynamics of the liquid is 
separated into (thermal) vibrations around inherent structures and 
(thermally activated) transitions between these. Presented with this scenario, 
one might very well
ask: How can a system with constant total energy (e.g. the system simulated
in the previous chapter) perform thermally activated processes? The answer
is that one should think of the transitions between inherent structures 
as involving only a few  (local) degrees of freedom, while the remaining 
degrees of freedom provides the heath bath. 

In his classic paper from 1969 Goldstein \cite{Goldstein69} argued 
that there exists a transition temperature, which we will term $T_x$,
 below which the flow of viscous 
liquids is dominated by potential barriers high compared to thermal 
energies, while  above $T_x$  this is no longer true.
Or in other words; below $T_x$ the dynamics is governed by the 
``vibrations plus transitions'' scenario given above. 
Goldstein gave as a (very) rough estimate, that the shear relaxation time at 
$T=T_x$ is on the order of  $10^{-9}$ seconds.
Later it was noted by Angell \cite{Angell88}, that 
experimentally it is  often found that the shear relaxation 
time is on the order of $10^{-9}$ seconds at the mode coupling  
temperature, $T_c$.
This lead to the argument that Goldstein's transition temperature, $T_x$,
is identical to the mode coupling temperature, $T_c$. A similar argument
was recently given by Sokolov \cite{Sokolov98}.
\begin{figure}
\epsfig{file=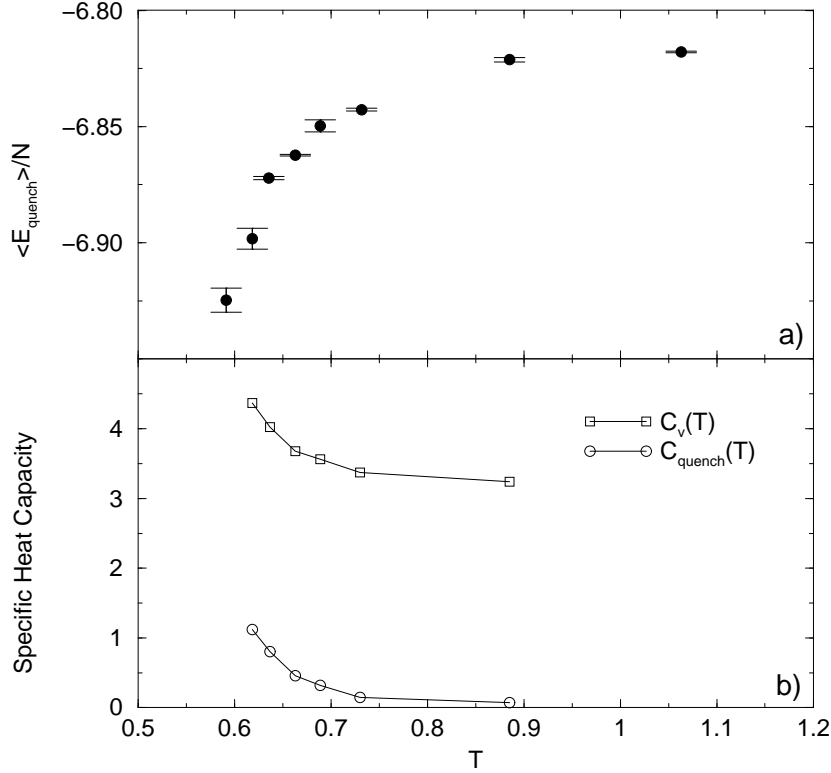, width=11.0cm}
\caption{
a) Mean value of the quenched energy, $\langle E_{quench} \rangle$ per 
particle, as a function of temperature.
b) The heat capacity at constant volume, $C_v(T)$,
and the corresponding ``quenched heat capacity'', $C_{quench}(T)$ (see text).
}
\label{fig:EqT}
\end{figure}

\label{sec:CvQ}
Recently, Sastry, DeBenedetti and Stillinger~\cite{Sastry98} 
 demonstrated that 
the onset of non-exponential relaxation ($\beta <1$) in
a simulated glass-forming
Lennard-Jones liquid is associated with a change in the system's
exploration of its potential energy landscape. In figure \ref{fig:EqT}a we
present data similar to the data presented 
in \cite{Sastry98}.  $\langle E_{quench} \rangle (T)$ is here
the mean value of the energy of inherent structures,
found by quenching configurations from a normal MD run equilibrated at
the temperature $T$. Like in \cite{Sastry98} we find
that this approaches a constant value at high temperatures, where the 
relaxation becomes exponential 
($\beta \approx$ 1, compare  fig. \ref{fig:FsP}). Similar results was 
found in \cite{Jonsson88} for the quenched enthalpy (letting the
volume change during the quench).

When plotting the total energy per particle, $E_{tot}(T)/N$ 
(in figure \ref{fig:EandPvsT}), 
we found  that the  specific  heat capacity at constant volume, 
$C_v(T)$, increases as the system is cooled.  In figure \ref{fig:EqT}b 
this is shown  explicitly, with $C_v(T)$ calculated from eq. 
\ref{Cv} (using central differences).  Also shown in 
figure \ref{fig:EqT}b is the ``quenched heat capacity'', $C_{quench}(T)$,
calculated in the same way, but with  $\langle E_{quench} \rangle (T)$
substituted for $\langle E_{tot} \rangle (T)$. 
Clearly the increase 
in $C_v(T)$ upon cooling is related to the increase in $C_{quench}(T)$.
  
\section{The Inherent Dynamics}

\begin{figure}
\epsfig{file=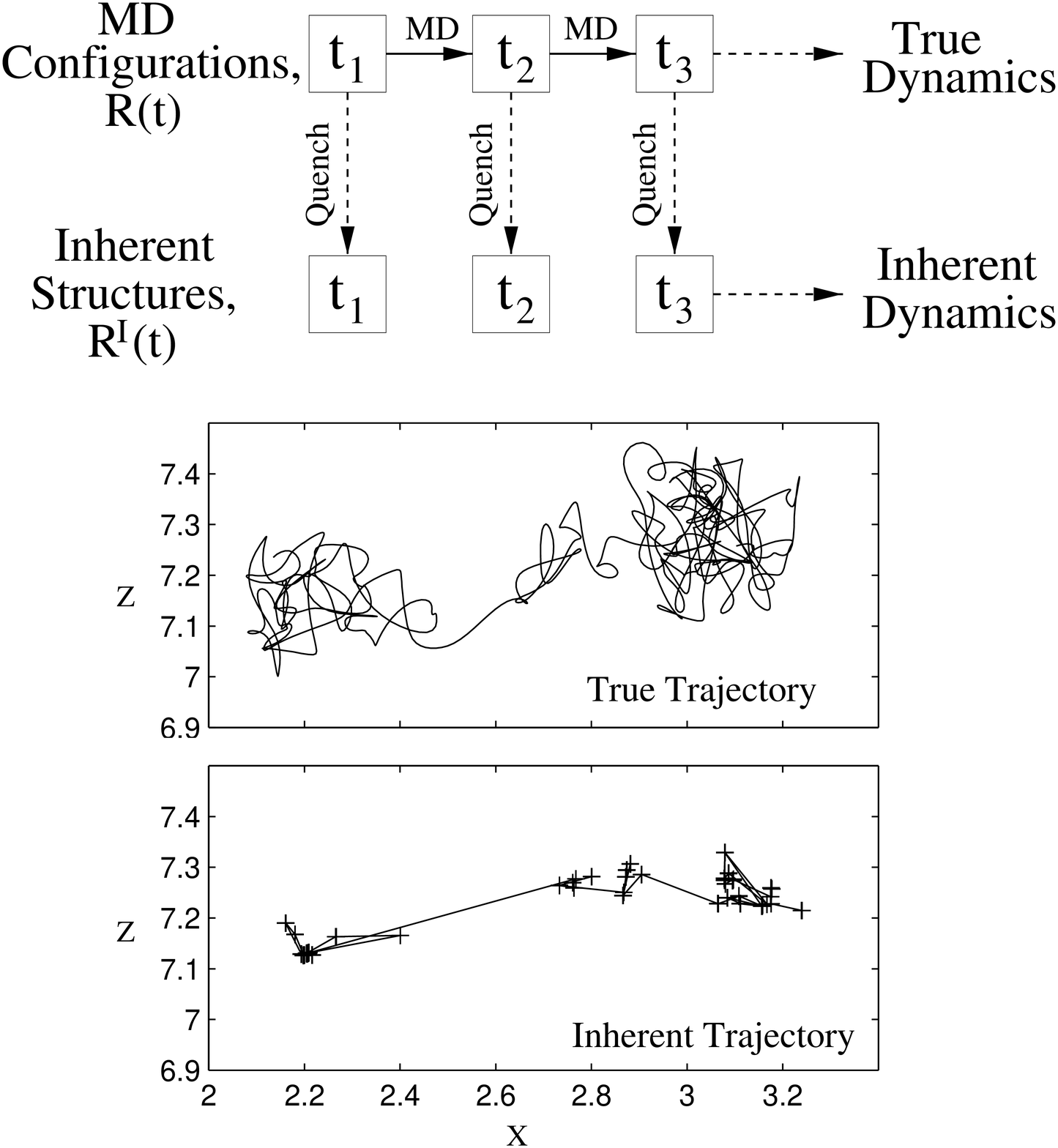, width=12.0cm}
\caption{
Schematic describing the principle of the ``inherent
dynamics'' approach. 
A time series of configurations ${\bf R}(t)$ of the
equilibrated liquid is generated using normal (Newtonian) MD.  
Configurations in  ${\bf R}(t)$ are quenched to produce their
corresponding inherent structures ${\bf R}^I(t)$.  Successive inherent
structures then form a time series which we use to calculate
``inherent dynamical'' quantities such as the inherent mean square 
displacement,  $\langle r^2(t)\rangle^I$, and the
inherent intermediate scattering function, $F_{s}^I(q, t)$.
Bottom part of the figure: Applying the inherent
dynamics  approach to the   trajectory of a hopping particle
(see text).}
\label{fig:Schematic}
\end{figure}

The basic idea of the ``inherent dynamics'' approach is the 
following (see figure \ref{fig:Schematic}); 
After equilibration at a given temperature, a time series of 
configurations,  ${\mathbf R}(t)$, is produced by a normal MD simulation.
Each of the configurations in  ${\mathbf R}(t)$ is now quenched\footnote{ 
In the present work this minimization
was done using the conjugate gradient method
\cite{Press}, which uses a succession of line minimizations in
configuration space to minimize the potential energy.},
to produce the corresponding time series of inherent structures, 
${\mathbf R}^I(t)$. We now have two ``parallel'' time series of configurations.
The time series  ${\mathbf R}(t)$ defines the ``true dynamics'', which is
simply the normal (Newtonian) MD dynamics as presented in the previous 
chapter, described by $\langle r^2(t)\rangle$, $F_{s}(q, t)$, etc. 
In a completely analogous 
way, we define the ``inherent dynamics'' as the dynamics described by
the time series ${\mathbf R}^I(t)$. In other words: If a function 
describing the true dynamics (eg. $\langle r^2(t)\rangle$ or $F_{s}(q, t)$)
is calculated by $f({\mathbf R}(t))$, then the corresponding 
function in the inherent dynamics (eg. $\langle r^2(t)\rangle^I$ or 
$F_{s}^I(q, t)$) is calculated in exactly the same way, except using the 
time series of inherent structures: $f({\mathbf R^I}(t))$.  

If the true dynamics can be separated into vibrations around inherent
 structures, 
and transitions between these, as stated in the ``vibrations plus transitions''
scenario given above, then the inherent dynamics can be thought of as what 
is left after the thermal vibrations is removed from the true dynamics.

In the bottom part of figure 
\ref{fig:Schematic} the inherent dynamics approach is 
applied to the trajectory of the hopping particle, which was 
shown in figure \ref{fig:hop}; All the configurations that 
were used to plot the true trajectory were quenched, and the
 position of the particle in the resulting time series of inherent structures
is plotted as the ``inherent trajectory''. The quenching
procedure is seen to remove the vibrations in the true
trajectory. The motion that seems to be left from the vibrations
(eg. a jump from $x \approx 2.2$ to $x \approx 2.4$) will be discussed
in section \ref{sec:transitions}.

\newpage 
Paper \VigoPaper ~and \QuenchPaper ~explores the possibilities 
of comparing the true and inherent dynamics of the model 
glass-former described in the previous chapter. The concept 
of inherent dynamics was first introduced in paper  \VigoPaper 
~(without using that name). In that paper the 
true and inherent versions of the mean square displacement and the
van Hove correlation function, was compared on 
a qualitatively level. Paper \QuenchPaper ~compares the 
true and inherent version of the self part of the intermediate 
scattering function, which can be done on a quantitative level 
by means of stretched exponentials (see section \ref{sec:Fs}).
This turns out to have important consequences, which is why we here
discuss this paper first.

\section{Paper \QuenchPaper}

In this paper the concept of inherent dynamics is applied to 
the self part of the intermediate scattering function, i.e. we 
compare $F_{s}({q_{max}}, t)$, with its inherent counterpart 
$F_{s}^I({q_{max}}, t)$. 
As explained above $F_{s}^I({q}, t)$ is defined in the 
same way as $F_{s}({q}, t)$, except that the 
normal particle coordinates, ${\mathbf r}_j(t)$, are substituted by 
the corresponding coordinates in the inherent structures, ${\mathbf r}^I_j(t)$
(compare equation \ref{FsDef})\footnote{
 ${\mathbf r}^I_j(t)$ is here the 3-dimensional vector describing the 
position of the $j$'th particle in the inherent structure ${\mathbf R}^I(t)$.}:
\begin{eqnarray}
    F_{s}^I({\mathbf q}, t) &\equiv& 
      \langle 
         \cos{{\mathbf q} ({\mathbf r}^I_j(t) - {\mathbf r}^I_j(0))} 
      \rangle \label{FsIDef}
\end{eqnarray}

\begin{figure}
\epsfig{file=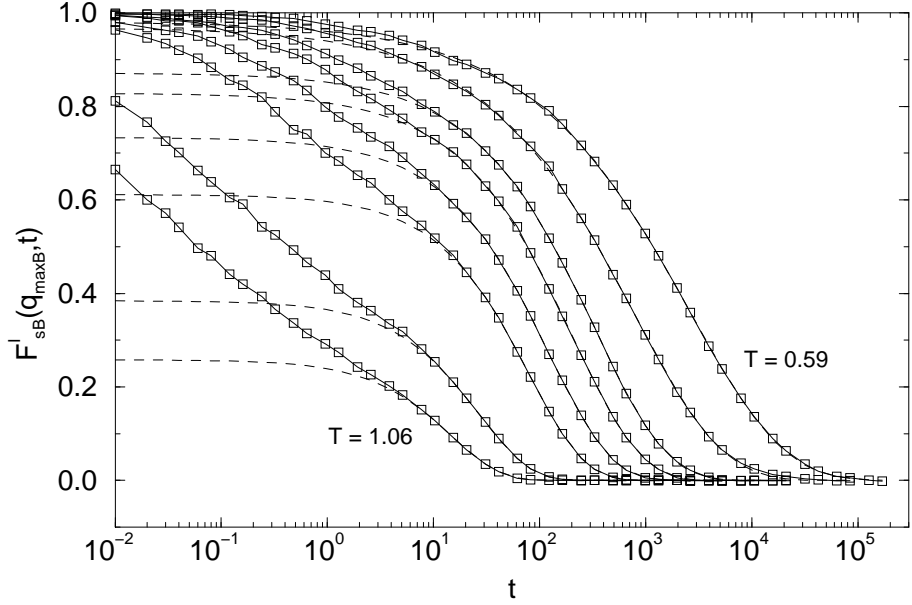, width=12cm}
\caption{ 
The self part of the inherent intermediate scattering function for the
B particles, $F_{sB}^I(q=8.1, t)$, corresponding to figure
3b in paper \QuenchPaper.  The dotted lines are fits to stretched exponentials:
$f^I(t) = f_c^I\exp(-(t/\tau^I_\alpha)^{\beta^I})$ 
}
\label{fig:FsIB}
\end{figure}

In figure 4b in paper \QuenchPaper ~$F_{s}^I({q_{max}}, t)$ is plotted 
for the A particles, at the 8 temperatures studied. 
Here we plot the similar data for the B particles in figure \ref{fig:FsIB}. 
In both cases we find that the 
long time relaxation of $F_{s}^I({ q_{max}}, t)$ can be fitted to 
stretched exponentials, like is the case for $F_{s}({q_{max}}, t)$. 
This is an important finding, since it enables
us to make an quantitative comparison between $F_{s}({ q_{max}}, t)$ and
$F_{s}^I({ q_{max}}, t)$, by comparing the two sets of fitting parameters.
In the following, the set of fitting parameters for $F_{s}({q_{max}}, t)$
is denoted $\{ \tau_\alpha, \beta, f_c \}$, while the corresponding set for  
$F_{s}^I({ q_{max}}, t)$ is denoted $\{ \tau_\alpha^I, \beta^I, f_c^I \}$

The key question now is: If the true dynamics follows the ``vibrations plus
transitions'' scenario, i.e. if the dynamics can be separated into 
vibrations around inherent structures, and transitions between these, how is 
$\{ \tau_\alpha^I, \beta^I, f_c^I \}$ expected to be related to  
$\{ \tau_\alpha, \beta, f_c \}$? To answer this question, we assume
that the initial relaxation in $F_{s}({ q}, t)$ is due to vibrations, 
which is the widely accepted explanation (see section \ref{sec:Fs}). 
If this is 
the case, then we expect the quenching procedure to remove the initial 
relaxation (since it removes the vibrations), which means that  
$F_{s}^I({ q}, t)$ can be thought of as $F_{s}({ q}, t)$ with the initial 
relaxation removed. This in turn means, that  $F_{s}^I({ q}, t)$ should be 
identical to the long time relaxation of $F_{s}({ q}, t)$, but rescaled 
to start at unity (by definition):  
$\{ \tau_\alpha^I, \beta^I, f_c^I \} = \{ \tau_\alpha, \beta, 1\}$.
 
The identity of the relaxation times, and the stretching parameters, 
$\{ \tau_\alpha^I, \beta^I \} = \{ \tau_\alpha, \beta\}$, will probably 
be true for any ``coarse-graining'' of the dynamics we might apply; if we 
eg. decide to add a small random displacement to all the particles (instead 
of quenching), we would still expect the long time relaxation to have the 
same shape and characteristic time, i.e. 
$\{ \tau_\alpha^I, \beta^I \} = \{ \tau_\alpha, \beta\}$. It is when we
find $f_c^I =1$ we know that the procedure we have applied removes the 
vibrations (or in more general terms: removes the part of the dynamics
responsible for the initial relaxation).

\begin{figure}
\epsfig{file=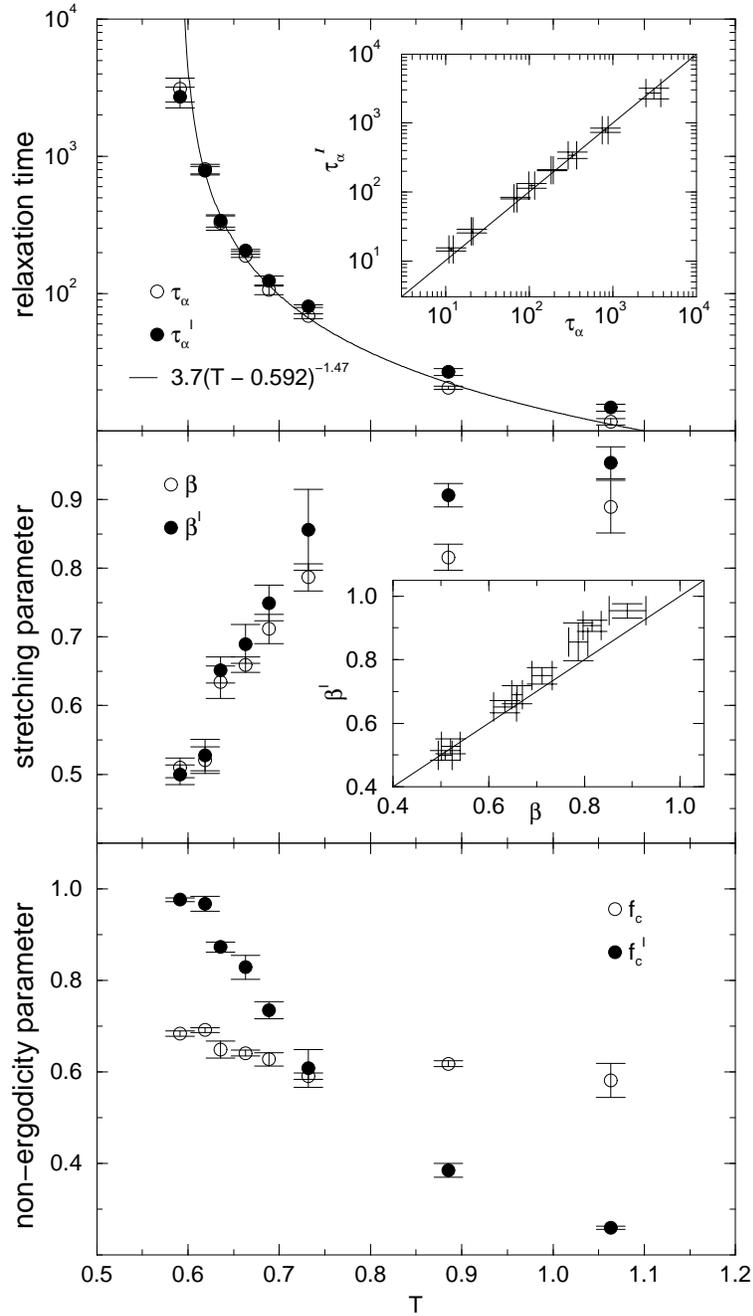, width=10cm}
\caption{ 
Parameters describing the fits of $F_{sB}(q=8.1, t)$ 
and  $F_{sB}^I(q=8.1, t)$  to
stretched exponentials; (a) The relaxation time
$\tau^I_{\alpha}$ vs. $T$. The solid line is the same power law fit as
in figure \ref{fig:FsP}b.  (b) The stretching parameter $\beta^I$ vs. $T$. (c)
The non-ergodicity parameter, $f_c^I$ vs. $T$.  
}
\label{fig:FsPIB}
\end{figure}

In figure 5 in paper \QuenchPaper ~ the fitting parameters for 
 $F_{s}({ q_{max}}, t)$ and   $F_{s}^I({ q_{max}}, t)$ are compared 
for the A particles. Here we show similar data for the B particles 
in figure \ref{fig:FsPIB}. In both cases we find for all temperatures
$\tau_\alpha^I \approx \tau_\alpha$. The stretching 
parameters are more difficult to compare at high temperatures, 
but they become identical at low temperatures, $\beta^I =\beta $ 
(within the error-bars). Whereas $ f_c$ is roughly constant as
a function of temperature, $ f_c^I$ is clearly seen to approach 
unity, as the system is cooled. We have thus found evidence, that the 
the vibrations plus transitions scenario holds below a transition 
temperature $T_x$, as argued by Goldstein. We estimate that $T_x$
is close to (or just below) the lowest of the temperatures simulated
($T = 0.591 \pm 0.002$).

At the transition we find $\tau_\alpha \approx 3\cdot 10^3$, which 
in ``Argon units'' corresponds to $\tau_\alpha \approx 10^{-9}$ seconds
(see section \ref{sec:method}), i.e. 
Goldstein's estimate of the shear relaxation time at $T_x$. This 
agreement is however probably ``to perfect''; As mentioned above 
Goldstein's estimate is very rough, and it regards the shear relaxation 
time and not $\tau_\alpha$. Consequently an agreement better than ``orders
of magnitude'' should probably be considered a coincidence.

Having found evidence that there exists a transition temperature, $T_x$, 
and estimated the (approximate) position of this, it is natural to 
proceed to check 
Angell's proposal, that $T_x \approx T_c$.  We find that both estimated 
values for $T_c$ ($0.592\pm 0.005$ from relaxation times
and $0.574\pm 0.005$ from diffusion) are in the 
temperature range where $f_c^I$  is approaching unity. To check the 
proposition that  $T_x \approx T_c$ on the system used here, 
is of course somewhat problematic, since it doesn't conform very well to 
the predictions of the mode coupling theory, as discussed in the 
previous chapter. It should be 
noted however, that the arguments given by Angell (and Sokolov), only relates 
to $T_c$ as the temperature where power-law fits to experimental data 
tends to break down, i.e. the ``usage'' of MCT in this argument
is similar to the way we have estimated $T_c$ in the previous chapter, 
and does not require e.g. time-temperature super-position.

At the end of paper \QuenchPaper ~we present results on the nature of transitions
between inherent structures. These will be discussed in section \ref{sec:transitions}.  

\section{Paper \VigoPaper}
\label{sec:VigoPaper}

In paper \VigoPaper ~the concept of inherent dynamics is applied
to the mean square displacement, $\langle \Delta r^2 (t) \rangle$, 
and the distribution of particle displacements, \\
$4\pi r^2G_{s}( r, t)$. 

Comparing $\langle \Delta r^2 (t) \rangle^I$ to 
$\langle \Delta r^2 (t) \rangle$ (figure 1 in paper \VigoPaper), we
find that the quenching procedure removes the plateau seen in  
$\langle \Delta r^2 (t) \rangle$. This is taken as (qualitative) 
evidence for the ``cage''-explanation for the plateau (see section 
\ref{sec:R2}). This conclusion is consistent with the conclusion 
we drew from $F_{s}^I(q, t)$ in the previous section, 
since the ``caging'' is what we 
call vibrations in configuration space.

Comparing  $4\pi r^2G_{s}^I( r, t)$ and $4\pi r^2G_{s}( r, t)$
(figure 2 in paper \VigoPaper), we find that the hopping peak is slightly 
sharper in $4\pi r^2G_{s}^I( r, t)$, and that the first peak is 
moved to the left. A point that is \emph{not} made in paper \VigoPaper
~is the following; if the particles only moved by either ``rattling''
in their local cages, or hopping approximately an inter-particle distance,
then one would expect the first peak seen in $4\pi r^2G_{s}( r, t)$
to be quenched to a delta-peak at $r=0$ in $4\pi r^2G_{s}^I( r, t)$. 
Figure 2b in paper \VigoPaper ~shows that this is clearly not the case. 
The reason for this will be discussed in the next section.

\section{Transitions between inherent structures}
\label{sec:transitions}

\begin{figure}
\epsfig{file=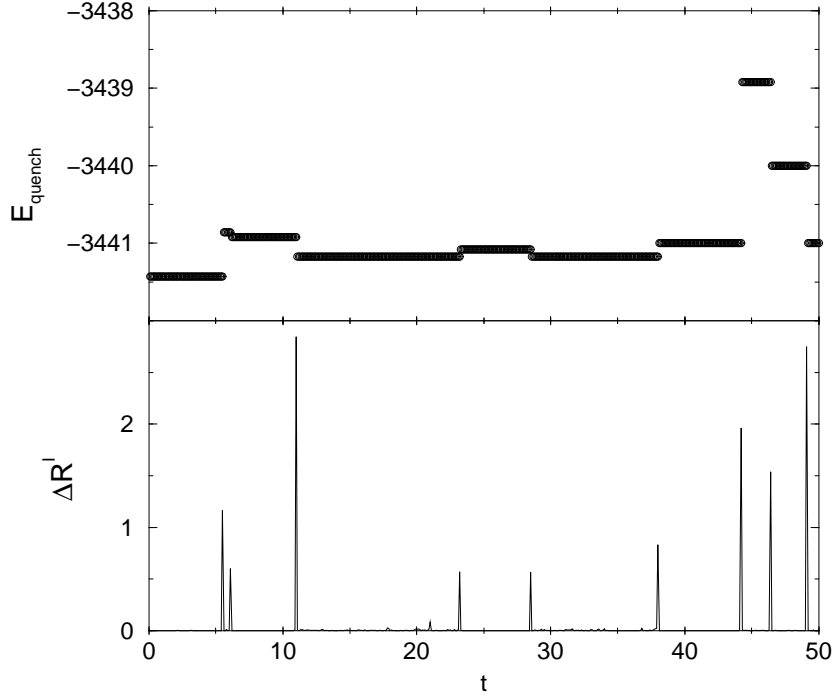, width=11cm}
\caption{
 Identifying transition among inherent structures; a) $E_{quench}$ vs. time.
b) $\Delta  R^I$ (equation \ref{DeltaR}) vs. time. Transitions between 
(the basin of attraction of) inherent structures is indicated by a jump in 
 $E_{quench}(t)$ and a corresponding peak in $\Delta R^I(t)$.
} 
\label{fig:IdentTrans}
\end{figure}

Having established that our lowest temperature (T=0.59) is close to $T_x$ (i.e.
the dynamics can be thought of as separated into vibrations around inherent 
structures and  transitions between these) we are now interested in  studying 
the nature of transitions between inherent structures at that temperature.
We have identified 440 such transitions, by  
quenching the true MD configurations every $0.1\tau$ (i.e. every 10 MD-steps), 
and looking for signatures of the system making a transition from one inherent
 structure to another. In figure \ref{fig:IdentTrans}a is shown 
the energy of the inherent structures as a function of time, 
$E_{quench}(t)$. As expected $E_{quench}(t)$ is found to be constant for some
time-intervals, and then jump to another level, i.e. the system has made a
transition from one inherent structure to another. 
In figure \ref{fig:IdentTrans}b is plotted as a function of time, the distance in
configuration space between two successive quenched configurations 
\cite{Ohmine95}:

\begin{equation}
   \Delta R^I(t) \equiv |\mathbf R^I(t+0.1) - \mathbf R^I(t)| \label{DeltaR}
   = \sqrt{\sum_{j=1}^{N}\left(\mathbf r^I_j(t+0.1) - \mathbf r^I_j(t)\right)^2}
\end{equation}

\begin{figure}
\epsfig{file=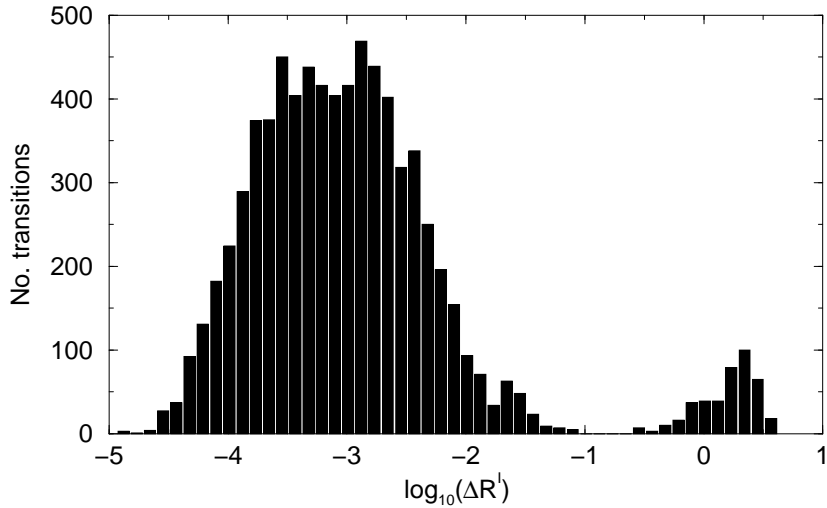, width=11cm}
\caption{
 Distribution of $\log_{10}( \Delta R^I(t) )$. This distribution is not 
 normalized, i.e. the y-axes indicates how many quenches gave values of 
 $\log_{10}( \Delta R^I(t) )$ in the interval on the x-axis. 
} 
\label{fig:DeltaR}
\end{figure}

Each jump in  $E_{quench}(t)$ is associated with a peak in 
$\Delta R^I(t)$, indicating the system has moved to a different 
inherent structure.
A transition might in principle occur between two  inherent structures
with exactly the same (quenched) energy. Such a transition 
would not be seen in   $E_{quench}(t)$, and consequently we use  
$\Delta R^I(t)$ to identify transitions. The distribution of 
$\log_{10}( \Delta R^I(t) )$ is shown in figure \ref{fig:DeltaR}.
The distribution  is seen to have two separated peaks. 
The peak to  the left
(centered around $\log_{10}( \Delta R^I(t) ) \approx {-3})$ is
due to  numerical uncertainties; two configurations within the same basin
of attraction is \emph{not} quenched to the exactly the same configuration. 
The peak on the right is the one containing  the 
physically interesting transitions, which is seen to have 
$\Delta R^I$ on the order of unity. In the following we 
use the condition $\Delta R^I > 0.1$ to identify the 
(physically interesting) transitions.

Figure \ref{fig:DeltaR} shows that most of the quenches results in 
\emph{not} finding a transition (left peak), and only a small fraction 
results in actually finding a transition (right peak). Or in actual 
numbers: Doing 7500 quenches we found 440 transitions. The quenching 
procedure takes a considerable amount of time 
(corresponding to approximately 1000 MD steps), which is why we haven't simply
continued this procedure to find a larger number of transitions. The 
data presented in the following is averaged over the 440 transitions
found using the ``brute force'' method described above, 
and should be considered preliminary. (NOTE: In paper  \QuenchPaper~
results from 12000 transitions are reported. The results are similar 
to the results presented here, except of course with less noise).

\begin{figure}
\epsfig{file=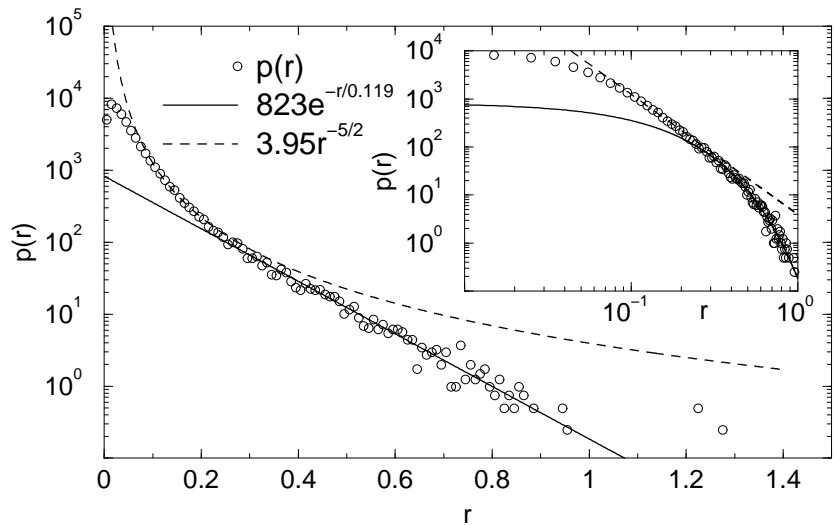, width=11cm}
\caption{
Distribution of particle displacements during transitions between 
consecutive inherent structures at T=0.59. 
Full curve is a fit to an exponential, for $r>0.2$.
The dotted curve is a fit to a power-law with exponent $-5/2$, 
as predicted by linear elasticity theory (see text).
} 
\label{fig:DisplDistr}
\end{figure}

For each transition, we monitor the displacements of the particles 
from one inherent structure to the other. The distribution of all 
particle displacements is shown in Fig. \ref{fig:DisplDistr}. 
While many particles move only a small distance ($r<0.2$)
during the transition, a number of particles move farther, and in
particular, we find that the distribution for $r>0.2$ is to 
a good approximation exponential. At present we have no explanation 
for this. The dotted curve is a fit to a power-law with exponent $-5/2$, 
which is a prediction from linear elasticity theory \cite{Dyre99a,Dyre99b}, 
describing the displacements of particles in the surroundings of
a local ``event''. This  power-law fit does not look very convincing 
by it self, but we note that the exponent was not treated as a fitting
parameter (i.e. only the prefactor was fitted), and the power-law is
\emph{expected} to break down at small displacements, since these corresponds
to distances far away from the local event, and is thus not seen in
our finite sample. From the change in behavior of $p(r)$ at $r \approx 0.2$,
it is reasonably to think of particles with displacements larger than $0.2$, 
as those taking part in the local event, and the rest of the particles as
being in the surroundings, adjusting to the local event. Using this 
definition it is found that on average approximately 10 particles participate 
in an event.

Figure \ref{fig:DisplDistr} has two important consequences with regards
to points discussed earlier in this thesis; 
i) The distribution of particle displacements during transitions 
shows no preference for displacement of the average inter-particle 
distance ($\approx 1$ in the used units). This 
shows that the hopping indicated by the secondary peak in 
$4\pi r^2 G_s(r,t)$ (figure \ref{fig:GsAt1} and \ref{fig:VHBt_star}) 
at low temperatures
is {\bf not} due to transitions over single energy barriers. A
(correlated) sequence of the these is needed, to ``build up'' the
secondary peak. This is consistent with the behavior seen in 
the inherent trajectory in figure \ref{fig:Schematic}; The jump does
not happen in one step, but through a number of ``intermediate'' inherent
structures.
ii) Particles in the surroundings of a local event are displaced 
small distances, to adjust to the larger displacements occurring
in the local region of the event itself. This kind of motion 
is very hard to detect in the true dynamics, since it is 
dominated by the thermal vibrations. Presumably this kind of motion 
 is the reason 
why the inherent trajectory in figure \ref{fig:Schematic} still 
retains some motion ``within'' the vibrations; as a consequence
of an event in the surroundings, the particle starts vibrating 
around a position that is slightly displaced. This view of the 
dynamics is also consistent with the fact, that the first peak
in $4\pi r^2 G_s^I(r,t)$ is not a delta function in $r=0$ 
(see discussion in section \ref{sec:VigoPaper}).

\begin{figure}
\epsfig{file=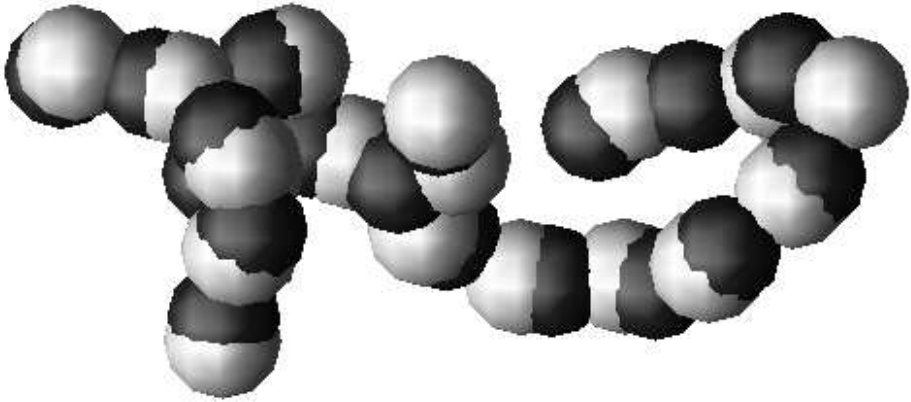, width=12.5cm}
\caption{Before (light) and after (dark) one typical transition, all the
particles which move a distance greater than 0.2. The 
cooperative, string-like  nature of the particle motions during the inter-basin
transition can be clearly seen.
}
\label{fig:Transition}
\end{figure}

From visual inspection of a number of the identified transitions
it was found, that these are cooperative and string-like in nature.
By visual inspection is here meant, that the position of  particles
moving more than $0.2$ during the transition, is plotted before
and after the transition, see figure \ref{fig:Transition}.
String-like motion has been found to be an important part of the
dynamics of glass-forming liquids. It is the natural consequence
of particles hopping to positions previous occupied by other particles
(as concluded from  figure \ref{fig:VHDBB}), and it was found (and quantified) 
in the Kob \& Andersen system, when looking at the ``mobile'' particles
\cite{Donati98}. In paper \VigoPaper ~strings was also found when looking
at how particles was displaced (in the inherent structures) during the 
time interval $t^*$ (figure 3 in  paper \VigoPaper). In paper \VigoPaper~
this was described as ``vacancy hopping''; one particle jumps, leaving room 
for another particle to jump, etc. The finding that also transitions
over energy barriers are  associated with strings, indicates that 
the vacancy hopping interpretation might not be correct; it seems to 
indicate that (at least some of) the strings are really cooperative in
nature, i.e the particles in the string move at the same time. 
Further (quantitative) investigations are obviously needed
to answer this question.

\section{Conclusion}

In this chapter we have presented results from analyzing the dynamics
of a model glass-forming liquid in terms of its potential energy landscape. We 
did so by introducing the new concept of ``inherent dynamics'', which 
can be thought of as a course-graining of the true dynamics, where 
the part of the dynamics related to vibrations around single inherent
structures is removed. 

Comparing the self intermediate scattering function, $F_s(q,t)$, with its
inherent counterpart, $F_s^I(q,t)$, we found direct numerical evidence
for the existence of a transition temperature, $T_x$, below which the 
true dynamics is separated into vibrations around inherent structures and
transitions between these (the ``vibrations plus transitions scenario''). 
We thus confirm the ``energy landscape'' picture, which is (at least)
30 years old. Given the fact the energy landscape \emph{does} exist 
(since its simply
the potential energy as function of the particle coordinates) and it 
\emph{does} have a number of local minima (inherent structures), 
it is not surprising that 
the dynamics becomes dominated by the energy barriers at sufficiently low
temperatures. 
What we have done here using the concept of inherent 
dynamics, is to provide direct numerical evidence for this, 
\emph{and} we have shown that this regime can be reached by 
(pseudo-) equilibrium molecular dynamics (for the particular system 
investigated here). To our knowledge this is the first time such
evidence has been presented.

The fact that we have been able to cool the system, under equilibrium
conditions, to temperatures where the separation between vibrations
around inherent structures and transitions between these is (almost)
complete, means that it now makes sense to study the individual transitions
over energy barriers, since these in this regime are ``significant''.
There is a lot of interesting questions to investigate regarding these 
transitions, and we have here only investigated a few of these. Specifically, 
we have not determined the energy barriers, but only compared the 
two inherent structures involved in a particular transition. This is 
an obvious point for further investigations.

\chapter{The Symmetric Hopping Model}

The symmetric hopping model is introduced, and  three analytical approximations
for calculating  the frequency dependent diffusion coefficient, $D(s)$, 
in the extreme disorder limit (low temperatures) of the model is described; 
the Effective Medium Approximation (EMA), the Percolation Path Approximation (PPA), and the 
Diffusion Cluster Approximation (DCA). DCA is a new  
approximation, developed by my supervisor Jeppe C. Dyre (See paper
\HopPRL ~and \HopRMP ).
Two numerical methods  for calculating $D(s)$ in the extreme disorder 
limit is discussed. 
The first method is derived from the mean square displacement 
and is equivalent with the method of the ac Miller-Abrahams 
(ACMA) electrical equivalent circuit. The second method (VAC)
is derived from the velocity auto correlation 
and is a new method. Numerical results using the VAC method are compared
to the three analytical approximations, and previous results from the 
ACMA method. 

The main results in this chapter are the development of the VAC method
(section \ref{sec:VAC}), and the numerical results
it leads to  (section \ref{sec:Ds} and \ref{sec:Dw}).
Results in this chapter are published in  paper \HopPRL ~and \HopRMP .

\section{The Symmetric Hopping Model}
\label{sec:SymHop}

\begin{figure}
\centerline{\epsfig{file=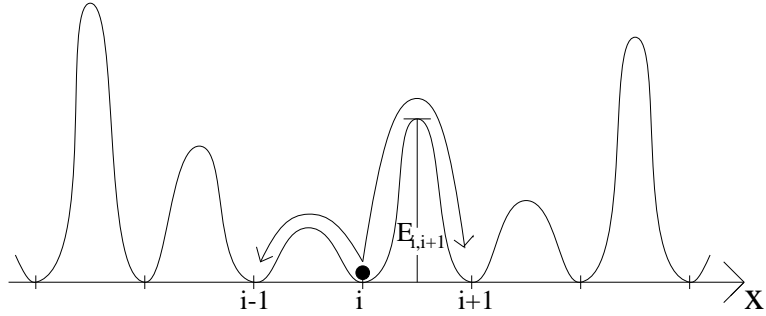, angle=-90,width=10cm}}
\caption{
   The symmetric hopping model in 1 dimension. A particle jumps between 
   nearest neighbor sites, by crossing energy barriers connecting these.
   The jump-rates are given by 
   $\Gamma(i \rightarrow i+1) = \Gamma_0\exp(-\beta E_{i,i+1})$, where
   $E_{i,i+1}$ is the energy barrier between the two sites.
} 
\label{fig:symhop}
\end{figure}

The symmetric hopping model \cite{Dyre88,Avramov93,Dyre94,Argyrakis95,Stein95,Dyre96} 
is  defined in the following way: 
A particle 'lives' on the sites of a $\mathcal D$-dimensional regular 
lattice (see figure \ref{fig:symhop} for a 1-dimensional
illustration), where all the sites has the same energy (which we set to 0).
The particle jumps over energy barriers connecting nearest neighbor sites, with 
jump rates (probability per unit time) 
given by $\Gamma(k \rightarrow i) = \Gamma_0\exp(-\beta E_{ki})$, 
where  $\Gamma_0$ is the (constant) ``attack-frequency'', 
$\beta \equiv (k_B T)^{-1}$ and $E_{ki}$ is the energy barrier 
between the two sites. The energy barriers are chosen randomly, 
from a probability distribution, $p(E)$, (to be specified). Jump rates between 
sites that are not nearest neighbors are zero. 
It follows from the above, that $\Gamma(i \rightarrow k) = 
\Gamma(k \rightarrow i)$, i.e. the jump-rates are  symmetric.
In the following we will use $\Gamma(E) \equiv \Gamma_0\exp(-\beta E)$,
and denote the lattice constant $a$.

If all the energy barriers are identical, i.e. $p(E) = \delta(E-E_0)$, the 
model describes diffusion in an ordered structure, and one finds 
normal diffusive behavior:
\begin{eqnarray}
  \langle { \Delta X}^2(t) \rangle = 2Dt \label{OrderedDiffusion}
\end{eqnarray}
where $\langle { \Delta X}^2(t) \rangle$ is the mean square displacement
along the x-direction, and $D=a^2\Gamma(E_0)$ is the diffusion 
coefficient \cite{Dyre94}.
The average in equation \ref{OrderedDiffusion} can either be a time-average
or a ensemble average. We will in the following use the  ensemble average, which is
characterized by all  the sites having the same probability (since 
they have the same energy). Instead of ensembles it is convenient to think
of a (large) number of independent particles moving around in the sample.

In a sample where the energy barriers are not identical, 
equation \ref{OrderedDiffusion} does not hold
at all time scales. Picture for example a 1-dimensional sample with 
mostly small energy barriers, $E_{small}$, and a few much larger energy barriers, 
$E_{large}$, (see eg. \cite{Pereyra94}). At small 
time scales most of the particles  will ``think'' they are  living on 
an ordered sample with only the small energy barriers (since they haven't 
yet encountered a large energy barrier), 
and the ensemble will follow equation \ref{OrderedDiffusion} with a large 
diffusion coefficient ($\propto \Gamma(E_{small})$). 
However, at long time scales the particles will 
start to ``feel'' the effect of the large energy barriers, which will slow down 
the diffusion. At very long time and length scales the sample will
appear homogeneous and the system again become diffusive, but now with 
a small diffusion coefficient ($\propto \Gamma(E_{large})$). 

The deviations from equation \ref{OrderedDiffusion} can be 
quantified using the time dependent diffusion coefficient \cite{Movaghar86}, $D(t)$,
or the frequency dependent diffusion coefficient \cite{Scher73}, $D(s)$ (with $s$
being the Laplace frequency,  $s=i\omega$):
\begin{eqnarray}
   D(t) &=& \frac{1}{2}  \frac{d}{dt}  \langle  { \Delta X}^2(t) \rangle \\
   D(s) &=& s \int_0^\infty e^{-st} D(t) dt =
    \frac{s^2}{2} \int_0^\infty e^{-st}
            \langle { \Delta X}^2(t) \rangle dt \label{DsX2}
\end{eqnarray}
Note, that for diffusion in a sample with identical energy barriers, equation 
\ref{OrderedDiffusion} leads to $D(s)=D(t)=D$. 

In a disordered sample 
one in general finds that the system becomes diffusive at long time scales 
(corresponding to low frequencies) when particles are moving much longer 
than the appropriate correlation length (assuming that such a correlation length exists); 
$D_0 \equiv D(s\rightarrow 0) = D(t\rightarrow \infty)$. In one dimension 
one finds \cite{Alexander81}: $D_0 = a^2 \langle \Gamma ^{-1} \rangle^{-1}$, where the 
average is over the distribution of jump rates, $\Gamma $. In higher dimensions no 
such simple expression exists, but for $\beta \rightarrow \infty$ one finds
\cite{Tyc89}: $D_0 \propto   \Gamma(E_c)$, where $E_c$ is the so-called
percolation energy (see section \ref{sec:PPA}).

At infinitely short time particles never jumps more than once, and one finds 
(where $\Gamma_L$ and   $\Gamma_R$ are respectively the jump rates to the left and to 
the right of a given site):
\begin{eqnarray} 
  \langle  { \Delta X}^2(\Delta t) \rangle 
  &=& \langle a^2 \Gamma_{L} \Delta t + a^2 \Gamma_{R} \Delta t \rangle \\
  &=& 2a^2\langle \Gamma \rangle \Delta t \label{Dinfty}
\end{eqnarray}
In the high-frequency limit, we thus find:
$ D_\infty \equiv D(s\rightarrow \infty)=D(t\rightarrow 0)= a^2\langle \Gamma \rangle$.  
 
In \cite{Dyre94} and paper \HopPaper, the symmetric hopping model was treated 
as a model for frequency dependent conduction in glasses. In that context the 
particles represent non-interacting charge carriers (ions or electrons), and the 
lattice sites represents the positions in the glass where the charge carriers
can reside.  The frequency dependent conductivity, $\sigma(s)$,
is related to $D(s)$ by the generalized Einstein relation \cite{Scher73}:
\begin{equation}
   \sigma(s) = \frac{e^2n}{k_BT}D(s) \label{SigmaS}
\end{equation}
where $n$ is the density of particles, and $e$ is the charge of the particles. 

In the present work we shift the attention slightly, and treat the symmetric 
hopping model more generally as a model for diffusion in disordered media. We
focus on the ``extreme disorder limit'', i.e. low temperatures, where the model 
is found to exhibit universal behavior; $D(s)$ becomes independent (suitably scaled) 
of the temperature \emph{and} the  chosen probability distribution for the energy 
barriers, $p(E)$. 
In the extreme disorder limit, it is 
\emph{not} feasible to simulate the model using standard Monte Carlo (MC) techniques
\cite{Avramov93,Argyrakis95,Horner95}; 
Due to the large difference in jump rates ($\propto e^{-\beta E}$), 
the particle will jump many times over small energy barriers, and only ``sample''
the rest of the lattice on much larger time scales. Although this reflects the
physics of the model in the extreme disorder limit, it makes MC methods unsuitable
in this limit. 

In the following we set the attack frequency, $\Gamma_0=1$, the lattice constant, $a=1$, 
and Boltzmann's constant, $k_B=1$, thus defining the scales for time, length,  and 
energy respectively. 

\section{The Master Equation}

The starting point for both the analytical and the  numerical 
methods, is the master equation \cite{Kimball78,Kampen81,Odagaki81,Derrida83,Boettger85} for the model;
Let $P(i, t| j, 0)$ denote the probability of
the particle being at site $i$ at time $t$, given that 
it was at site $j$ at $t=0$. The master equation
for the system is then:
\begin{equation}
   \frac{dP(i, t|j, 0)}
        {dt} 
       = -\gamma_i  P(i, t|j, 0)+ 
         \sum_{k}\Gamma(k \rightarrow i)
            P(k, t|j, 0)
   \label{Mastereq}
\end{equation}
where $\gamma_i \equiv \sum_{k}\Gamma(i \rightarrow k)$. 
The first term on the right hand side is the probability flow out of site $i$, 
and the second term is the flow into site $i$.
Defining the ($N^{\mathcal D}$x$N^{\mathcal D}$) matrix $\mathbf P(t)$ with the components 
$\mathbf P_{ij}(t) = P(i, t|j, 0)$, we write the master 
equation on matrix form:

\begin{equation}
   \frac{d}{dt}{\mathbf P}(t) = {\mathbf {HP}}(t)
   \label{MatrixMaster}
\end{equation}
$\mathbf H$ is here a ($N^{\mathcal D}$x$N^{\mathcal D}$) matrix containing
the jump rates; $\mathbf H_{ii} = -\gamma_i$ and $\mathbf H_{i\ne k} = \Gamma(k \rightarrow i)$.
Note that only $2\mathcal D + 1$ elements in each row in  $\mathbf H$ are different
from zero (the diagonal and the $2\mathcal D$  elements corresponding to nearest neighbors);
$\mathbf H$ is sparse. This is an essential feature in the numerical methods; without it
we wouldn't be able to treat large enough sample sizes.

Taking the Laplace transform of equation \ref{MatrixMaster}
we get: 
\begin{equation}
   s{\mathbf G}(s) - \mathbf P(t=0) = {\mathbf {HG}}(s)
   \label{LapMaster}
\end{equation}
where the $\mathbf G_{ij}(s)$ is the Green's function, i.e. the 
Laplace transform of $\mathbf P_{ij}(t)$:
$   \mathbf G_{ij}(s)  \equiv
   \int_0^\infty \mathbf P_{ij}(t)
       e^{-st}dt $. 
The initial condition, $\mathbf P(t=0)$, is given by the 
identity matrix, $\mathbf I$, and we thus find:

\begin{equation}
   (s\mathbf I - \mathbf H){\mathbf G}(s) = \mathbf I 
   ~~~\Leftrightarrow ~~~
   {\mathbf G}(s) = (s\mathbf I - \mathbf H)^{-1}
   \label{Greens}
\end{equation}

\begin{figure}
\centerline{\epsfig{file=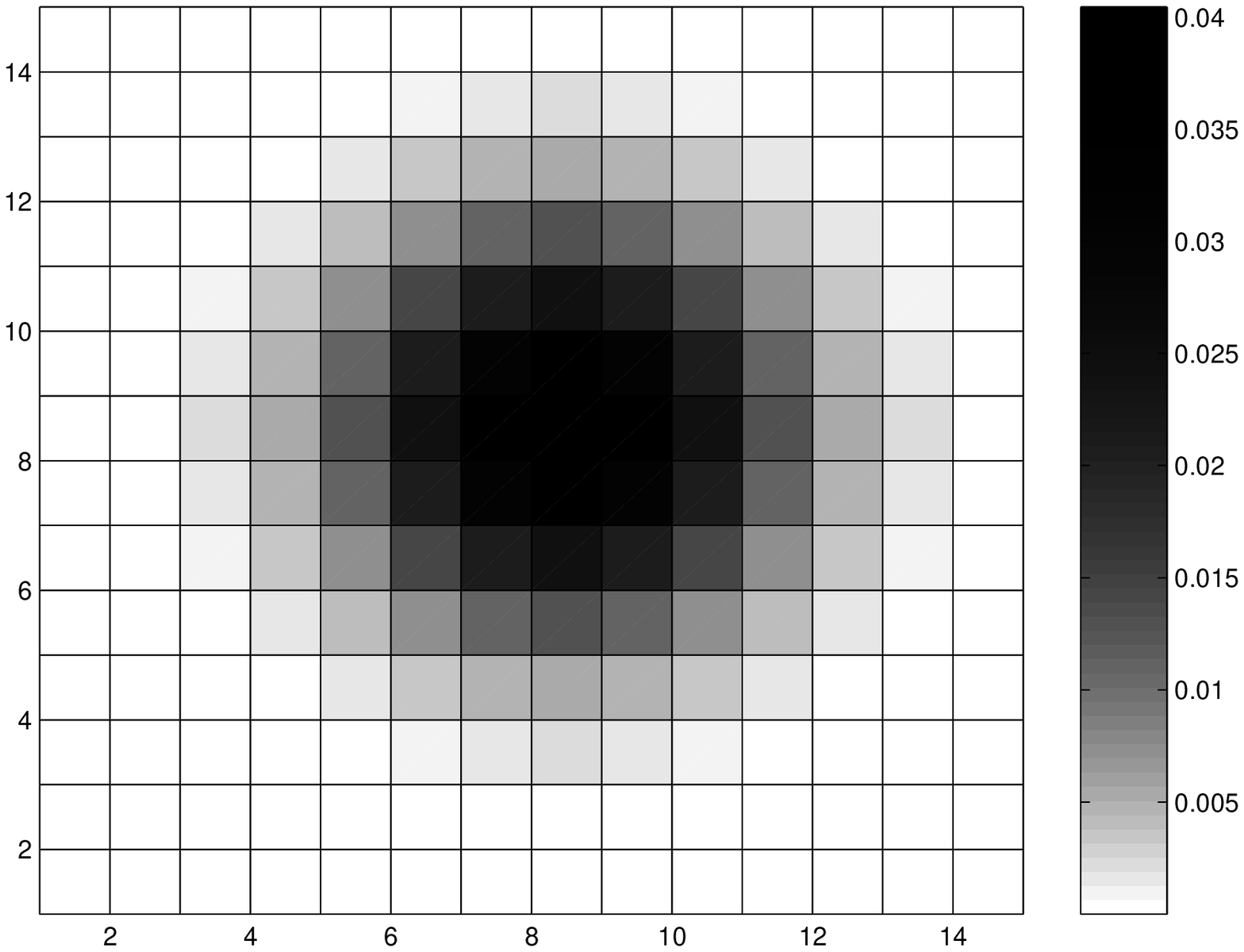, angle=0,width=10cm}}
\vspace{.5cm}
\centerline{\epsfig{file=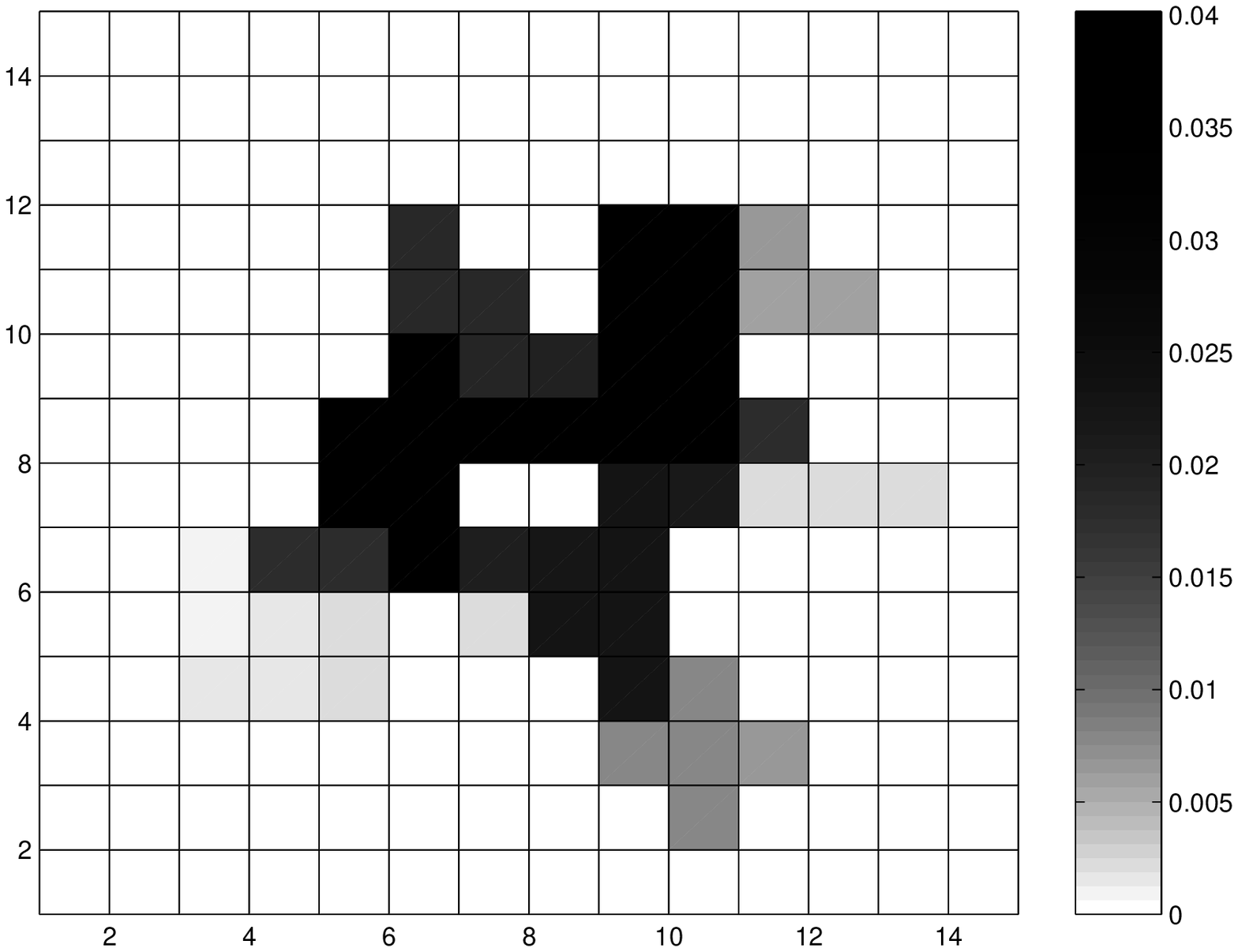, angle=0,width=10cm}}
\caption{
${\mathbf P}(t)$ for a 2 dimensional realization of the symmetric hopping
model, with a box distribution of energy-barriers. Each site is represented
by a square, and this is colored a shade of grey according to the value of
${\mathbf P}(t)$ at the given site. a) ${\mathbf P}(t=2)$ for $\beta = 0$.
b) ${\mathbf P}(t=10^5)$ for $\beta = 40$.
} 
\label{fig:SymPreview}
\end{figure}

Before proceeding to discuss the analytical approximations and the numerical 
methods, we present a ``preview'' of the dynamics of the symmetric hopping
model in figure \ref{fig:SymPreview}. A 2 dimensional sample was set up 
using a box-distribution of barrier energies; $p(E)=1, 0\le E \le1$.
The time evolution of ${\mathbf P}(t)$ was determined by a  discrete version  of
the master equation (eq. \ref{MatrixMaster}):
\begin{equation}
   {\mathbf P}(t + \Delta t) ~=~ {\mathbf P}(t) + \Delta t{\mathbf {HP}}(t)
                             ~=~ ({\mathbf I} + \Delta t{\mathbf H}){\mathbf P}(t)
   \label{DiscreteMatrixMaster}
\end{equation}
A time step of $\Delta t=0.1$ was used, and as initial condition a particle was placed at
a particular site. In figure \ref{fig:SymPreview}a ${\mathbf P}(t=2)$ is 
shown for $\beta = 0$, i.e. infinitely high temperature. In this limit all
the jump rates are identical (equal to unity), and as expected the probability 
distribution spreads in a symmetric manner. In figure \ref{fig:SymPreview}b 
${\mathbf P}(t=10^5)$ is shown for $\beta = 40$, with the same starting 
position, and same energy landscape (i.e. same set of energy barriers) as in 
figure \ref{fig:SymPreview}a. The time at which the two temperatures are compared
was chosen so that the probability of finding the particle at the starting point
was the same ($\approx 0.04$). The qualitative difference between the dynamics at high and
low temperature is evident; At the low temperature (figure \ref{fig:SymPreview}b)
the spread of the probability distribution ${\mathbf P}(t)$ is highly irregular, as a 
consequence of the particles ``preferring'' low energy barriers and avoiding high 
energy-barriers. The particular structure of ${\mathbf P}(t)$ at the low 
temperature depends strongly on where the particle was started. If it was started 
in one of the two enclosed empty sites (white in figure \ref{fig:SymPreview}b)
it would be stuck there on the time scale used here, since these must be connected with 
high energy-barriers to the surroundings (otherwise they would not be empty in
figure \ref{fig:SymPreview}b).

\newpage
\section{Analytical approximations}
\label{sec:AnalAprrox}

In this section we'll briefly describe the 3 analytical approximations
which will be compared to the numerical results, in section \ref{sec:Ds}
and  \ref{sec:Dw}.

\subsection{Effective Medium Approximation (EMA)}

In EMA \cite{Odagaki81,Dyre94} the disordered sample is replaced by an ordered sample
(the ``effective medium''), 
where all the jump rates are replaced by an ``effective'' jump-rate, 
$\Gamma_{E}(s)$. The value of $\Gamma_{E}(s)$ is determined 
by a self-consistency condition; A single jump rate in the effective medium 
is replaced by its value in the disordered sample, and an average is performed 
over the distribution of jump rates. The condition that this average should 
be equivalent with the effective medium leads to the following condition
 \cite{Odagaki81,Dyre94,SchroederSpecialePhys}:

\begin{eqnarray}
   \left< 
      \frac{D - \Gamma}{\mathcal DD-(D - \Gamma) 
         (1-s \tilde G   )} \right>_\Gamma  &=& 0 
  \label{EMA2}
\end{eqnarray}
where $\tilde G$ is the diagonal element of the Green's function for the effective
medium, which depends on the spatial dimension, $\mathcal D$. In the extreme 
disorder limit ($\beta \rightarrow \infty$) one finds \cite{Dyre94} for $\mathcal D\ge 2$:
\begin{eqnarray}
   \tilde D \ln( \tilde D) =  \tilde s
   \label{EMA}
\end{eqnarray}
where $\tilde D \equiv D(s)/D_0$, and $\tilde s \equiv s/s_c$, where $s_c$ is a suitably 
defined characteristic frequency. The EMA thus predicts universality in the extreme 
disorder limit; Properly scaled $D(s)$ becomes independent of temperature \emph{and}
the distribution of energy barriers, $p(E)$. 

In \cite{Dyre94} the predictions of EMA was compared with numerical 
results using the ACMA method (described in section \ref{sec:ACMA}) in 2 
dimensions. The existence of universality was confirmed by the numerical
results (for 4  different energy-distributions), but the shape of $\tilde D(\tilde s)$ 
was found to deviate from 
the one predicted by EMA. Or to be more specific: EMA predicts a shape 
of the universal $\tilde D(\tilde s)$ that is growing to rapidly at 
the onset of frequency dependence; it is to ``sharp'' (figure 5 in \cite{Dyre94}). 
The EMA prediction for  $\langle { \Delta X}^2(t) \rangle$ is discussed in 
\cite{Dyre95}.

\subsection{Percolation Path Approximation (PPA)}
\label{sec:PPA}

In the extreme disorder limit, 
the low-frequency diffusion is expected to be dominated by percolation 
\cite{Boettger85,Bunde91,Hunt95,Horner95}; when  diffusing over long distances (i.e. in the
low-frequency limit) the particles ``chose'' to do so by jumping over 
energy-barriers that are as small as possible. The largest energy 
barrier that a particle has to cross to move through the sample 
is given by the ``percolation energy'', $E_c$, defined by \cite{Dyre94}:
\begin{eqnarray}
   \int_0^{E_c} p(E) dE = p_c \label{Ec}
\end{eqnarray}
where $p_c$ is the percolation threshold for bond percolation 
(for an introduction to percolation theory, see \cite{Stauffer}).
In a 2 dimensional regular lattice $p_c$ equals $1/2$ (exact) and in a 
3 dimensional cubic sample $p_c$ equals $0.2488$ (approximated).
Equation \ref{Ec} can be interpreted as follows; In a given sample 
mark the energy-barriers (bonds) starting with the smallest energy-barrier,
then the next smallest etc.
At some point
the marked energy-barriers will ``percolate'' i.e. make a cluster that stretches
through the whole sample (the ``percolation cluster''). 
The highest energy 
needed to get percolation is (for an infinite sample) $E_c$. For the box distribution
of energy barriers used in the present work ($p(E)=1, 0\le E \le1$), 
the percolation energy equals the (bond) percolation threshold: $E_c = p_c$.
The percolation cluster is a fractal, with  fractal dimension 1.9 and 2.5 in 2 and 3 
dimensions respectively \cite{Stauffer}.

In paper \HopPaper  ~we argue, that the reason why  EMA does
not predict the shape of the universal curve $\tilde D(\tilde s) $ in the 
extreme disorder limit  
correctly might be, that it replaces the disordered sample by an effective
media of the \emph{same} dimension, whereas the actual low frequency 
diffusion  happens on a cluster of lower dimensionality. We will here term this
cluster ``the diffusion cluster'', and its fractal dimension $D_f$ 
(this is \emph{not} the same as the percolation cluster, see below). 
PPA can be considered  as a first attempt at trying to incorporate the fractal dimension
of the diffusion cluster into an analytical approximation. 

The percolation cluster contain ``dead-ends'', which contributes 
little to the low frequency diffusion. Removing dead-ends from the
percolation cluster leaves us with the ``backbone'',  with 
fractal dimension 1.6 and 1.7 in 2 and 3 dimensions respectively 
\cite{Stauffer}. In paper
\HopPaper  ~we argue, that $D_f$
is even lower, since the  backbone contains loops, 
where one of the branches usually will be preferred (since the jump-rates depend
strongly on the energy barriers in the extreme disorder limit). In PPA
the extreme view that $D_f=1$ is taken, i.e. that the (low frequency) diffusion
takes place on 1-dimensional ``percolation paths''. 
With this assumption we in  paper \HopPaper  ~arrive at the PPA approximation, given by:

\begin{eqnarray}
   \sqrt{\tilde s\tilde D} \ln \left(1 + \sqrt{\tilde s\tilde D}\right) =  \tilde s
   \label{PPA}
\end{eqnarray}

PPA thus predicts universality in the extreme disorder limit, like is the 
case for EMA. In paper \HopPaper ~, we find that PPA agrees better
with the numerical results than EMA. This is especially true in 3 dimensions.
The agreement is however not perfect, and a number of problems remain
unanswered (see last section in paper \HopPaper ). 
The PPA prediction for  $\langle { \Delta X}^2(t) \rangle$ is discussed in 
\cite{Dyre96b}.

\subsection{The Diffusion Cluster Approximation (DCA)}

The Diffusion Cluster Approximation  can 
be thought of as a refinement of PPA (paper \HopPRL ~and \HopRMP ); 
instead of setting $D_f=1$, this 
is in DCA left as a parameter in the model. Using the approach of EMA, 
but with an fractal effective medium with fractal dimension $D_f$, one finds: 
\begin{eqnarray}
   \ln( \tilde D) =  \left( \frac{ \tilde s}{\tilde D} \right) ^{{D_f}/{2}}
   \label{DCA}
\end{eqnarray}
This expression is limited to $1<D_f<2$. For $D_f\ge 2$ DCA reduces to the EMA
prediction (eq. \ref{EMA}), and for $D_f\le 1$ it is undefined. 
Since we do not have an independent estimate of $D_f$ we will in the present
work treat it as a fitting parameter. 
This must obviously be taken into 
consideration, when comparing with EMA and PPA, since these have no fitting 
parameters. 

From the arguments given above, we expect $D_f$ to be limited above
by the fractal dimension of the backbone. Furthermore we expect $D_f$ to 
limited below by the fractal dimension of the so-called ``red bonds'', which
are the singly connected bonds on the backbone, i.e. if a red bond is 
removed the backbone is broken into 2 parts. The fractal dimension of
the red bonds are $3/4$ and $1.14$ in 2 and 3 dimensions respectively. 

\newpage
\section{The ACMA Method}
\label{sec:ACMA}

In this section we briefly describe the ACMA method, which was used to 
obtain the numerical results reported in \cite{Dyre94} and paper \HopPaper.
 
Defining $X_i$ as the x-coordinate of site $i$ we can 
write the frequency dependent diffusion coefficient (equation \ref{DsX2}):

\begin{eqnarray}
   D_x(s) 
        &=&
      \frac{s^2}{2}\int_0^\infty e^{-st}
         \frac{1}{N^d} \sum_{i,j}
            (X_i - X_j)^2
            \langle P(i, t| j, 0) \rangle
          dt  \label{msd}\\
      &=& \frac{s^2}{2N^d} \sum_{i, j}
         (X_i - X_j)^2
            \langle G(i, s| j) \rangle   
     \label{Dtilde}
\end{eqnarray}
The subscript $x$ in
$D_x(s)$ is here used to emphasize that the diffusion coefficient
here is calculated from the mean square displacement in the 
x-direction. In more than 1 dimension similar expressions 
obviously hold for other directions.

Equation \ref{Greens} and equation \ref{Dtilde} together
constitutes a method for computing D(s); calculate
$\langle G(i, s| j) \rangle$ by inverting $(s\mathbf I - \mathbf H)$
(equation \ref{Greens}) and insert the result in equation \ref{Dtilde}
\cite{Scher73}.


\begin{figure}
\centerline{\epsfig{file=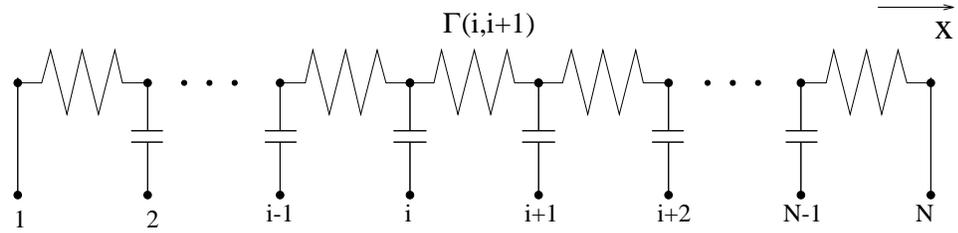, angle=-90,width=12.5cm}}
\caption{
   The ac Miller-Abrahams (ACMA) electrical equivalent circuit for
   the symmetric hopping model. The admittance of the resistor between two sites
   is equal to the corresponding jump rate. The admittance of the capacitors are $s$.  
} 
\label{fig:ACMAnet1D}
\end{figure}

In my master thesis \cite{SchroederSpecialePhys,SchroederSpecialeDat}
it was demonstrated, that the method described above  is equivalent to 
the method of the ac Miller-Abrahams (ACMA) electrical
equivalent circuit \cite{Dyre94}. In this method a large network
of linear electric components are set up (see figure \ref{fig:ACMAnet1D} for
an illustration of the 1-dimensional case). 
After eliminating all the 'internal sites' (those without numbers in
figure \ref{fig:ACMAnet1D}) by the so-called
generalized Star-Mesh transformation, $\sigma(s)$ ($\propto D(s)$, 
see equation \ref{SigmaS}) is calculated 
as a weighted sum over the effective conductances between the 
'external sites' (those with numbers in figure \ref{fig:ACMAnet1D}). 
The elimination process described above is equivalent with Gauss-Jordan 
elimination on the matrix $(s\mathbf I - \mathbf H)$ \cite{SchroederSpecialeDat}, 
and the Fogelholm algorithm \cite{Fogelholm80a} 
for the order in which the internal sites are 
eliminated is equivalent with the minimum degree pivoting algorithm 
\cite{Duff89}: At each step eliminate the site/row with the smallest
 number of connections/elements.

A number of 'tricks' is used in the ACMA method, to improve
efficiency over a straight forward computation of first  
equation \ref{Greens} and then equation \ref{Dtilde}. The 
most important trick is that the (dense) matrix 
$\mathbf G(s)$ is not explicitly calculated, but instead 
a 'condensed' version where columns corresponding to the 
sites with the same x-coordinate are added together. Thus
instead of solving for $N^{\mathcal D}$ right-hand sides (inverting
$s\mathbf I - \mathbf H$) only $N$ right-hand sides are 
used.

In the simulations  using the  ACMA method presented in
\cite{Dyre94} and paper \HopPaper ~'perfect electrodes' were used in the 
x-direction (see figure \ref{fig:ACMAnet1D}). This ensures that $D(s)$ 
goes to a constant value as $s \rightarrow 0$, 
corresponding to $\langle { \Delta X}^2(t) \rangle$ being proportional 
to $t$, as $t \rightarrow \infty$, i.e. the system being diffusive. 
However, with this modification the model 
does \emph{not} correspond directly to the solution of a master 
equation. This is evident from the fact that the mean square 
displacement as calculated in equation \ref{msd} is finite for 
any finite sample, and thus it can not be proportional to $t$
at long times. 

To what extent the 'perfect electrodes' mimics the effect of real 
electrodes when experimentally measuring the frequency dependent conductivity, 
$\sigma(s)$, will not be discussed here. What we are after here, is to calculate 
the \emph{bulk} value of the frequency dependent diffusion coefficient $D(s)$ 
in the extreme disorder limit.

\section{The Velocity Auto Correlation (VAC) method}
\label{sec:VAC}

In this section the Velocity Auto Correlation (VAC) method is developed. 
Its main advantages over the ACMA method is, that it can be (and is) 
used with periodic boundary conditions, and still give a 
diffusive regime for $s\rightarrow 0$ (i.e. $t \rightarrow \infty$).
Another advantage over the ACMA method (as it was implemented in the previous 
work), is that the VAC method 
is formulated in terms of sparse matrices, which means that standard methods for solving
these can be used\footnote{The numerical results  presented here
was done using Matlab version 5.3.0. 
}. The differences between the two methods will be discussed further  in section 
\ref{sec:ACMAvsVAC}.

To derive the VAC method we express the diffusion coefficient in terms
of  the velocity auto correlation function \cite{Scher73}:
\begin{equation}
  D_x(s) = \int_0^\infty \left<
           v(t)v(0) \right>_x e^{-st} dt \label{Dvac} 
\end{equation}
$v(t)$ is here the velocity (in the x-direction) of the particle at time t.
The motion of the particle in the symmetric hopping model is instantaneous;
when it jumps from one site to another, this happens in a infinitesimally 
small time-interval, $\Delta t$. We may thus assign the constant velocity
$\pm a/\Delta t$ (where we briefly reintroduce the lattice constant, $a$) 
to the particle in the time interval, $\Delta t$, ensuring that it 
moves one lattice constant, $a$, either to the left or to the right. 
$v(t)$ thus takes on the value $a/\Delta t$
in a time interval $\Delta t$ when the particle jumps to the right, 
the value $-a/\Delta t$
in a time interval $\Delta t$ when the particle jumps to the left, and
zero the rest of the time (It is \emph{not} important whether or
not this is a physically reasonable choice for $v(t)$. The only requirement 
for eq. \ref{Dvac} to hold is that $x(t) = \int_o^t v(t) dt + x(o)$). 
With this choice of $v(t)$,  the function
$v(t')v(0)$ has the value $a^2/\Delta t^2$ if the particle jumps in the 
same direction at $t=t'$ and $t=0$,  the value $-a^2/\Delta t^2$ if the particle 
jumps in opposite directions at $t=t'$ and $t=0$, and zero otherwise (i.e 
if the particle does not jump \emph{both} at  $t=t'$ and $t=0$).

The probability, $P_{RR}$, that the particle jumps to the right both at $t=t'$ and
$t=0$ is (for now we assume that $t' \ne 0$):
\begin{eqnarray}
 P_{RR}  &= & \frac{1}{N^d} \sum_{i,j} 
           \Gamma_R(j)\Delta t \left< P(i,t|j+1,0) \right>\Gamma_R(i)\Delta t \label{Prr1}\\
   &= & \frac{\Delta t^2}{N^d} \sum_{i,j} 
           \Gamma_L(j) \left< P(i,t|j,0) \right>\Gamma_R(i) \label{Prr2}
\end{eqnarray}
Equation \ref{Prr1} can be read from the left as follows; If the particle starts at site
$j$ the probability of jumping to the right at $t=0$ is $\Gamma_R(j)\Delta t$. This means that 
it is now at site $j+1$, and the probability that it moves to site $i$ in the 
time $t$ is $\left< P(i,t|j+1,0) \right>$, and finally the probability that it 
jumps to the right from site $i$ is $\Gamma_R(i)\Delta t$. In equation 
\ref{Prr2} $j$ is substituted with $j-1$, and $\Gamma_R(j-1) = \Gamma_L(j)$
is used.
Calculating in the same way the probability of the other events that contributes to 
the velocity auto correlation function, $\left< v(t)v(0) \right>_x$ ($P_{RL}$, $P_{LR}$, and
$P_{LL}$), we can now write  $\left< v(t)v(0) \right>_x$,
in terms of the Green's function (where again use $a=1$, and the delta function 
takes care of the special case $t=0$, see below): 
\newpage
\begin{eqnarray}
  \left< v(t)v(0) \right>_x &=& C\delta(t) + \frac{1}{\Delta t^2}
                               \sum_{i,j} (P_{RR} + P_{LL} - P_{RL} - P_{LR})  \\
          &=& C\delta(t) +  \frac{1}{N^d} \sum_{i,j}
             \left< P(i,t|j,0) \right> \label{VVx}  \\
          &&  \left[\Gamma_L(j)\Gamma_R(i) + \Gamma_R(j)\Gamma_L(i) - 
              \Gamma_L(j)\Gamma_L(i) - \Gamma_R(j)\Gamma_R(i) \right]_{~}\nonumber\\
 &=&C\delta(t)^{~}  + \label{VVx2} \\
 & &\frac{1}{N^d} \sum_{i,j} (\Gamma_R(i) - \Gamma_L(i) )
             \left< P(i,t|j,0) \right> (\Gamma_L(j) - \Gamma_R(j) ) \nonumber  
\end{eqnarray}
Taking the Laplace transform of equation \ref{VVx2} we get:

\begin{equation}
  D_x(s) = C +
  \frac{1}{N^d} \sum_{i,j} (\Gamma_R(i) - \Gamma_L(i) )
             \left< G(i,s|j) \right> (\Gamma_L(j) - \Gamma_R(j) )
\end{equation}

In the high- frequency limit, $s\rightarrow \infty$, we find from 
equation \ref{Greens} \\ $\left< G(i,s|j) \right> \rightarrow 0$. 
We thus get  $C=D_\infty=\left< \Gamma \right>$ (see equation \ref{Dinfty}):

\begin{equation}
  D_x(s) = \left< \Gamma \right> -
  \frac{1}{N^d} \sum_{i,j} (\Gamma_R(i) - \Gamma_L(i) )
             \left< G(i,s|j) \right> (\Gamma_R(j) - \Gamma_L(j) )
  \label{DsNumProb}
\end{equation}
In the case where all the jump-rates are identical, 
$\Gamma_R(i) = \Gamma_L(i)$, equation  \ref{DsNumProb} leads
to $D_x(s) = \left< \Gamma \right> = D_\infty$, as expected 
(see section \ref{sec:SymHop}).
 
Equation \ref{DsNumProb} can be used together with 
equation \ref{Greens} to calculate $D_x(s)$.
However this method poses a serious numerical problem at low 
temperatures where $D_0$ is very small compared 
to $D_\infty$. Calculating $D_x(s)$ at low frequencies from 
equation \ref{DsNumProb} in this limit amounts to calculating a small difference 
between two large numbers, which leads to large uncertainties in the result. 

In the following we will derive a  version
of the VAC method, which does not have the numerical problems
described  above. We do this first in 1 dimension (section \ref{sec:VAC1D}), followed 
by a general derivation in $\mathcal D$ dimensions (section \ref{sec:VACDD}).

\subsection{The VAC method in 1 Dimension}
\label{sec:VAC1D}

The idea behind the derivation of the VAC method in 1 
dimension is to rewrite the problem in terms 
of the variable: 

\begin{eqnarray}
  J_R(i,t|j,0) \equiv \Gamma_R(i) \left[P(i,t|j,0) - P(i+1,t|j,0)\right]
\end{eqnarray}
which can be interpreted as the particle current to the right of
site $i$ at time $t$ given that the particle started at site
$j$ at $t=0$.
The master equation (\ref{Mastereq}) can then be written as:

\begin{eqnarray}
   \frac{dP(i,t|j,0)}{dt} = J_R(i-1,t|j,0) - J_R(i,t|j,0)
\end{eqnarray}
We now define two new matrices; $\mathbf \Gamma_R$ is a diagonal 
matrix containing the jump rates to the right of each site: 
$(\mathbf \Gamma_R)_{ij} = \delta (i,j) \Gamma_R(i)$. 
$\mathbf A$ is a matrix with $+1$ on the diagonal, $-1$ on
the sub-diagonal and zero everywhere else: 
$(\mathbf A)_{ij} = \delta (i,j) - \delta (i-1,j)$. In more 
general terms; $\mathbf A_{ij}$ equals $-1$ if site $j$ is the left
neighbor to site $i$. Here
(as in the rest of this chapter) periodic boundary conditions 
are implicit, i.e. $(\mathbf A)_{1N} = -1$.
We can now write the 1 dimensional master equation as: 

\label{SecH1D}
\begin{eqnarray}
   \frac{d}{dt}\mathbf P(t) &=& -\mathbf A \mathbf J_R(t), ~~~
   \mathbf J_R(t) = \mathbf \Gamma_R \mathbf A^T \mathbf P(t)
   \label{StrMatMaster}
\end{eqnarray}
These equations provides us with an
an explicit expression for the matrix $\mathbf H$ in the master 
equation (\ref{MatrixMaster}):
$\mathbf H = - \mathbf A\mathbf \Gamma_R \mathbf A^T$.
What we are after here is however an equation for $\mathbf J_R$. 
To get this we take the Laplace transform
of equation \ref{StrMatMaster}:
\begin{eqnarray}
   s\mathbf G(s) + \mathbf A \mathbf J_R(s) = \mathbf P(t=0) = \mathbf I 
\end{eqnarray}
where $\mathbf J_R(s)$ is the Laplace transform of $\mathbf J_R(t)$.
By multiplying from the left by $\mathbf A^T$ and using
$ \mathbf J_R(s) = \mathbf \Gamma_R \mathbf A^T \mathbf G(s) $ we get:
\begin{eqnarray}
   & & (s\mathbf \Gamma_R^{-1} + \mathbf A^T \mathbf A) \mathbf J_R(s) = 
       \mathbf A^T  \Leftrightarrow \\
   & & \mathbf J_R(s) = (s\mathbf \Gamma_R^{-1} + 
                     \mathbf A^T \mathbf A)^{-1} \mathbf A^T  
\end{eqnarray}
Here we assume that the diagonal matrix $\mathbf \Gamma_R$ is invertible, 
i.e. that all the jump rates are different from zero.
We are now ready to derive an equation for D(s) in 1 dimension. To
do this we rewrite equation \ref{DsNumProb} using 
$\Gamma_L(i)=\Gamma_R(i-1)$:
\begin{eqnarray}
  D(s) = \left< \Gamma_R \right> -
  \frac{1}{N} \sum_{i,j} (\Gamma_R(i) - \Gamma_R(i-1) )
             \left< G(i,s|j) \right> (\Gamma_R(j) - \Gamma_R(j-1) )
             \nonumber
\end{eqnarray}

This can be written as (where $\mathbf 1$ is a column vector
containing all 1's):
\begin{eqnarray}
   D(s)N &=& \mathbf 1^T \mathbf \Gamma_R \mathbf 1
            -  \mathbf 1^T \mathbf \Gamma_R \mathbf A^T 
                         \mathbf G(s) \mathbf A \mathbf \Gamma_R \mathbf 1
                          \label{DsStr1} \\
         &=&    \mathbf 1^T \mathbf \Gamma_R \mathbf 1
            -  \mathbf 1^T \mathbf J_R(s) \mathbf A \mathbf \Gamma_R \mathbf 
            1 \\
         &=&\mathbf 1^T \mathbf \Gamma_R \mathbf 1
            -  \mathbf 1^T (s\mathbf \Gamma_R^{-1} + 
                     \mathbf A^T \mathbf A)^{-1} \mathbf A^T 
                     \mathbf A \mathbf \Gamma_R \mathbf 1 \\
         &=&\mathbf 1^T (s\mathbf \Gamma_R^{-1} + \mathbf A^T \mathbf A)^{-1}
            \left[(s\mathbf \Gamma_R^{-1} + \mathbf A^T \mathbf A) - 
                   \mathbf A^T \mathbf A \right] \mathbf \Gamma_R \mathbf 
                   1 \nonumber \\ 
          &=&s\mathbf 1^T (s\mathbf \Gamma_R^{-1} + \mathbf A^T \mathbf 
          A)^{-1}
                   \mathbf 1 \\
          &=&s\mathbf 1^T \mathbf x, 
              ~~~ (s\mathbf \Gamma_R^{-1} + \mathbf A^T \mathbf A) 
                   \mathbf x = \mathbf 1
\end{eqnarray}
We have now avoided the problematic subtraction (in equation
\ref{DsNumProb} and \ref{DsStr1}), and reduced the problem 
to finding the vector $\mathbf x$ by solving a sparse system 
of linear equations. 

\subsection{The VAC method in $\mathcal D$ Dimensions}
\label{SecH2D}
\label{sec:VACDD}

In this section we generalize the VAC method to $\mathcal D$
dimensions. In analogy with $\mathbf J_R(t)$ used in the 1-dimensional 
version, we define $\mathbf J_k(t)$ as the particle current in the 
$k$'th direction:
\begin{eqnarray}
  \mathbf J_k(t) = \mathbf \Gamma_k \mathbf H_k^T \mathbf P(t) \label{Jk} 
\end{eqnarray}
where $\mathbf \Gamma_k$ and $\mathbf H_k$ are generalization of the matrices
$\mathbf \Gamma_R$ and $\mathbf A$ used in the previous section. $\mathbf \Gamma_k$
is a diagonal matrix, which for each site contains the jump rate to the ``right''
in the $k$'th direction. $(\mathbf H_k)_{ij}$ equals $-1$ if site $j$ is the 
``left'' neighbor to site $i$ along direction $k$, and it has $1$ on the 
diagonal and zero everywhere else. The structure of $\mathbf H_k$ depends
on the numbering scheme chosen for the sites, but the resulting physics
is obviously independent of this. 
An explicit expression for $\mathbf H_k$ (and the corresponding 
numbering of the sites) is given in appendix \HkIndexAPP.

The goal is now to set up a master-equation for the currents, and from that 
calculate $D(s)$, like it was done in the 1-dimensional case. 
Obviously it is not possible to solve for $\mathbf J_k(s)$ without at the same
time solving for the currents in the other directions. We thus generalize 
equation \ref{Jk} to 
\begin{eqnarray}
  \mathbf J_*(t) = \mathbf \Gamma_* \mathbf H_*^T \mathbf P(t) \label{Jstar}
   ~~ \Leftrightarrow ~~  \mathbf J_*(s) = \mathbf \Gamma_* \mathbf H_*^T \mathbf G(s)
\end{eqnarray}
where we have defined the following block-matrices:
\begin{eqnarray}
    \mathbf J_*(t) \equiv 
     \left(
       \begin{array}{c}
          \mathbf J_1(t) \\
         \mathbf J_2(t) \\
         \vdots \\
         \mathbf J_{\mathcal D}(t) \\
   \end{array}
  \right)
   ,~  
    \mathbf \Gamma_* \! \! \! &\equiv&  \! \! \!
    \left(
    \begin{array}{c c c c}
       \mathbf \Gamma_1 & \mathbf 0  & &  \mathbf 0 \\
       \mathbf 0 & \mathbf \Gamma_2  & &  \mathbf 0 \\ 
                 &                   &\ddots&         \\
       \mathbf 0 & \mathbf 0    & &  \mathbf \Gamma_{\mathcal D} \\ 
   \end{array}
  \right),~
   \mathbf H_* \equiv \left( \mathbf H_1, 
    \dots,  \mathbf H_{\mathcal D} \right) \nonumber \\
     &&{}
\end{eqnarray}
and $\mathbf J_*(s)$ is the Laplace transform of  $\mathbf J_*(t)$.   

The master equation for $\mathbf P(t)$ is similar to the one in 1-dimension 
(eq. \ref{StrMatMaster}), but the change in the probability at a given site now 
has contributions from all $\mathcal D$ directions:
\begin{eqnarray}
  && \frac{d}{dt}\mathbf P(t) = - \sum_{k=1}^{\mathcal D} \mathbf H_k \mathbf J_k(t)
    = -\mathbf H_* \mathbf J_*(t)  =  
  -\mathbf H_*\mathbf \Gamma_* \mathbf H_*^T \mathbf P(t)  ~~ \Leftrightarrow ~~~~~{} \\
  && s\mathbf G(s) + \mathbf H_* \mathbf J_*(s) = \mathbf P(t=0) \label{MasterSD}
\end{eqnarray}

By multiplying equation \ref{MasterSD} from the left with $ \mathbf \Gamma_* \mathbf H_*^T$, 
and substituting for $\mathbf J_*(s)$ (equation \ref{Jstar}), 
 we arrive at the (Laplace transformed) master equation for $\mathbf J_*(s)$:
\begin{eqnarray}
&&   \left(s\mathbf I + \mathbf \Gamma_*\mathbf H_*^T\mathbf H_*\right) \mathbf J_*(s) = 
      \mathbf \Gamma_*\mathbf H_*^T\mathbf P(t=0) ~~ \Leftrightarrow ~~\\
&&   \mathbf J_*(s) = 
      \left(s\mathbf I + \mathbf \Gamma_*\mathbf H_*^T\mathbf H_*\right)^{-1}
      \mathbf \Gamma_*\mathbf H_*^T\mathbf P(t=0)
\label{MasterSBlok}
\end{eqnarray}

To derive the final result for $D_k(s)$ (i.e. the diffusion coefficient 
as calculated from velocity correlations in direction $k$) we 
define $\mathbf 1_k$ as a column vector with $\mathcal DN^{\mathcal D}$ elements, where the 
$N^{\mathcal D}$ elements corresponding to currents in the $k$'th direction
is $1$ and the rest is zero (so that eg. 
$\mathbf 1_k^T  \mathbf \Gamma_* \mathbf 1_k = \mathbf 1^T  \mathbf \Gamma_k \mathbf 1$):
\begin{eqnarray}
    \mathbf 1_k \equiv 
     \left(
       \begin{array}{c}
          \mathbf 0 \\
          \vdots \\
          \mathbf 1 \\
         \vdots \\
         \mathbf 0 \\
	   \end{array}
      \right)
\end{eqnarray}	

From equation \ref{DsNumProb} we finally get (using $\mathbf P(t=0) = \mathbf I$):
\begin{eqnarray}
   D_k(s)N^{\mathcal D} 
    &=& \mathbf 1^T \bigg( 
           \mathbf \Gamma_k - \mathbf J_k(s) \mathbf H_k \mathbf \Gamma_k 
        \bigg) \mathbf 1 \\
    &=& \mathbf 1_k^T \bigg( 
           \mathbf I  - \mathbf J_*(s) \mathbf H_*
        \bigg)  \mathbf \Gamma_*  \mathbf 1_k \\
    &=& \mathbf 1_k^T \bigg( 
           \mathbf I  - \left(s\mathbf I + \mathbf \Gamma_*\mathbf H_*^T\mathbf H_*\right)^{-1}\mathbf \Gamma_*\mathbf H_*^T\mathbf H_*  
        \bigg) \mathbf \Gamma_* \mathbf 1_k \\
    &=& \mathbf 1_k^T \left(s\mathbf I + \mathbf \Gamma_*\mathbf H_*^T\mathbf H_*\right)^{-1}
        \bigg( 
          \left(s\mathbf I + \mathbf \Gamma_*\mathbf H_*^T\mathbf H_*\right)  - \mathbf \Gamma_*\mathbf H_*^T\mathbf H_*  
        \bigg) \mathbf \Gamma_* \mathbf 1_k \nonumber \\
 &=& s \mathbf 1_k^T 
        \left(s\mathbf I  + \mathbf \Gamma_*\mathbf H_*^T\mathbf H_*\right)^{-1}
     \mathbf \Gamma_* \mathbf 1_k \\
 &=& s \mathbf 1_k^T 
        \left(s\mathbf \Gamma_*^{-1} + \mathbf H_*^T\mathbf H_*\right)^{-1}
     \mathbf 1_k \\
 &=& s \mathbf 1_k^T \mathbf x~,~~~  
        \left(s\mathbf \Gamma_*^{-1} + \mathbf H_*^T\mathbf H_*\right)
        \mathbf x =  \mathbf 1_k  \label{VACfinal}      
\end{eqnarray}

Like in the 1-dimensional case in the previous section, we have now
reduced the problem to finding the vector $\mathbf x$ by solving a sparse 
system of linear equations. 
In $\mathcal D$ dimensions the matrix involved in this is a 
($\mathcal D N^{\mathcal D}$x$\mathcal D N^{\mathcal D}$) matrix.

For real $s>0$ the matrix 
$\left(s\mathbf \Gamma_*^{-1} + \mathbf H_*^T\mathbf H_*\right)$
is positive definite\footnote{The matrix $\mathbf A$ is symmetric and 
positive definite if $\mathbf y ^T \mathbf A \mathbf y > 0$ for any 
$\mathbf y$.}, which ensures the numerical stability when doing
Gaussian elimination \cite{Duff89}:
\begin{eqnarray}
  \mathbf y ^T \left(s\mathbf \Gamma_*^{-1} + \mathbf H_*^T\mathbf H_*\right)  \mathbf y
 &=& s \mathbf y ^T \mathbf \Gamma_*^{-1} \mathbf y + \mathbf y ^T \mathbf H_*^T\mathbf H_* \mathbf y \\ &=&  s \mathbf y ^T \mathbf \Gamma_*^{-1} \mathbf y  + (\mathbf H_* \mathbf y)^T (\mathbf H_* \mathbf y)
\end{eqnarray}
The second term is greater than or equal to zero, and
the first term is greater than zero, since $\mathbf \Gamma_*^{-1}$ is a diagonal 
matrix with only positive values on the diagonal. When representing the problem 
in finite precision in the computer this last statement is violated for 
very low $s$ values; The elements corresponding to small energy barriers 
(large jump-rates) are effectively set to zero when the two terms are 
added (all elements in $\mathbf H_*$ are O(1)). 
This means that the matrix in practice is \emph{not} positive 
definite and the attempt to solve it fails. To avoid this numerical problem 
a small constant $\delta $ ($=10^{-14}$) is added to the diagonal ensuring that 
the matrix is positive definite and thus can be solved. In physical terms
this corresponds to applying a minimum value for the energy barriers, 
ensuring that all jump-rates are finite ($< s/\delta$). It was checked 
numerically (by varying $\delta$) that the effect of this procedure is 
negligible (relative error $\lesssim 10^{-4}$), as expected from our
physical understanding of the model (see section \ref{sec:PPA}).

Like in paper \HopPaper ~ the jump rates corresponding to energy barriers
larger than $E_c + K/\beta$ ($K=6.4$) was set to zero, to speed up the
calculations (and reduce the amount of memory needed). By varying $K$ the 
relative error introduced by this procedure was estimated to be less 
than 1\%.
 
Equation \ref{VACfinal} was solved using Cholesky factorization for real
$s$-values and LU factorization for imaginary $s$-values. In both cases
pivoting was done using the minimum degree algorithm, and all computations 
was done using Matlab version 5.3.0.


\section{Numerical results, $D(s)$}
\label{sec:Ds}

In this section we report the results of numeric calculations
of $D(s)$ in 3 dimensions using the VAC method. Like in \cite{Dyre94} and
paper \HopPaper ~ the calculations are done with for 
real Laplace frequencies, $s$, i.e. corresponding to 
imaginary values of $\omega$ ($s = i\omega$). When comparing the numeric
results and analytical approximations for real values of $s$, we are 
relying on the principle of analytical continuation \cite{Arkfen95}; 
If two complex functions coincide on a line in the complex plane, they 
coincide everywhere in the complex plane (where they are both well-defined). 

In section \ref{sec:Dw} we present the corresponding results
for real values of $\omega$. Besides providing results directly 
comparable with experimental results for $\sigma(\omega)$, it will be clearly 
demonstrated that we can indeed trust the principle of 
analytical continuation when comparing analytical approximations
and numerical results.

It has already been demonstrated in \cite{Dyre94} and
paper \HopPaper ~ that the symmetric hopping model becomes
universal in the extreme disorder limit, i.e. that $\tilde D(\tilde s)$
(or equivalently $\tilde \sigma(\tilde s))$ becomes independent of 
temperature and the distribution of energy-barriers. Here we will focus
on the shape of $\tilde D(\tilde s)$, and only present results 
for the box-distribution of energy barriers: $p(E)=1, 0\le E \le 1$.

\begin{figure}
\epsfig{file=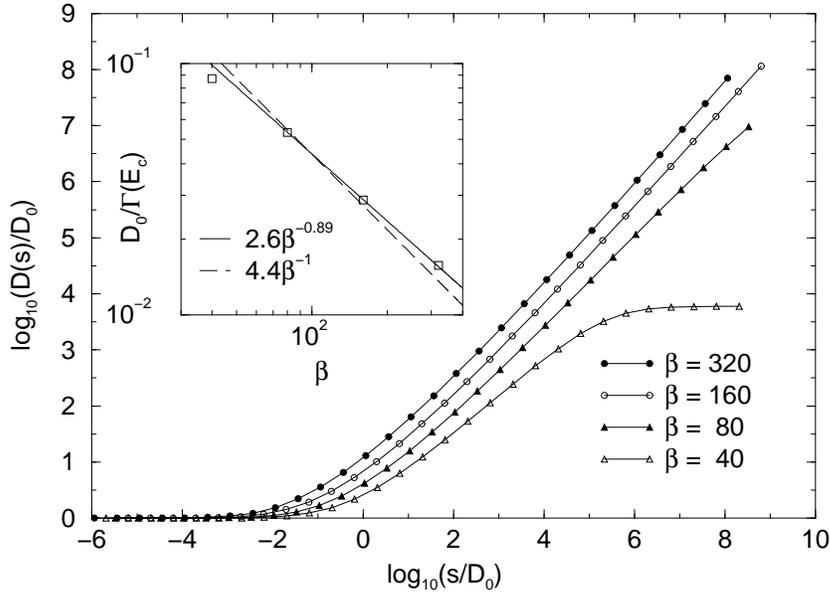, width=11cm}
\vspace{-.5cm}
\caption{The frequency dependent diffusion coefficient, $D(s)$,  vs. 
the (real) Laplace frequency, $s$, both scaled by $D_0$. 
Inset: $D_0/\Gamma(E_c)$ vs. $\beta$, a power-law fit, and a power-law
with exponent $-1$ for reference.
}
\label{fig:DsvsSoDo}
\end{figure}

Figure \ref{fig:DsvsSoDo} show the frequency dependent
diffusion coefficient for real Laplace frequencies, $D(s)$, in 
a log-log plot, for 4 $\beta$-values. Both axis in figure 
\ref{fig:DsvsSoDo} are scaled by the DC level, 
$D_0 \equiv D(s\rightarrow 0)$. 
In the inset of 
figure \ref{fig:DsvsSoDo} is shown $D_0$ scaled by $\Gamma(E_c)$
vs. $\beta$. For $\beta \ge 80$ $D_0$ is seen to be 
well approximated by:
\begin{equation}
   D_0 \propto \beta ^{-\gamma} \Gamma(E_c), ~~~ \gamma = 0.89 \pm 0.01   
\end{equation}
Note that the dominant $\beta$-dependence of $D_0$ is given by 
$\Gamma(E_c)$, which changes by 30 orders of magnitude for
the $\beta$-values used (see table \ref{GammaEcEtc}). In
the simulations reported in paper \HopPaper ~, 
we found: $\gamma = 0.81 \pm 0.04$ \cite{SchroederSpecialePhys}.

 
\begin{table}
\begin{tabular}{|c|c|c|c|c|} \hline
$\beta$       & $40$      & $80$       & $160$     & $320$ \\ \hline \hline
$N$           & $24$      & $32$       & $ 64$     & $ 96$ \\ \hline  
$\Gamma(E_c)$ & $4.8\cdot 10^{-05}$ & $2.3\cdot 10^{-09} $ & $5.1\cdot 10^{-18}$ & $2.6\cdot 10^{-35}$ \\ \hline
$D_0        $ & $4.2\cdot 10^{-06}$ & $1.2\cdot 10^{-10} $ & $1.5\cdot 10^{-19}$ & $4.2\cdot 10^{-37}$ \\ \hline
$s_c$         & $6.3\cdot 10^{-06}$ & $7.1\cdot 10^{-11} $ & $3.4\cdot 10^{-20}$ & $3.7\cdot 10^{-38}$ \\ \hline
\end{tabular}
\caption{Various parameters as a function of the $\beta$-values used.
The results reported are from averages over 100 NxNxN samples, where
N depends on $\beta$ as shown above.
}
\label{GammaEcEtc}
\end{table}

\begin{figure}
\epsfig{file=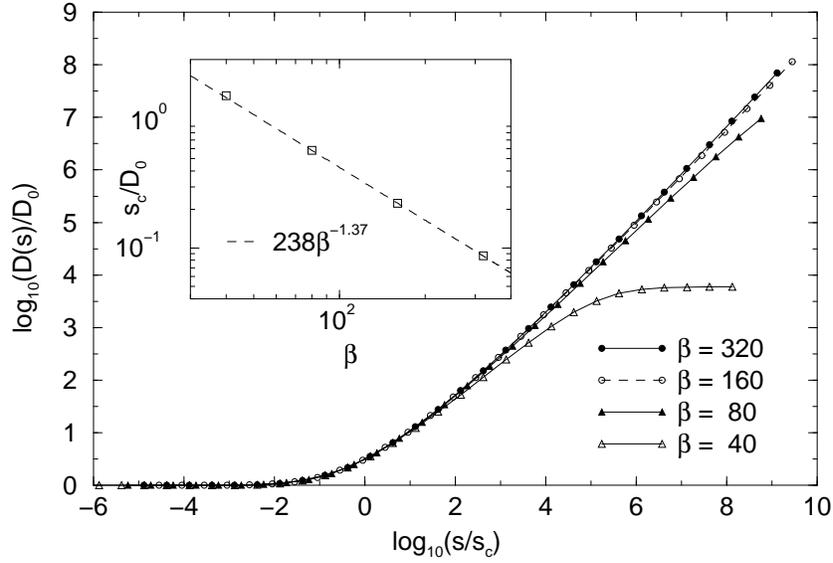, width=11cm}
\caption{Same data as in fig. \ref{fig:DsvsSoDo} but scaled on the 
frequency axis, to agree at 
$\log_{10}(D(s_c)/D_0) = 0.5$ Approach to 
universal curve is evident. Inset: The scaling parameter, 
$s_c$ divided by $D_0$ vs. $\beta$, and a fit to a power-law.   
}
\label{fig:DsvsSoSc}
\end{figure}

To illustrate how  $D(s)$ approaches a 
universal curve (suitable scaled) at high $\beta$-values, 
the data presented in figure \ref{fig:DsvsSoDo} was scaled
by the characteristic frequency, $s_c$, defined here by
$D(s_c)/D_0 = \sqrt{10}$. In figure \ref{fig:DsvsSoSc}
the result of scaling the data in this way is shown.
It is clearly seen that $\tilde D(\tilde s)$ approaches
a universal curve, as the $\beta$-values increases. 
Or in other words: the way the system approaches (long time)
diffusion becomes universal as the temperature is decreased.
The universal curve is estimated to be close to the data for $\beta = 320$,
and the frequency regime for which the data follows the universal 
curve is seen to increase as $\beta$ increases. 

In the inset of
figure \ref{fig:DsvsSoSc} is shown $s_c$ scaled by $D_0$
vs. $\beta$. For $\beta \ge 80$ $s_c$ is seen to be
well approximated by:
\begin{equation}
   s_c \propto \beta ^{-\eta} D_0, ~~~ \gamma = 1.37 \pm 0.01
\end{equation}
The dominant $\beta$-dependence of $s_c$ is given by
$D_0$, which changes by 31  orders of magnitude for
the $\beta$-values used (see table \ref{GammaEcEtc}). In
the  simulations reported in paper \HopPaper , the power law was less well
defined, and $\eta$ was estimated to be $1.48 \pm 0.06$ 
(depending on the $\beta$-range used in the power-law fit) 
\cite{SchroederSpecialePhys}.

For both of the scaling parameters, $D_0$ and $s_c$, the 
$\beta$-dependence given by the power-laws reported are 
very small compared to the $\beta$-dependence given by 
the other involved quantities. In the simulations presented 
here, the main focus has not been on determining the scaling 
exponents, $\gamma$ and $\eta$, precisely,  but on the \emph{shape} of the 
universal curve seen in figure  \ref{fig:DsvsSoSc}.

Before proceeding to compare the numeric results with the analytical approximations, 
we will briefly discuss the possibilities of computing $\langle  { \Delta X}^2(t) \rangle$ 
from $D(s)$. As discussed above we will compare numerical results and analytical approximations
for $D(s)$, and consequently we are calculating  $\langle  { \Delta X}^2(t) \rangle$ 
\begin{figure}
\epsfig{file=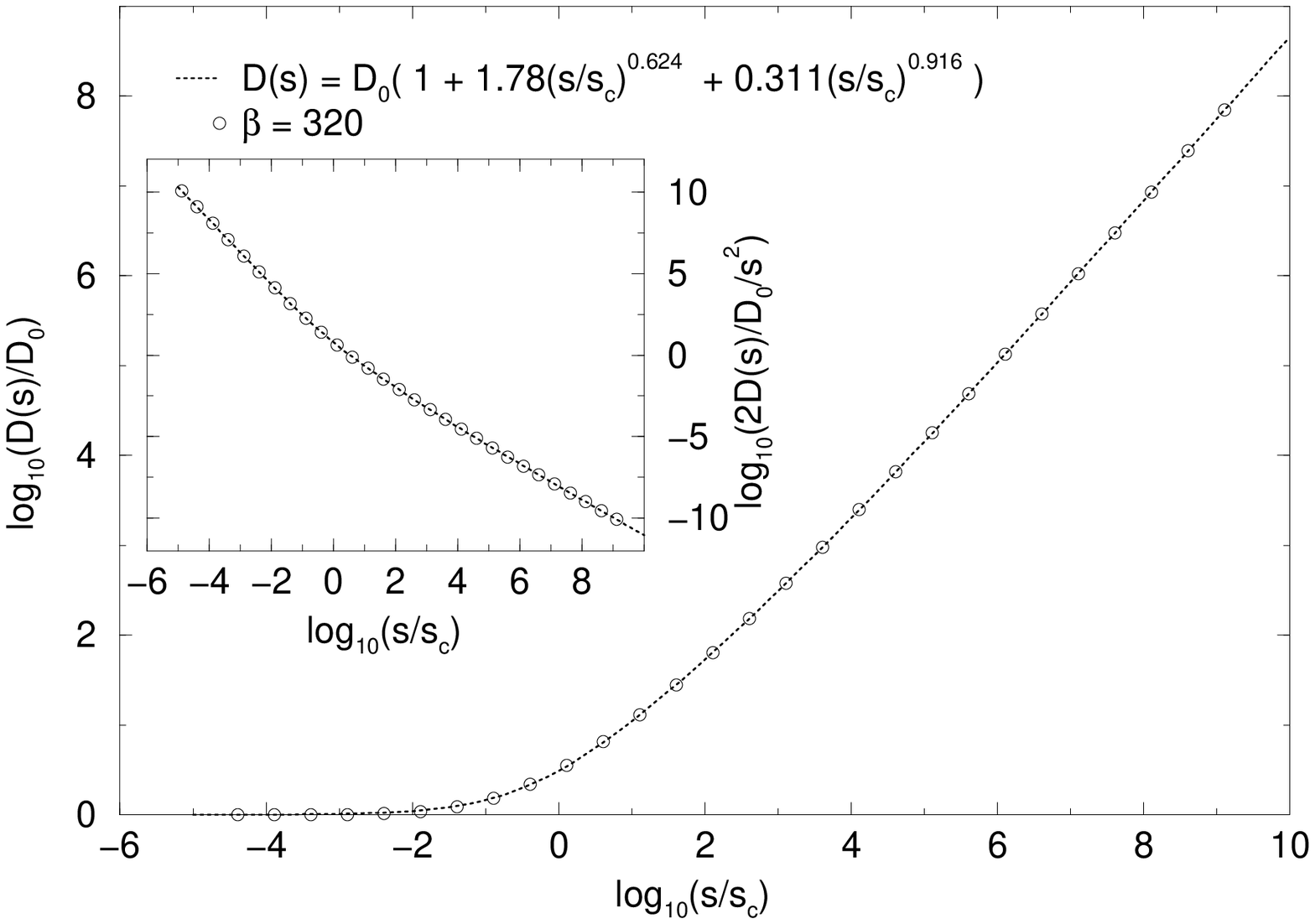, width=12cm}
\epsfig{file=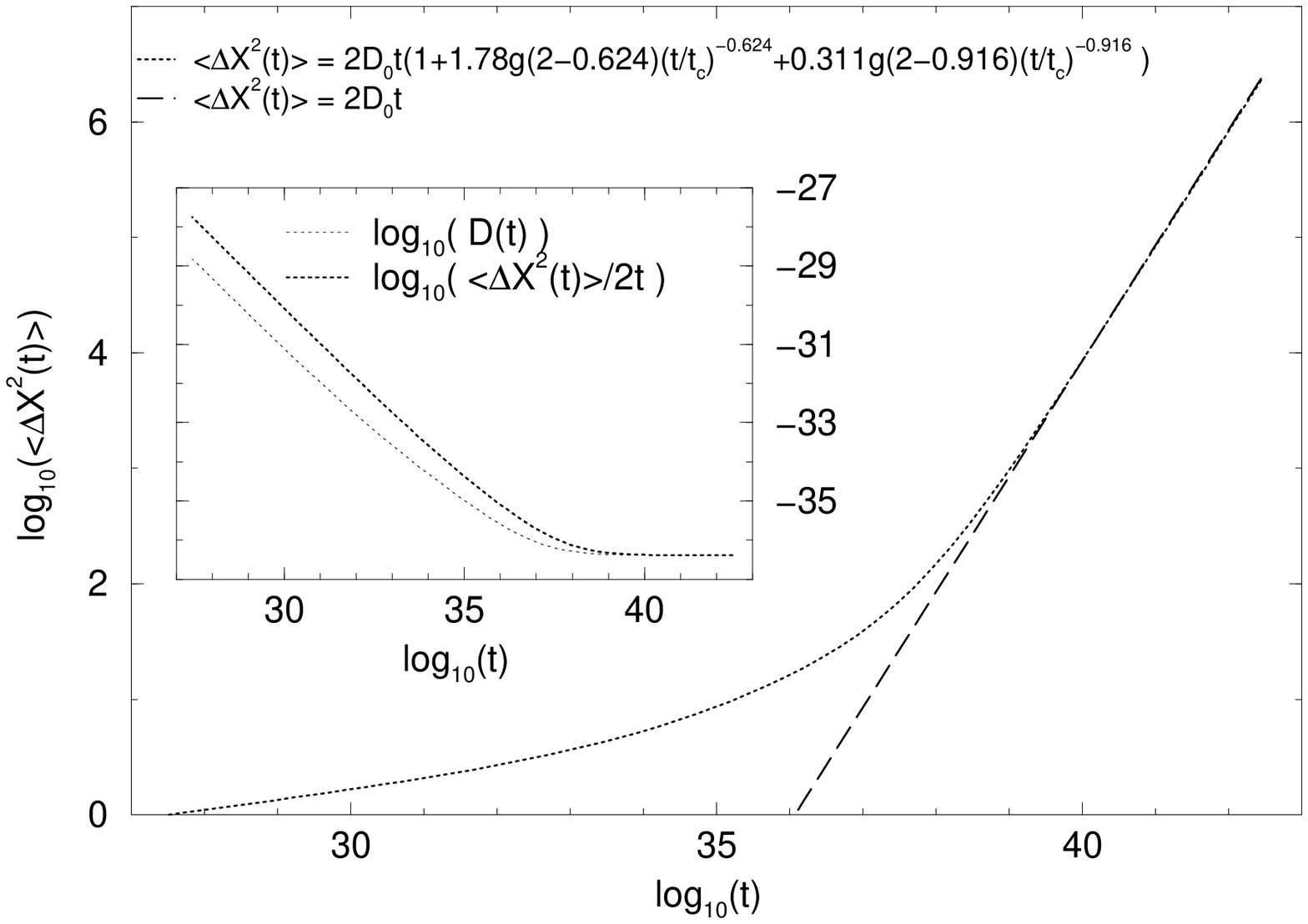, width=12cm}
\caption{
Estimating  $\langle  { \Delta X}^2(t) \rangle$ from $D(s)$. 
a) Fit to $D(s)$ to functional form 
so that $2D(s)/s^2$ (inset) has known inverse Laplace transform  b) Resulting estimate for 
$\langle  { \Delta X}^2(t) \rangle$, and (in inset) $D(t)$ and 
$\langle  { \Delta X}^2(t) \rangle/2t$ ($g(x)$ is here the 
gamma-function, normally denoted $\Gamma (x)$). 
}
\label{fig:R2t}
\end{figure}
\noindent only for
illustrative purposes. From equation \ref{DsX2} this is in principle straight forward; 
$\langle {\Delta X}^2(t) \rangle$ is the inverse Laplace transform of $2D(s)/s^2$. 
The general inverse Laplace transform is however in practice problematic \cite{Arkfen95}, 
and here we furthermore have the problem of the many decades of frequency/time involved.
Instead we use the following trick; we  fit $D(s)$ in such a
way that $2D(s)/s^2$ has a known inverse Laplace transform. This is done for $\beta=320$ in figure
\ref{fig:R2t}a, where $D(s)$ is fitted to a sum of power-laws. Note that this is a
purely empirical fit, and it does not necessarily have anything to do with the true 
functional form of $D(s)$. The inset in figure \ref{fig:R2t}a shows $2D(s)/s^2$ and
the corresponding fit, i.e. the quantity to which the inverse Laplace transform is applied.
The resulting   $\langle{\Delta X}^2(t) \rangle$ is shown in figure \ref{fig:R2t}b,
 without using the scaling parameters, so that the actual size of the quantities is seen 
($t_c \equiv 1/s_c$). Note that the system becomes diffusive at very long time scales; 
$t\approx 10^{40}$. This illustrates the impossibility of using traditional MC techniques, 
since this would require at least $10^{40}$ time steps. The inset in figure \ref{fig:R2t}b 
shows $D(t)$ and $\langle{\Delta X}^2(t)/2t \rangle$ as resulting from the 
procedure described above.

\begin{figure}
\epsfig{file=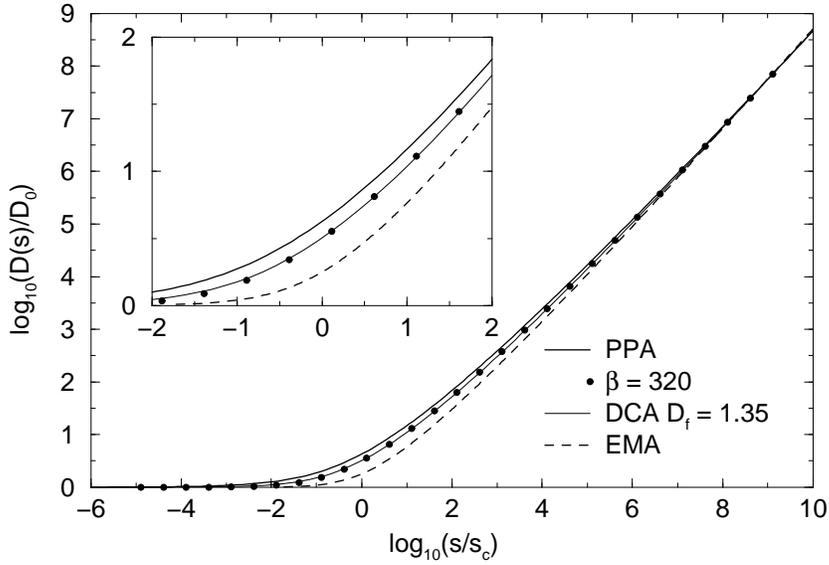, width=11cm}
\vspace{-.5cm}
\caption{Comparing the universal curve for $\tilde D(\tilde s)$ ($\beta=320$), 
with the 3 analytical approximations. 
The approximations were
scaled to fit the data at the highest frequencies. The fractal dimension,
$D_f = 1.35$  in the Diffusion Cluster Approximation (DCA) was
determined by fitting (see fig. \ref{fig:PotensApr}).
}
\label{fig:DsvsSoScApprox}
\end{figure}

In figure \ref{fig:DsvsSoScApprox} the universal curve for $D(s)$
(represented by the data for $\beta=320$) is compared
with the 3 analytical approximations described in section 
\ref{sec:AnalAprrox}. The fractal dimension $D_f$ used in 
the Diffusion Cluster Approximation (DCA) was treated as a fitting parameter. The 
fitting procedure will be explained when discussing figure
\ref{fig:PotensApr}. The fit of DCA is seen to be almost 
perfect, and clearly better than both PPA and EMA. This 
is perhaps not surprising given that DCA has a fitting 
parameter which PPA and EMA has not, and the value of
fit achieved by DCA thus greatly depends on whether
an independent argument can be given for the value of $D_f$.
For now we only note, that the value $D_f=1.35$ is in the expected range: $1.14 < D_f < 1.7$
(see section \ref{sec:AnalAprrox}).

\begin{figure}
\epsfig{file= 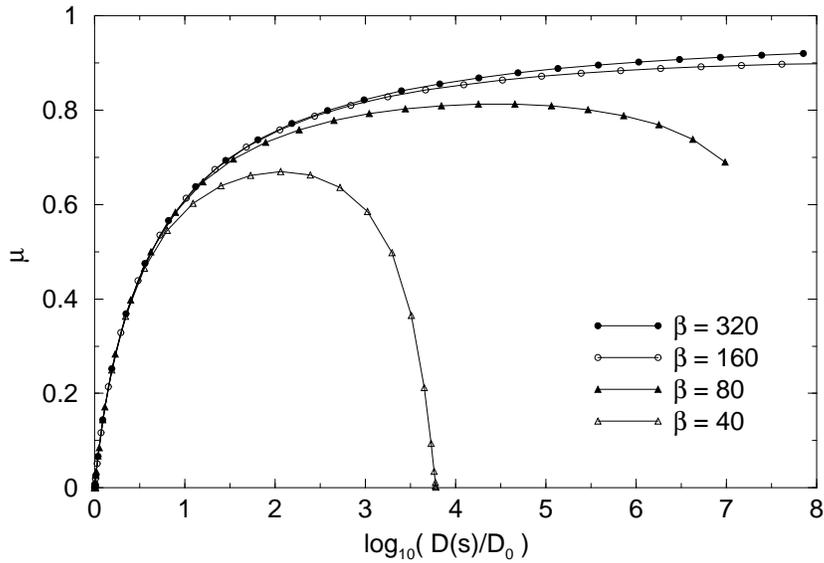, width=11cm}
\vspace{-.5cm}
\caption{
The exponent $\mu$ (eq. \ref{eq:Mu}) vs. $\log_{10}(D(s)/D_0)$. 
}
\label{fig:PotensBeta}
\end{figure}

\begin{figure}
\epsfig{file=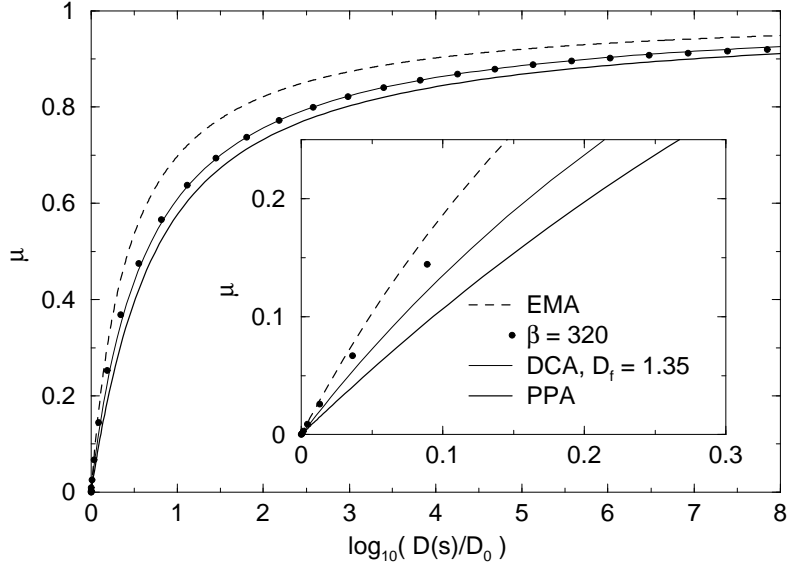, width=10.5cm}
\vspace{-.5cm}
\caption{Comparing $\mu$ for the universal curve (approximated 
by $\beta=320$) with the three analytical approximations. The fractal 
dimension, $D_f=1.35$, in DCA was fitted (by hand) to the 
numerical data in this plot.   
}
\label{fig:PotensApr}
\end{figure}

Like in paper \HopPaper, we now proceed to 
focus on the shape of $\tilde D (\tilde s)$, by calculating 
the apparent power-law exponent, 
i.e. the logarithmic derivative of $\tilde D (\tilde s)$:
\begin{equation}
   \mu \equiv \frac{d \log_{10}(D)}{d \log_{10}(s)}
   \label{eq:Mu}
\end{equation}
When plotting $\mu$ as a function of $\log_{10}(D(s)/D_0)$
we can compare the shapes of the data presented above, 
without relying on an empirical scaling of the frequency axis. 
This is done in figure \ref{fig:PotensBeta}, for the same data 
as shown in \ref{fig:DsvsSoSc}. Plotting the data in this 
way is seen to be more sensitive; The small difference 
between the data for $\beta=160$, and $\beta=320$ in figure 
\ref{fig:DsvsSoSc} is more pronounced in figure \ref{fig:PotensBeta}.

In figure \ref{fig:PotensApr} we compare the universal curve 
for $\mu$ (approximated by the data for $\beta=320$) with 
the 3 analytical approximations. As in figure \ref{fig:DsvsSoScApprox}
the data from the simulations is seen to lie between EMA and PPA, 
and DCA is seen to give a good fit. The fitting of DCA
was done by hand using this figure. 
The inset in figure \ref{fig:PotensApr} shows the same data, 
but focusing on the low values of $\tilde D$,
corresponding to low values of $\tilde s$ and $\mu$. In this 
limit, the data is seen to deviate from DCA, and seems to
approach EMA.

\begin{figure}
\epsfig{file= 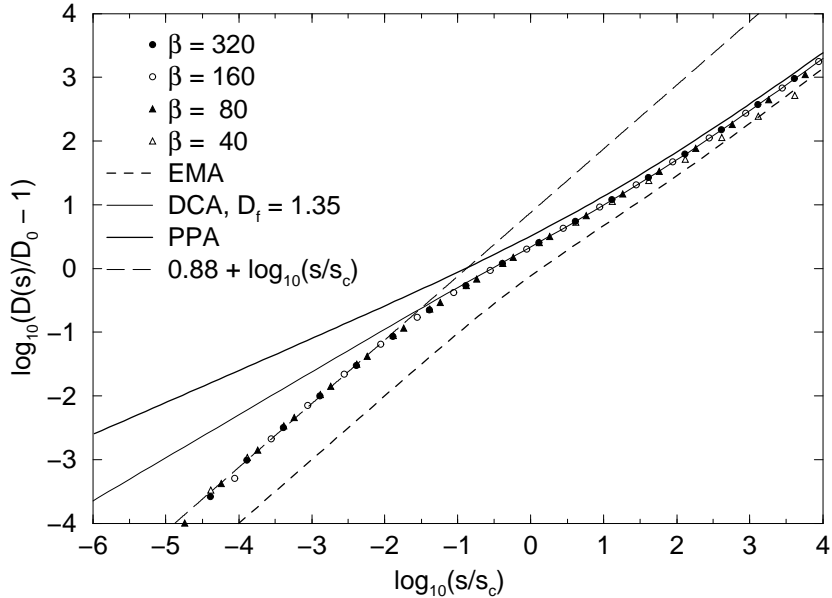, width=11cm}
\vspace{-.5cm}
\caption{Focusing on the very low frequencies: $D(s)/D_0 - 1$ vs. 
$s/s_c$. The universality for the numerical data holds even at the 
very low frequencies. The numerical data and EMA are well approximated
by $D(s) = D_0(1 + As)$, thus agreeing with the first two terms of 
eq. \ref{LongTT}. This does not hold for PPA and DCA.
}
\label{fig:DLowS}
\end{figure}

The behavior of $\tilde D(\tilde s)$ at low frequencies,
i.e. the departure from the DC level ($D_0$) is better
investigated by plotting $\tilde D(\tilde s) - D_0 $
as its done in figure \ref{fig:DLowS}. The universality for the 
numerical data is seen to hold even at the very low frequencies
in this plot. This demonstrates that the universality seen 
at low frequencies in $\tilde D(\tilde s)$ (figure \ref{fig:DsvsSoSc}),
is \emph{not} just a consequence of the DC level being 
dominant. On the other hand the apparent reasonable agreement 
between $\tilde D(\tilde s)$ and DCA seen in figure 
\ref{fig:DsvsSoScApprox}  breaks down when the DC level 
is subtracted. 

The low-frequency expansion of $D(s)$ is known to be \cite{Boettger85}:
\begin{eqnarray}
   \tilde D(\tilde s) = 1 + A\tilde s + B\tilde s^{3/2} + ... \label{LongTT}
\end{eqnarray}
In figure \ref{fig:DLowS} the numerical data and EMA is seen
to agree with the first two terms of this expansion.

\section{Numerical Results, $D(\omega)$
}
\label{sec:Dw}

In this section we present numerical results for $D(\omega)$, i.e.
for imaginary Laplace-frequency, $s=i\omega$. In this case 
the diffusion coefficient is a complex quantity, and we can 
compare both the real and imaginary part with the analytical 
approximations. 

\begin{figure}
\epsfig{file= 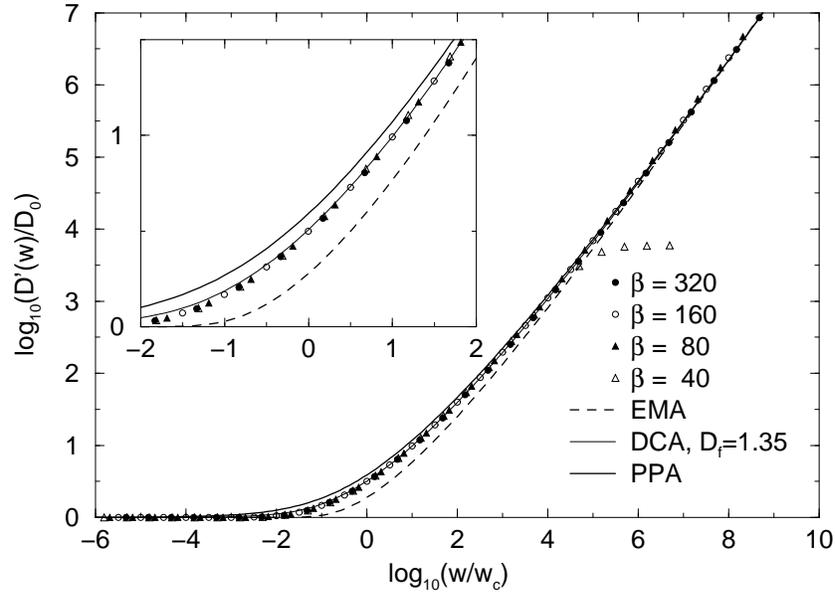, width=11cm}
\vspace{-.5cm}
\caption{Real part of $D(\omega)$ vs. $\omega$, both scaled
(see fig. \ref{fig:ImagScale}). $D_f=1.35$ was used in DCA, like for 
real Laplace frequencies (fig. \ref{fig:PotensApr}), i.e. it was not
fitted to these data. 
}
\label{fig:RealDw}
\end{figure}

In figure \ref{fig:RealDw} is shown the (scaled) real part of $D(\omega)$
vs. the frequency (scaled). The scaling parameters, $D_0$ and $\omega_c$
is shown in figure \ref{fig:ImagScale}. Like we found in the previous 
section for  real Laplace frequencies the  data is here seen to approach 
a universal curve which agrees well with DCA, with $D_f=1.35$. This 
is the value found in the previous section from $D(s)$; it was not found to be 
necessary to make a new fit for $D(\omega)$.

\begin{figure}
\epsfig{file= 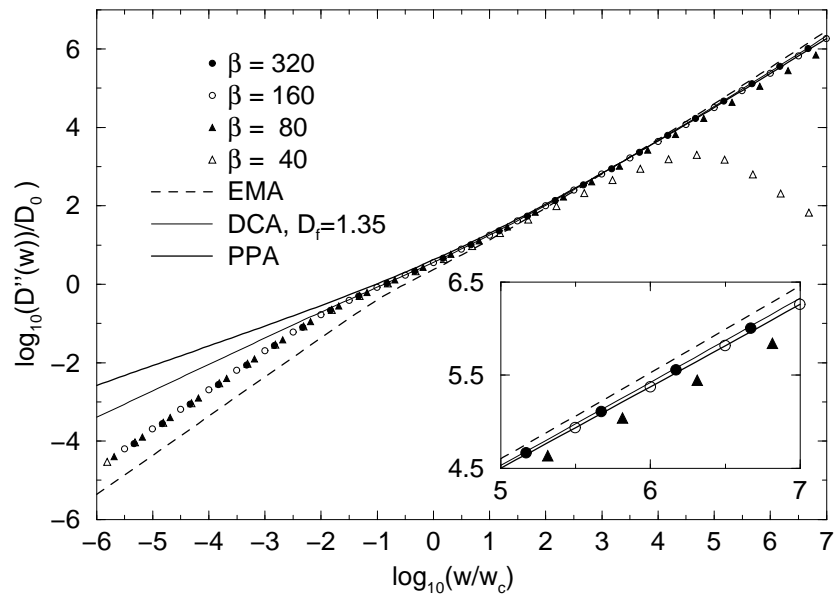, width=11cm}
\vspace{-.5cm}
\caption{Imaginary part of $D(\omega)$ vs. $\omega$, both scaled as
in fig. \ref{fig:ImagDw}.  The numerical data and EMA agrees 
 with the first two terms of eq. \ref{LongTT}.
}
\label{fig:ImagDw}
\end{figure}

\begin{figure}
\epsfig{file= 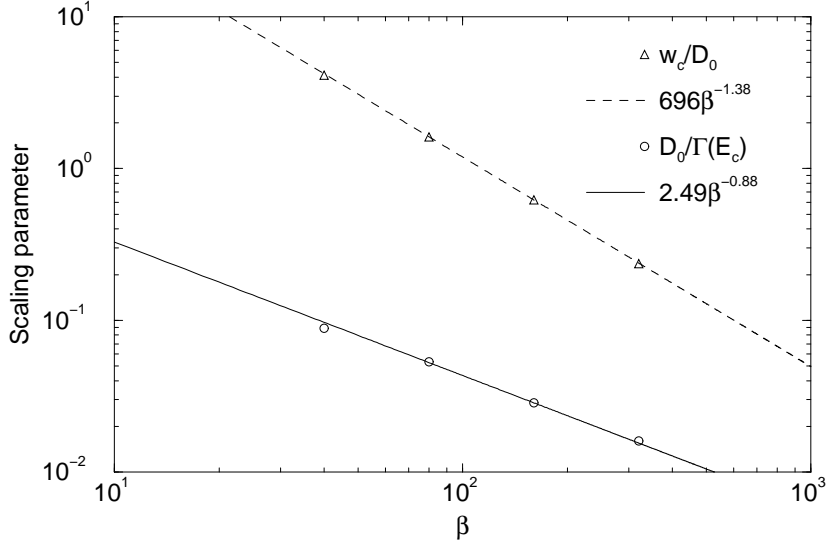, width=11cm}
\vspace{-.5cm}
\caption{Scaling parameters used for data with imaginary 
Laplace frequencies, $s=i\omega$. Note the agreement  with the 
scaling parameters used for real $s$ 
(
 fig. \ref{fig:DsvsSoDo}, and 
fig. \ref{fig:DsvsSoSc}).
}
\label{fig:ImagScale}
\end{figure}

Figure \ref{fig:ImagDw} shows the imaginary part of $D(\omega)$
corresponding to the data in  figure \ref{fig:RealDw}, and using the
same scaling parameters (and $D_f$).  Universality is evident at 
low frequencies (agreeing with EMA, apart from scaling),
and is approached at higher frequencies. 

\begin{figure}
\epsfig{file= 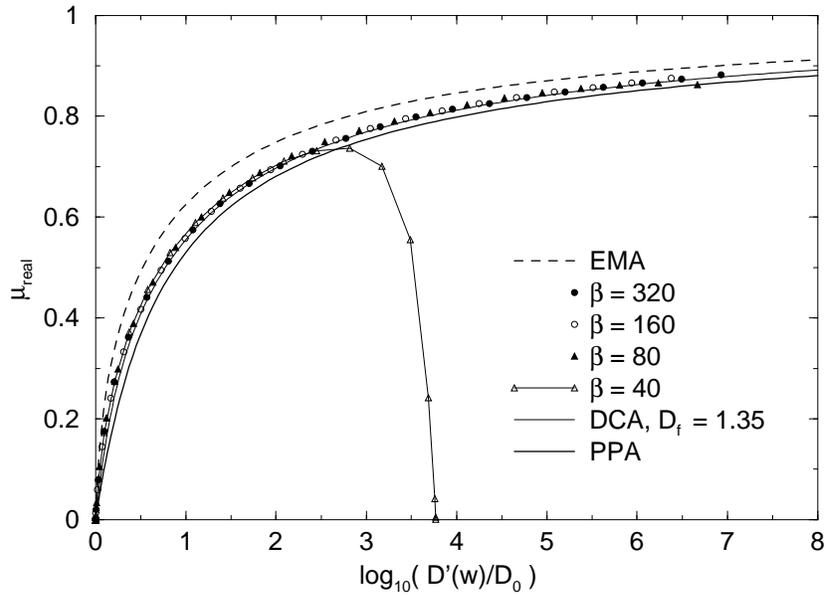, width=11cm}
\vspace{-.5cm}
\caption{$\mu_{real}$ vs. the real part of $D(\omega)$. 
}
\label{fig:MuReal}
\end{figure}

As an analogue to the (apparent) exponent $\mu$ (eq. \ref{eq:Mu}) used for $D(s)$ in the 
previous section, we can define an exponent for the real part of $D(\omega)$:
\begin{equation}
   \mu_{real} \equiv \frac{d \log_{10}(D'(\omega))}{d \log_{10}(\omega)}
   \label{eq:MuReal}
\end{equation}
$\mu_{real}$ is plotted in figure \ref{fig:MuReal} as a function of $\omega$.
Note that the convergence toward universality is more ``abrupt'' than 
it was found for $D(s)$ (fig. \ref{fig:PotensBeta}); Only the data
for $\beta=40$ deviates significantly from DCA.

\begin{figure}
\epsfig{file= 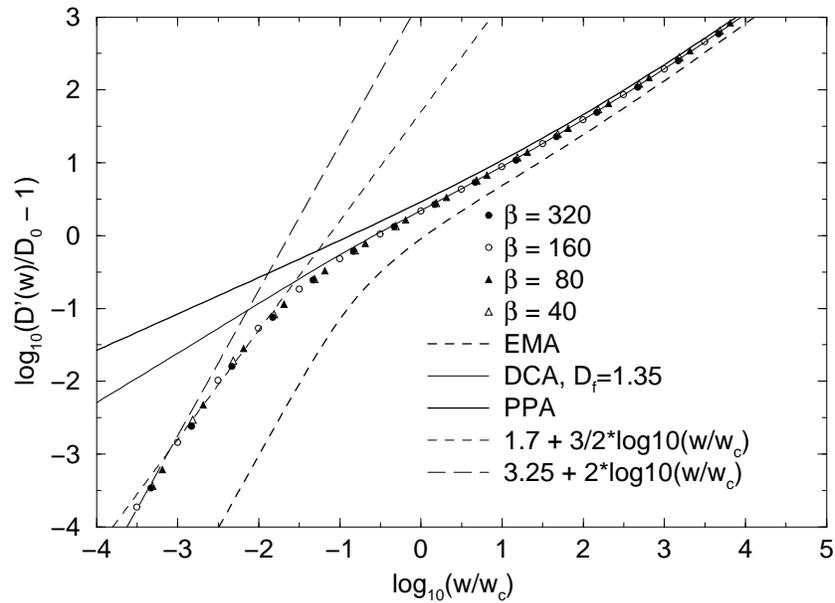, width=11cm}
\vspace{-.5cm}
\caption{Focusing on low frequencies: $D'(w)/D_0 - 1$ vs $\omega/\omega_c$.
Dashed thin curve: low frequency expansion (exponent
$3/2$, eq. \ref{LongTT}). 
}
\label{fig:DwM1}
\end{figure}

In figure \ref{fig:DwM1} we focus on the low frequency behavior
of $D'(\omega)$ by subtracting the DC level, which gives us the chance
to check the agreement with third term in the low-frequency expansion 
(Eq. \ref{LongTT}), i.e. the exponent $3/2$.
Neither of the approximations agrees with this, which in particular means that 
EMA only agrees with the first 2 terms in the low-frequency expansion.
The numerical data seems to agree better with the exponent $3/2$, 
although it is difficult to judge the numerical significance of this.

\begin{figure}
\epsfig{file= 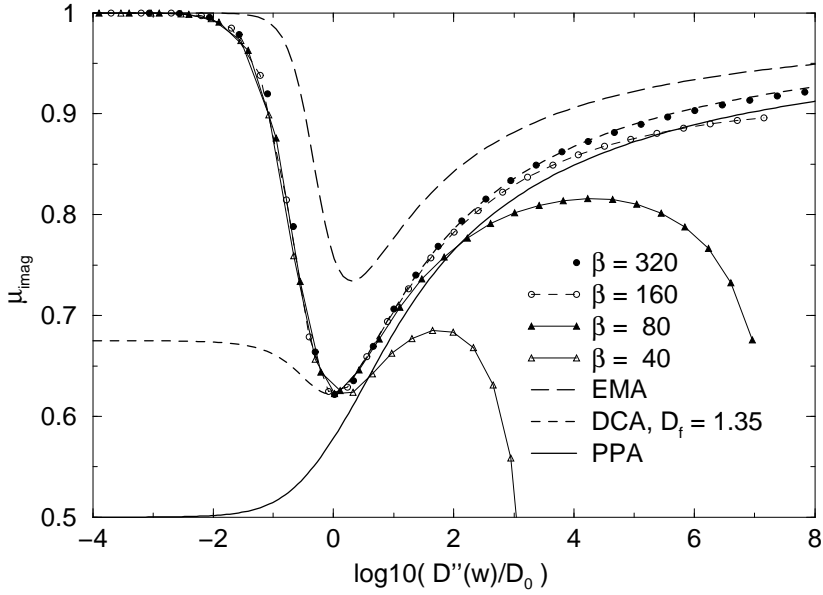, width=11cm}
\vspace{-.5cm}
\caption{The exponent $\mu_{imag}$ (eq. \ref{MuImag}). 
}
\label{fig:MuImag}
\end{figure}

We define the apparent exponent for the imaginary part of $D(\omega)$
in the following way:
\begin{equation}
   \mu_{imag} \equiv \frac{d \log_{10}(D''(\omega))}{d \log_{10}(\omega)}
   \label{MuImag}
\end{equation}
This is plotted as a function of $D''(\omega)$ in figure \ref{fig:MuImag}.
EMA has the right 
value at the very low frequencies, as seen in fig. \ref{fig:ImagDw}.
DCA has the right behavior from $D''(\omega)\ge D_0$, with small 
deviations at the highest values (notice the y-axis being different
from the other exponent-plots).

\begin{figure}
\epsfig{file= 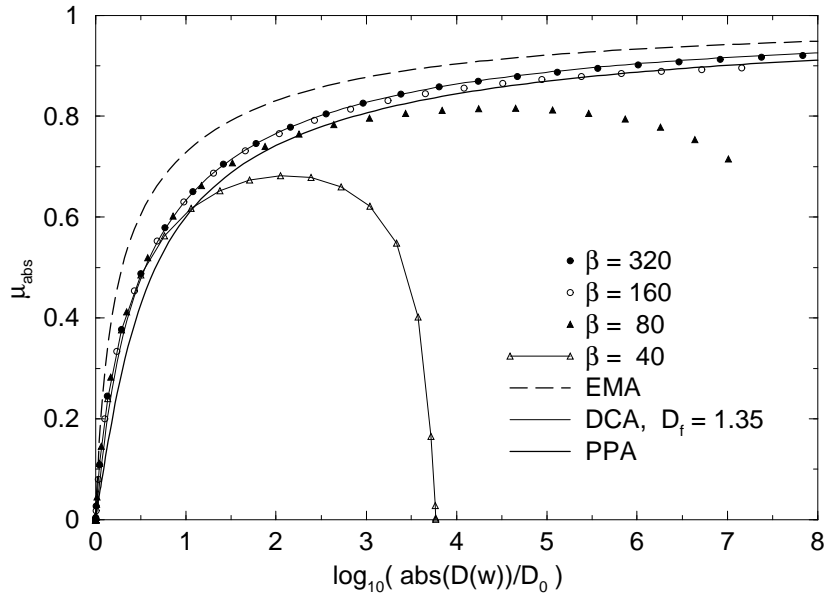, width=11cm}
\vspace{-.5cm}
\caption{The exponent $\mu_{abs}$ (eq. \ref{eq:MuAbs}). The 
approach to universality is almost identical to the one seen 
for real Laplace frequencies (fig. \ref{fig:PotensApr}).
}
\label{fig:MuAbs}
\end{figure}

In the data presented  so far we have found much the same behavior in 
$D(\omega)$ as for $D(s)$ with regards to how the numerical data 
agrees with the analytical approximations; at high frequencies the 
numerical data falls between EMA and PPA and is well fitted by 
DCA with $D_f=1.35$, whereas the very low frequencies is governed 
by the low frequency expansion (eq. \ref{LongTT}), 
which (for the first 2 terms) agrees with EMA. In contrast the approach to   
 universality seems to different for especially $D(s)$ and $D'(\omega)$.
The quantity that approaches universality in a similar
manner as $D(s)$ is the absolute value of $D(\omega)$.
This is illustrated in figure \ref{fig:MuAbs} where we plot the apparent 
exponent for the absolute value of $D(\omega)$:
\begin{equation}
   \mu_{abs} \equiv \frac{d \log_{10}(|D(\omega)|)}{d \log_{10}(\omega)}
   \label{eq:MuAbs}
\end{equation}

\section{VAC vs. ACMA}
\label{sec:ACMAvsVAC}

\begin{figure}
\epsfig{file= 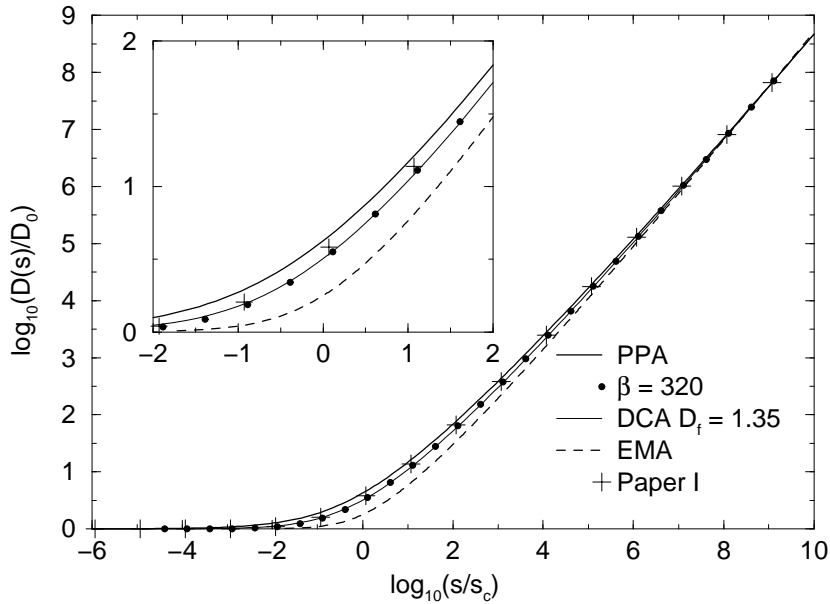, width=11cm}
\vspace{-.5cm}
\caption{Comparing $D(s)$ for $\beta=320$ calculated by  the VAC 
method (fig. \ref{fig:DsvsSoScApprox}) and the ACMA method 
(fig. 1 in paper \HopPaper). A small but significant difference is found.
}
\label{fig:PotensACMA}
\end{figure}

In figure \ref{fig:PotensACMA} the universal curve for $D(s)$ as 
calculated from the VAC method (fig. \ref{fig:DsvsSoScApprox})
and the ACMA method (fig. 1 in paper \HopPaper) is compared. There is a small but
significant difference between the results from the two methods, as
can also be seen by comparing the apparent exponents from the two
methods (fig. \ref{fig:PotensBeta} and fig. 2 in paper \HopPaper).
The main difference between the two methods is the boundary 
conditions, so the differences in these are the ``main suspect'' 
for the (slightly) different results. At infinitely large samples
we would expect the results of both methods to converge to the bulk-limit.
The way the two methods converge to the bulk limit might however be different;
In the ACMA method there is some fraction of the sites that are close
to an electrode, and thus give a wrong contribution to $D(s)$. As the 
sample size increases this fraction decreases, and the results converges
to the true bulk-limit. In the VAC method there are no sites that 
are ``worse'' than the other, and the finite size effects are more subtle;
they arise from particles traveling trough the sample to where they 
started, and thus experiencing false correlations.

\begin{figure}
\epsfig{file= 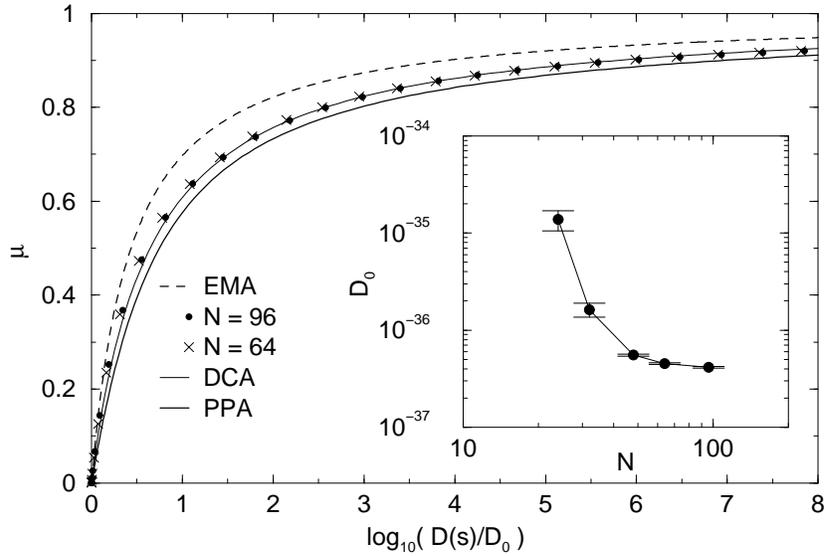, width=11cm}
\vspace{-.5cm}
\caption{$\mu$ for $\beta=320$ with $N=96$ (Average of 100 samples)
and with $N=64$ (Average of 600 samples). Inset: N-dependence
of $D_0$.
}
\label{fig:VAC_N}
\end{figure}

In figure \ref{fig:VAC_N} we compare the apparent exponent $\mu$ for $\beta=320$
as calculated with $N=96$ (i.e. the data in fig. \ref{fig:PotensBeta}) and 
$N=64$ (averaged over 600 samples). The agreement is seen to be excellent, 
with the largest error at the very low frequencies. The inset in figure \ref{fig:VAC_N}
shows the N-dependence of $D_0$, with error-bars estimated from the standard
deviation. The tendency to converge to a constant as $N$ increases is evident. 

In figure \ref{fig:VAC_N} is shown $\mu$ calculated  independently 
for each of the 100 (96x96x96) samples at $\beta=320$, i.e. without
averaging. 
The noise shows that the samples are not 
self-averaging, i.e. it \emph{is} necesarry to average over a number of samples. 
The data points are distributed evenly around 
DCA, showing that the averaging over (non self-averaging) 
samples do not introduce 
systematic errors (compare fig. \ref{fig:PotensApr})

\begin{figure}
\epsfig{file= 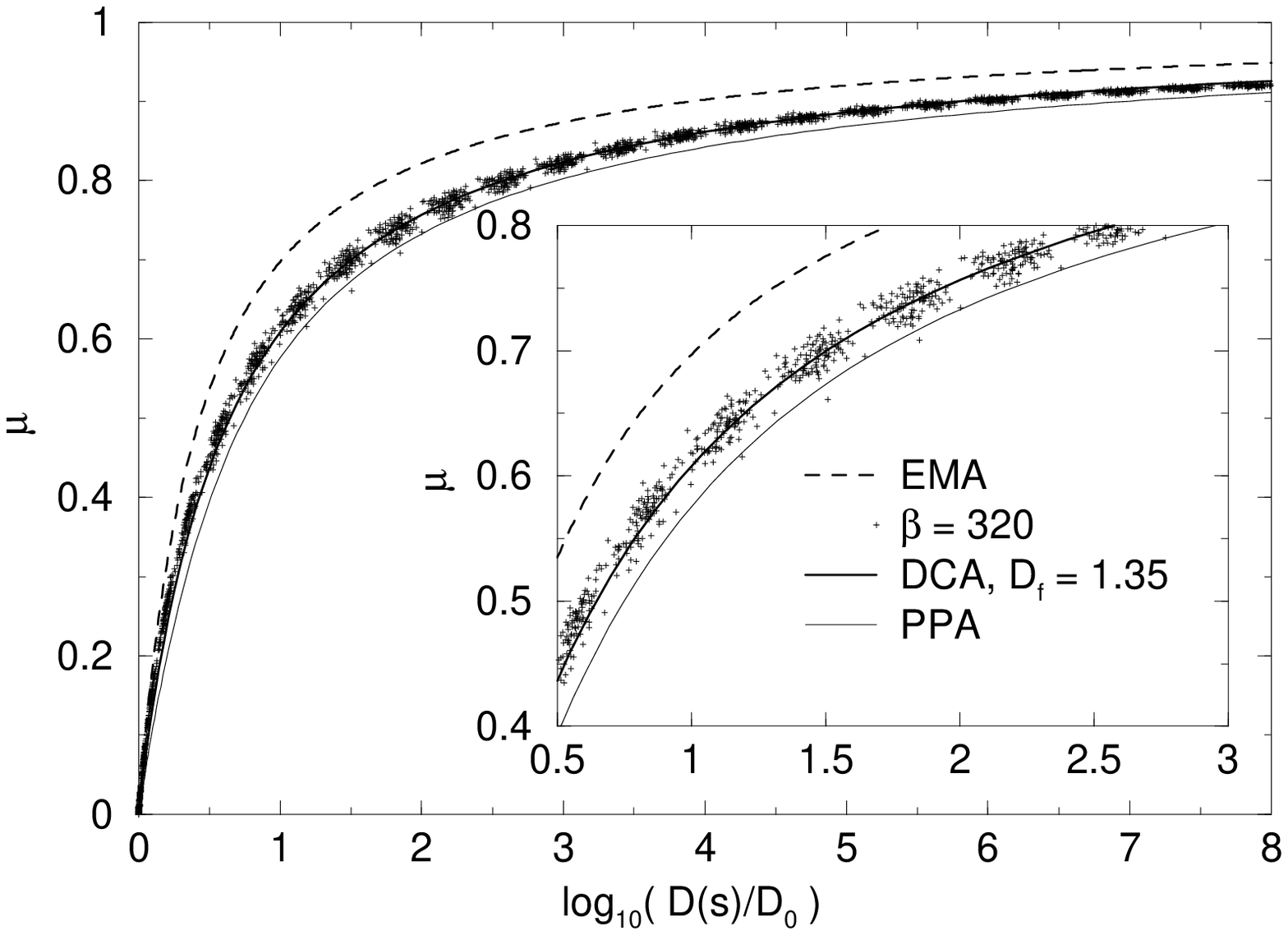, width=11cm}
\vspace{-.5cm}
\caption{$\mu$ calculated independently for each of the 100 
(96x96x96) samples at $\beta=320$. 
}
\label{fig:fig1}
\end{figure}

\section{Conclusions}
The main result in this chapter is the development of the 
Velocity Auto Correlation (VAC) method. At first it might seem 
strange to work in terms of velocities in a hopping model, as it
is evident from the following quote from Scher \& Lax \cite{Scher73}:

\begin{quote}
 The relation as it stands [ i.e. eq. \ref{Dvac} in the present work, 
relating $D(s)$ to the velocity auto correlation function ] is inconvenient 
to use in a hopping model since it refers to velocities rather than 
positions.
\end{quote}

It should be evident by now, that it is worth to suffer the (initial)
inconvenience of working with velocities; The VAC method is clearly 
to prefer to the ACMA method, since it can be used with periodic 
boundary conditions, and still give a diffusive regime. The physical 
reason for this being possible is that the diffusive regime is 
characterised by loss of correlations, which is possible in a finite 
sample (as opposed to the mean square displacement being proportional to 
time). 

The numerical results was found to be in excellent agreement with 
the Diffusion Cluster Approximation, except for the very low 
frequencies where the low frequency expansion holds. The agreement with DCA  was 
achieved by a (single) fit to the fractal dimension of the 
diffusion cluster, $D_f=1.35$, which is within the limits expected
from percolation arguments:  $1.14 < D_f < 1.7$ (see section \ref{sec:AnalAprrox}).
Two questions needs to be answered, to decide whether the agreement between 
the numerical data and DCA signals that the picture behind the approximation 
is correct, or if it is simply a result of fitting:

\begin{itemize}
\item Is DCA a good approximation of diffusion on a diffusion cluster 
      with fractal dimension $D_f$? This corresponds to the check of
      PPA in 1 dimension shown in figure 2a in paper \HopPaper . 
\item Is the diffusion cluster  (in 3 dimensions) a fractal with fractal dimension (close to) 
      1.35?
\end{itemize}
If either of the answers to these questions are negative, then the agreement
found between numerical data and DCA is  a coincidence, and 
DCA is merely a convenient way to describe the data. It is left as future work
to answer these questions.

\appendix
\chapter{Numbering scheme for VAC method}

The sites in the $\mathcal{D}$-dimensional regular lattice
are numbered by:

\begin{eqnarray}
   \mbox{SiteIndex} = \sum_{j=1}^{\mathcal{D}}(C_j - 1)N^{j-1} + 1
\end{eqnarray}
where $C_j$ is the coordinate  (counted from 1 to N) 
of the site in the $j$'th direction. 


The ($N^{\mathcal{D}} \times N^{\mathcal{D}}$) matrix $\mathbf H_k$ used in 
section \ref{sec:VACDD} (eq. \ref{Jk}) describes the connectiviy in the lattice;
$(\mathbf H_k)_{ij}$ equals $-1$ if site $j$ is the 
``left'' neighbor to site $i$ along direction $k$, and it has $1$ on the 
diagonal and zero everywhere else. 
With the numbering scheme given above,  $\mathbf H_k$ is given by
$\mathbf I \otimes \mathbf I \otimes ...   \mathbf A  ... 
\otimes \mathbf I \otimes \mathbf I$, 
where $\mathbf I$ is the  ($N \times N$) identity matrix, $\otimes$
is the direct (Kronecker) multiplication, and the matrix 
$\mathbf A$ is at the $j$'th position from the right\footnote{
The  ($N \times N$) matrix $\mathbf A$ is here defined as in section \ref{sec:VAC1D}:
$(\mathbf A)_{ij} = \delta (i,j) - \delta (i-1,j)$ (with periodic boundary conditions)}.

\newpage
In 2 dimensions with N=3 we eg. have:
\begin{eqnarray}
    \mathbf H_1 &\equiv& \mathbf I  \otimes  \mathbf A =
     \left(
       \begin{array}{c c c}
          \mathbf A & \mathbf 0 & \mathbf 0\\
          \mathbf 0 & \mathbf A & \mathbf 0 \\
         \mathbf 0 & \mathbf 0 & \mathbf A \\
   \end{array}
  \right)\\ &=& 
     \left(
       \begin{array}{r r r r r r r r r}
          1 & 0 & -1 & 0 & 0 & 0 & 0 & 0 & 0 \\
          -1 & 1 & 0 & 0 & 0 & 0 & 0 & 0 & 0 \\
          0 & -1 & 1 & 0 & 0 & 0 & 0 & 0 & 0 \\
          0& 0& 0&1 & 0 & -1 & 0 & 0 & 0 \\
          0& 0& 0&-1 & 1 & 0 & 0 & 0 & 0 \\
          0& 0& 0&0 & -1 & 1 & 0 & 0 & 0 \\
          0& 0& 0&0& 0& 0&1 & 0 & -1  \\
          0& 0& 0&0& 0& 0&-1 & 1 & 0  \\
          0& 0& 0&0& 0& 0&0 & -1 & 1  \\
       \end{array}
  \right) \\
    \mathbf H_2 &\equiv& \mathbf A  \otimes  \mathbf I =
     \left(
       \begin{array}{r r r}
          \mathbf I &  \mathbf 0 & -\mathbf I\\
          -\mathbf I &  \mathbf I &  \mathbf 0 \\
          \mathbf 0 & -\mathbf I &  \mathbf I \\
   \end{array}
  \right)\\ &=& 
     \left(
       \begin{array}{r r r r r r r r r}
          1 & 0 & 0 & 0 & 0 & 0 &-1 & 0 & 0 \\
          0 & 1 & 0 & 0 & 0 & 0 & 0 &-1 & 0 \\
          0 & 0 & 1 & 0 & 0 & 0 & 0 & 0 &-1 \\
         -1 & 0 & 0 & 1 & 0 & 0 & 0 & 0 & 0 \\
          0 &-1 & 0 & 0 & 1 & 0 & 0 & 0 & 0 \\
          0 & 0 &-1 & 0 & 0 & 1 & 0 & 0 & 0 \\
          0 & 0 & 0 &-1 & 0 & 0 & 1 & 0 & 0  \\
          0 & 0 & 0 & 0 &-1 & 0 & 0 & 1 & 0  \\
          0 & 0 & 0 & 0 & 0 &-1 & 0 & 0 & 1  \\
       \end{array}
  \right) 
\end{eqnarray}

\bibliography{Thesis}
\bibliographystyle{unsrt}

\end{document}